\documentclass[twocolumn,twocolappendix,times,tighten,trackchanges]{aastex631} 
\usepackage[version=3]{mhchem}
\usepackage{comment}
\usepackage{chngcntr}

\newcommand{\mum}{$\rm \mu \mathrm{m}~$}

\newcommand{\OIII}{[\textsc{Oiii}]\,}
\newcommand{\OII}{[\textsc{Oii}]\,}
\newcommand{\HII}{H{\sc ii}~}
\newcommand{\HI}{H{\sc i}~}
\newcommand{\CII}{[\textsc{Cii}]\,}
\newcommand{\NII}{[\textsc{Nii}]\,}
\defcitealias{Sugimura:2024}{S24}

\usepackage{color}

\shorttitle{Multi-phase Gas Emission in the First Galaxies}
\shortauthors{Nakazato et al.}

\newcommand{\figdir}{./}

\begin{document}

\title{Unveiling the Ionized and Neutral ISM at $z > 10$ : The Origin of \OIII/\CII Ratios from a Sub-parsec Resolution Radiative Transfer Simulation
}

\correspondingauthor{Yurina Nakazato}
\email{ynakazato@flatironinstitute.org}
\author[0000-0002-0984-7713]{Yurina Nakazato}
\affiliation{Center for Computational Astrophysics, Flatiron Institute, 162 5th Avenue, New York, NY 10010}
\affiliation{Department of Physics, The University of Tokyo, 7-3-1 Hongo, Bunkyo, Tokyo 113-0033, Japan}
\author[0000-0001-7842-5488]{Kazuyuki Sugimura}
\affiliation{Faculty of Science, Hokkaido University, Sapporo, Hokkaido 060-0810, Japan}
\author[0000-0002-7779-8677]{Akio K. Inoue}
\affiliation{Waseda Research Institute for Science and Engineering, Faculty of Science and Engineering, Waseda University, 3-4-1 Okubo, Shinjuku, Tokyo 169-8555, Japan}
\affiliation{Department of Physics, School of Advanced Science and Engineering, Faculty of Science and Engineering, Waseda University, 3-4-1, Okubo, Shinjuku, Tokyo 169-8555, Japan}
\author[0000-0003-4223-7324]{Massimo Ricotti}
\affiliation{Department of Astronomy, University of Maryland, College Park, MD 20742, USA}

\begin{abstract}
Recent multi-wavelength observations by JWST and ALMA are unveiling both ionized and neutral ISM components in high-redshift ($z>6$) galaxies. In this work, we investigate the origin of rest-frame far-infrared \OIII88\mum and \CII158\mum emission by performing zoom-in cosmological simulations of dwarf-galaxy progenitors at $z=9-13$. Our simulations incorporate on-the-fly radiative transfer at sub-pc ($\sim$ 0.1 pc) resolution, allowing us to resolve the multi-phase ISM. We compute emission lines on a cell-by-cell basis, taking into account local temperature, density, metallicity, radiation field strength, column density, and spectral hardness of radiation bins.
We find that \OIII predominantly arises from centrally located ionizing bubbles with temperatures of $\sim (1-5)\times 10^4\,\mathrm{K}$ and high ionization parameters of \(\log U_{\mathrm{ion}} \simeq -1.5\). In contrast, \CII is produced in the surrounding dense neutral regions at $\sim 5\times 10^3\,\mathrm{K}$, which are heated by strong FUV radiation ($G/G_0 \sim 10^{3-5}$) from the central stellar clusters. This spatial arrangement leads to large local variations in \OIII/\CII, ranging from $\sim$ 100 to 0.01. Our galaxy reproduces the global ratio $\OIII/\CII \sim5-30$, consistent with recent ALMA detections at $z>6$ without invoking enhanced O/C abundance ratios. 
We further derive that \OIII/\CII linearly scales with the mass and density ratios of ionized to neutral gas, $M_{\rm HII}/M_{\rm HI}$ and $n_{\rm HII}/n_{\rm HI}$ and show that the \OIII/\CII ratio typically changes from 5.7 to 0.3 from high-z to low-z. For future synergies of JWST and ALMA, we derived $M_{\rm HII}/M_{\rm HI}$ for observed $z >6$ galaxies using ${\rm H}\beta$ and \CII and show the validity of our scaling relations. 
\end{abstract}


\section{Introduction} \label{sec:intro}
The interstellar medium (ISM) plays a crucial role in galaxy formation and evolution by regulating star formation, controlling chemical enrichment, and mediating feedback processes. Although studies of local galaxies have built the basics for ISM physics, understanding how the ISM behaves at $z>6$ is essential for investigating the earliest stages of galaxy assembly during/before the Epoch of Reionization. High-redshift galaxies often exhibit extreme conditions (e.g., lower metallicities, intense radiation fields) that can strongly affect the physical and chemical states of the ISM.

Among the various ISM tracers, the far-infrared (FIR) lines \OIII 88\mum and \CII 158\mum have been used most significantly for high-redshift studies. The \OIII line traces ionized gas surrounding massive star-forming regions, whereas the \CII line originates both in neutral photodissociation regions (PDRs) and mildly ionized gas, 
though it is thought to mostly originate in PDRs \citep{Stacey:2010, Gullberg:2015, Vallini:2015, Lagache:2018, Cormier:2019}. Observations of these FIR lines with the Atacama Large Millimeter/Submillimeter Array (ALMA) have opened a new era for investigating the physical conditions in star-forming regions of $z>6$ galaxies.

Recent ALMA observations have successfully detected \OIII 88\mum and \CII 158\mum in galaxies at $z>6$ \citep[e.g.,][]{Inoue:2016}, and even up to $z>10$ for some JWST spectroscopically confirmed targets \citep{Schouws:2024, Carniani:2024, Zavala:2024}. Interestingly, these galaxies exhibit high \OIII/\CII ratios of 1-20, while local galaxies tend to show ratios of 0.04-1 \citep{Howell:2010, De_Looze:2014, Diaz-Santos:2017}. In the highest-redshift galaxy GS-z14-0, non-detection of \CII implies a lower limit of \OIII/\CII $> 3.5$ \citep{Schouws:2025}.

These findings suggest that high-redshift galaxies might host unique ISM conditions. Several explanations have been proposed, such as lower PDR covering fractions, higher ionization parameters, lower metallicities, and an enhanced oxygen-to-carbon abundance ratio \citep[e.g.,][]{Harikane:2020,  Katz:2022, Nyhagen:2024}. However, the origin of the observed line ratios remains under debate.

Theoretical studies have attempted to explain these puzzling observations by modeling line emission with semi-analytical suites \citep{Lagache:2018, Popping:2019}, cosmological zoom-in simulations focusing on small galaxy samples (snapshots) \citep[e.g.,][]{Vallini:2015, Vallini:2017, Katz:2019, Olsen:2017, Pallottini:2019, Arata:2020, Schimek:2024, Nyhagen:2024}, and simulations with large samples of galaxies \citep{Katz:2022, Pallottini:2022, Ramod-Padilla:2023}. However, most simulations do not solve on-the-fly radiative transfer (RT) coupled with chemical networks, forcing us to make a lot of assumptions to model multi-phase ISM. A few simulations solve on-the-fly RT, but their spatial resolutions are $\sim$ 10-20 pc scale, which may overlook small-scale ISM, affecting their ionizing state. Despite the variety of theoretical modeling approaches, not all models have been successful at reproducing the high \OIII/\CII ratios observed at $z > 6$. 
Indeed, the typical size of \HII regions around a single OB type star can be estimated by the Str\"omgren radius,
\begin{equation}
    R_{\rm S} \;=\;
    1.85\,\mathrm{pc}\,\left(\frac{n_{\rm HII}}{500\,\mathrm{cm^{-3}}}\right)^{-2/3}
    \left(\frac{Q(\mathrm{H})}{5\times10^{49}\,\mathrm{s^{-1}}}\right)^{1/3}, \label{eq:stromgren_raius}
\end{equation}
where $n_{\rm HII}$ and $Q(\mathrm{H})$ is the gas number density of \HII region and the ionizing photon number per unit time emitted from the source stars, respectively. Eq. (\ref{eq:stromgren_raius}) shows that for typical densities observed in the ISM at high-z, sub-pc resolution simulations are needed to resolve the multi-phase ISM accurately. 

In this paper, we present a new set of zoom-in simulations at 0.1\,pc scale resolution with on-the-fly RT, enabling us to accurately capture the multi-phase ISM and compute emission lines from rest-frame UV to rest-frame FIR. We focus on redshifts $z > 10$ to identify the origin of the \OIII/\CII ratio in high-redshift galaxies. We investigate the spatial distribution of \OIII/\CII,  time evolution of \OIII/\CII, and an analytical toy model of multi-phase gas. By examining the synergy between ALMA FIR lines (\OIII, \CII) and JWST optical lines (e.g., H$\beta$, \OIII 4959, 5007\AA), we demonstrate how multi-phase gas conditions can be inferred from observational data, offering insights into the physical and feedback processes governing the early Universe.

The remainder of this paper is structured as follows. In Section \ref{sec:method}, we describe our simulation setup and line calculation methods. Section \ref{sec:result} presents the resulting emission line properties and compare these results with observations and other theoretical works. In Section \ref{sec:discussion}, we discuss the redshift evolution of \OIII/\CII from high-z to low-z and investigate the synergy with JWST rest-frame optical lines. Finally, Section~\ref{sec:summary} summarizes our conclusions and outlines future prospects.

\section{Methods} \label{sec:method}
\subsection{Zoom-in simulation}\label{subsec:zoom_in_simulation}
The simulations employed in this work were originally introduced in \citet{Garcia:2023,Garcia:2025} and \citet{Sugimura:2024} (hereafter \citetalias{Sugimura:2024}). We run a cosmological zoom-in radiation-hydrodynamics simulation targeting a dwarf galaxy. Our calculations use the \textsc{RAMSES-RT} code \citep{Teyssier:2002, Rosdahl:2013}, which employs Adaptive Mesh Refinement (AMR) and solves radiative transfer using the M1 closure scheme. The M1 implementation in \textsc{RAMSES-RT} can reproduce shadowing structures reasonably well, as demonstrated by \citet[][Figs. 16-18]{Rosdahl:2013}.

We begin with a low-resolution, dark-matter-only simulation within a comoving volume of $35\,h^{-1}\,\mathrm{Mpc}$ on a side. The initial conditions are generated at $z = 127$ with \textsc{MUSIC} \citep{Hahn_Abel:2011}. 
From this simulation, we select a dark matter halo with $M_{\rm vir} = 8.8\times 10^9\, M_\odot$ at $z=0$. This halo grows to a mass of $10^8\, M_\odot$ by $z\sim 10$, eventually becoming a dwarf galaxy in the Local Group by $z=0$ \citep{Ricotti:2022}. 

We then perform a zoom-in simulation focusing on this halo. The refined region covers $\sim 300\,h^{-1}\,\mathrm{ckpc}$ on each side, with an initial resolution of $2\,h^{-1}\,\mathrm{ckpc}$, and dark matter particles of mass $800\,M_\odot$. We use the \textsc{ROCKSTAR} halo finder \citep{Behroozi:2013} to identify halos. We impose a requirement of at least 300 particles per halo, which can resolve the minimum halo mass of $2.4\times10^5\,M_\odot$, comparable to minihalos expected to host the first star formation \citep[e.g.,][]{Yoshida:2003}. We evolve the simulation down to $z=9.5$.

AMR refinement takes place on the fly. A cell is flagged for refinement if it contains more than eight dark matter or star particles. Additionally, the Jeans length must be resolved by $N_{\rm J}$ cells. \citet{Truelove:1997} argued that the Jeans length should be resolved by at least four cells $(N_{\rm J} = 4)$ to prevent artificial fragmentation. Specifically, for cells with $2.4 \, {\rm pc} [(1+z)/10]^{-1} \leq \Delta x \leq 77\, {\rm pc} [(1+z)/10]^{-1}$, we set $N_{\rm J} = 8$, and for cells with $0.3 \, {\rm pc} [(1+z)/10]^{-1} \leq \Delta x \leq 1.2\, {\rm pc} [(1+z)/10]^{-1}$,
we set $N_{\rm J} = 4$. At the highest refinement level ($l = 25$), the minimum physical cell size is therefore $\Delta x_{\rm min}=0.15 \,{\rm pc}[(1+z)/10]^{-1}$.

We split the radiation into four energy bins: Lyman-Werner (FUV, $11.2$--$13.6\,\mathrm{eV}$), H-ionizing (EUV1, $13.6$--$24.6\,\mathrm{eV}$), He-ionizing (EUV2, $24.6$--$54.4\,\mathrm{eV}$), and He$^+$-ionizing (EUV3, $54.4$--$200\,\mathrm{eV}$). While our four-bin setup (three EUV $+$ one FUV) is similar to those used in other on-the-fly RT cosmological simulations \citep[e.g.,][]{Kannan:2022, Pallottini:2022, Katz:2024}, this binning may still introduce mild spectral hardening effects, as discussed by \citet{Baumschlager:2024}. However, such effects become important primarily near He\textsc{iii} region boundaries where the optical depth transitions sharply. Since our analysis focuses on galaxy-integrated line luminosities, the overall impact of spectral hardening on our conclusions should be negligible. For dust attenuation, we employ a metallicity-dependent cross section, $\sigma_{\rm d, eff}=4\times10^{-21} \, {\rm cm^2} (Z/Z_\odot)$ assuming the constant dust-to-metal ratio (DTM). We note that the exact DTM values at $z > 6$ are not yet constrained observationally, though some theoretical models provide predictions \citep[e.g.,][]{Dayal:2022, Tsuna:2023, Toyouchi:2025}. However, even if the DTM value were varied by a factor of a few, the FUV emission would remain optically thin to the gas and would not significantly change the resulting line luminosities.

Photoionization and photoheating are modeled using a seven-species, non-equilibrium chemical network that tracks $
\mathrm{H},\;\mathrm{H^{+}},\;\mathrm{e^{-}},\;\mathrm{H_{2}},\;\mathrm{He},\;\mathrm{He^{+}},\;\mathrm{He^{++}}$,
following \citet{Park:2021}. We assume that $\mathrm{H^-}$ remains in equilibrium with these species. We also incorporate absorption of FUV radiation in self-shielded cells \citep{Park:2021} and include dust photoelectric heating \citep{Kimm:2017}. Lyman-Werner self-shielding in dense regions is treated using the fitting formula of \citet{Draine_Bertoldi:1996}. Metal and dust cooling/heating rates are calculated as in \citet{Katz:2017}, using tables from \textsc{Grackle} \citep{Smith:2017}. Unlike \citet{Katz:2017}, we do not apply a sub-grid clumping factor for $\mathrm{H_2}$ formation on dust, assuming this process is directly captured by our higher spatial resolution.

Star formation occurs in cells at the maximum refinement level ($l=25$) whose gas density satisfies
\begin{equation}
n \;>\; n_{\rm crit} 
= 5.0 \times 10^4\, \mathrm{cm}^{-3}
\left(\frac{T}{100\, \mathrm{K}}\right)\!
\left(\frac{1+z}{10}\right)^{2}\!
\left(\frac{N_{\rm cr}}{4}\right),
\label{eq:ncrit}
\end{equation}
which arises from the Jeans criterion $\Delta x_{\rm min} <\lambda_{\rm J}/N_{\rm cr}$ in physical units. This corresponds to densities on the order of $10^{4-5}\,\mathrm{cm}^{-3}$ and adequately resolves gravitational collapse of star-forming gas (see Fig.~9 of \citetalias{Sugimura:2024} for the density--temperature phase diagram). Once a cell is marked for star formation, the following procedure is used to generate star particles.
\begin{figure*}[t]
    \centering
    \includegraphics[width = \linewidth, clip]{\figdir/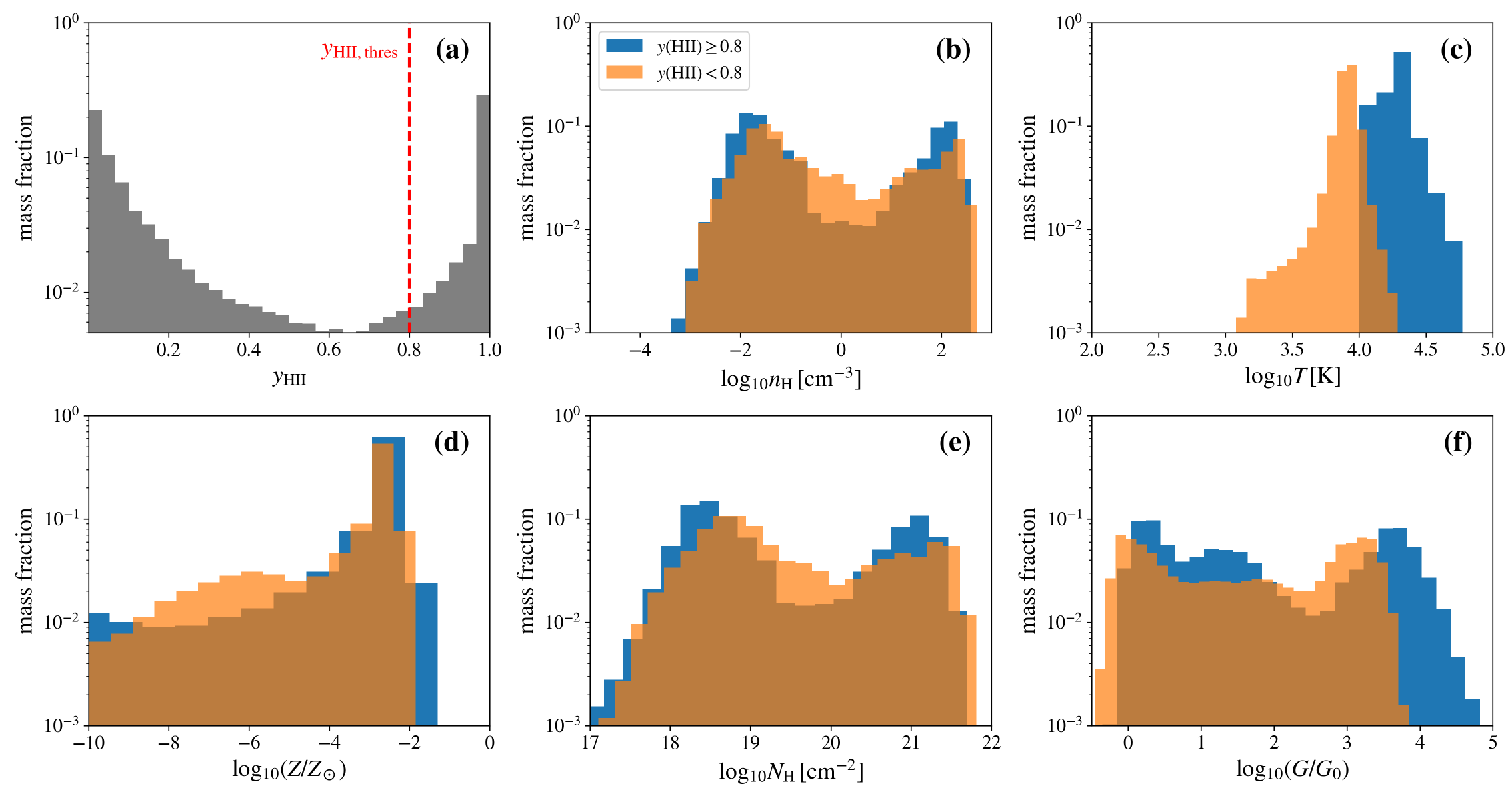}
    \caption{ Histograms of various physical properties of simulated cells at $z=10.5$ (snapshot~208), which marks the largest SFR during the burst of star formation. Panels show (a) ionization fraction, (b) gas number density, (c) gas temperature, (d) gas-phase metallicity, (e) gas column density, and (f) FUV radiation field. For panels~(b)--(f), the orange histograms correspond to PDR cells ($y_{\rm HII} < 0.8$), and the blue histograms correspond to \HII cells ($y_{\rm HII} \geq 0.8$). The histograms are weighted by gas mass. Panel (a) is normalized by the total gas mass within the virial radius, whereas in panels (b)-(f), the histograms for ionized and neutral gas are weighted by the total ionized and neutral gas mass, respectively.
}
    \label{fig:parameter_histgram}
\end{figure*}
\begin{figure}
    \centering
    \includegraphics[width = 0.8\linewidth, clip]{\figdir/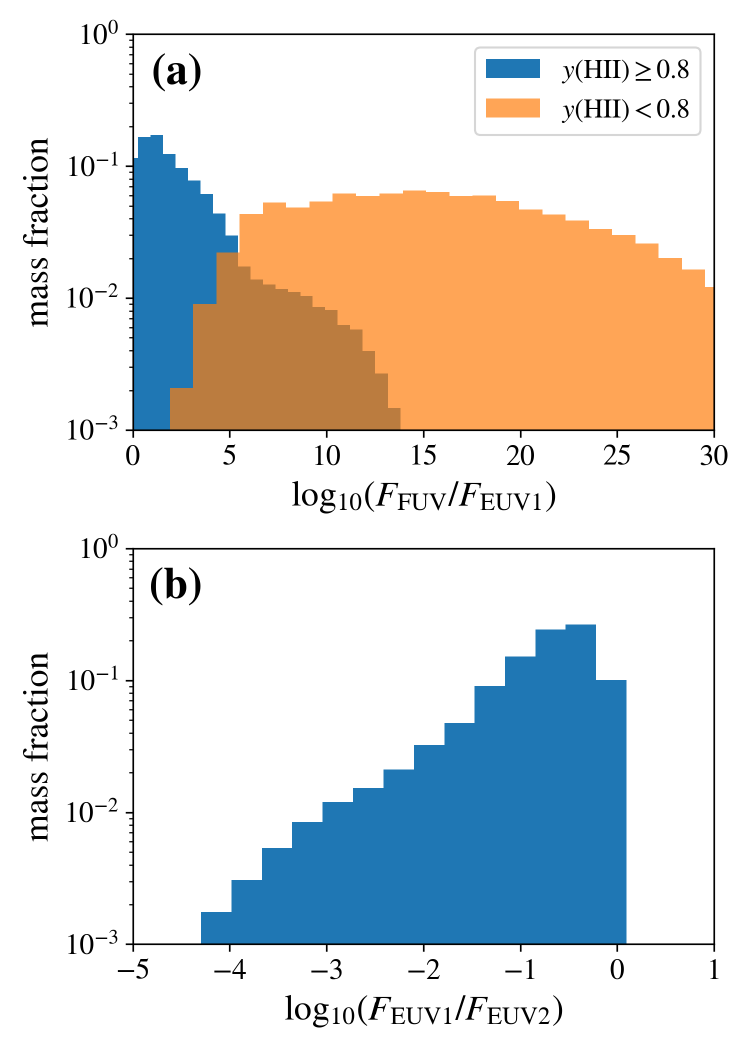}
    \caption{ Histograms of the radiation field at $z=10.5$ (snapshot~204). 
(a) The flux ratio of FUV to EUV1. Orange and blue histograms correspond to 
PDR and \HII cells, respectively. We find that the FUV flux in PDR cells 
exceeds the EUV flux by $\sim 20$ orders of magnitude, making the EUV 
component negligible. (b) The flux ratio of EUV1 to EUV2 for \HII cells. As in Figure \ref{fig:parameter_histgram}, the distributions are weighted by gas mass and normalized by the total ionized and neutral gas mass, respectively.
}
    \label{fig:radiation_histgram}
\end{figure}

If the metallicity is below $Z_{\rm crit} = 10^{-5}\, Z_\odot$, we create a single PopIII star particle representing a binary of $40\,M_\odot$ and $80\,M_\odot$ stars, as recent radiation hydrodynamic simulations suggest that Pop III stars form in massive binary systems \citep{Sugimura:2020,Sugimura:2023}. Both the $40\, M_\odot$ and $80\, M_\odot$ stars emit UV photons following \citet{Schaerer:2002} for 4~Myr. The $40\, M_\odot$ star eventually undergoes a hypernova, releasing $3\times10^{52}\,\mathrm{erg}$ of thermal energy and leaving a $20\,M_\odot$ black hole  \citep{Woosley:2002, Wise:2012}. The $80\,M_\odot$ star collapses directly into a black hole without an explosion, resulting in the formation of a $100\,M_\odot$ black hole binary modeled as a single sink particle. We track subsequent black hole growth via Bondi-Hoyle-Lyttleton accretion but neglect mechanical and X-ray feedback from these BHs. Thermal or kinetic energy from BHs can suppress star formation, while X-ray emission may enhance star formation by boosting ${\rm H_2}$ formation \citep[e.g.,][]{Ricotti:2016}. In our simulation, however, the BH mass reaches only $\sim 100\, M_\odot$ at $z = 9$ \citep[see Figure 17 of ][]{Sugimura:2024}, which is too small for the corresponding mechanical or X-ray feedback to play a significant role.

If $Z > Z_{\rm crit}$, we form PopII star clusters by spawning multiple $100\,M_\odot$ PopII star particles, each of which represents an unresolved stellar population following Salpeter IMF \citep{Salpeter:1955} in the range $1$--$100\,M_\odot$. The total mass of a new cluster, $M_{\rm cluster}$, is computed as $M_{\rm cluster} = f_*\, M_{\rm cloud}$, where $M_{\rm cloud}$ is the gas mass in cells with $n > 10^{-3} n_{\rm peak}$ around the density peak ($n_{\rm peak}$ is the peak density). The star formation efficiency $f_*$ is given by
\begin{align}
    f_* &= {\rm min} [
    0.8, 0.004\left(\frac{Z_{\rm cloud}}{10^{-3} Z_\odot}\right)^{0.25} \left(\frac{M_{\rm cloud}}{10^4\, M_\odot}\right)^{0.4} \notag \\
    &\times \left(1 + \frac{n_{\rm cloud}}{100\, {\rm cm}^{-3}}\right)^{0.91} ]
 , \label{eq:SFE}
\end{align}
where $Z_{\rm cloud}$ and $n_{\rm cloud}$ are the average metallicity and density of the cloud. This prescription is based on the results of radiation hydrodynamic simulations of star cluster formation \citet{He:2019}. For radiation from PopII stars, we use the stellar population synthesis models of \citet{Bruzual_Charlot:2003}. Each PopII star particle (of $100\,M_\odot$) contains approximately one massive star ($\gtrsim 8\,M_\odot$), which eventually explodes as a supernova (SN). We assign the SN event randomly between 4~Myr and 40~Myr \citep{Leitherer:1999}, with an ejecta mass of $10\,M_\odot$, a thermal energy of $10^{51}\,\mathrm{erg}$, and a metal mass $M_{\rm metal, II}=0.5\, M_\odot$ \citep{Kimm:2015}.

Regarding convergence of resolution, lowering the resolution (i.e., 1, 5 pc) would prevent us from resolving the gravitational collapse of star-forming gas clouds, which is essential for the star formation model adopted in our simulation. This would therefore introduce inconsistencies that are likely to affect the results. A detailed justification of the resolution dependence, including direct comparisons with higher-resolution runs, will be presented in a forthcoming paper (Sugimura et al., in prep.). For the AMR resolution, the typical cell size is $\sim$ 0.2-4 pc in dense gas with $n_{\rm gas} \gtrsim 100 \, {\rm cm^{-3}}$ and $\sim$ 4-100 pc in diffuse gas with $n_{\rm gas} \gtrsim 1\, {\rm cm^{-3}}$. A more detailed description of our simulation is found in \citetalias{Sugimura:2024}. 

\subsection{Line calculation} \label{subsec:line_calculation}
\begin{table*}[htbp]
\centering
\caption{Parameter ranges for PDR cells ($y_{\rm HII} < 0.8$) and \HII cells ($y_{\rm HII} \ge 0.8$) used in the CLOUDY calculations.}
\label{table:cloudy_parameter}
\begin{tabular}{lcc}
\hline
& \textbf{PDR cells} ($y_{\rm HII} < 0.8$) & \textbf{\HII cells} ($y_{\rm HII} \ge 0.8$) \\
\hline\hline
$\log (G/G_0)$ & 0, 1, 2, 3, 4 & 1, 2, 3, 4, 5 \\
$\log (n/\mathrm{cm}^{-3})$ & -2, -1, 0, 1, 2, 3, 4 & -2, -1, 0, 1, 2, 3 \\
$\log (Z/Z_\odot)$
 & -4, -3, -2, -1, 0
 & -4, -3, -2, -1, 0 \\
$\log (T/\mathrm{K})$
 & 2, 2.5, 3, 3.5, 4
 & 4, 4.5, 5 \\
$\log (N_{\rm H}/\mathrm{cm}^{-2})$
 & 17, 18, 19, 20, 21, 22
 & 17, 18, 19, 20, 21, 22 \\
$\log (F_\mathrm{FUV}/F_\mathrm{EUV1})$
 & 
 (considered only FUV radiation)
 & 0, 1, 2, 3, 4, 5 \\
$\log (F_\mathrm{EUV1}/F_\mathrm{EUV2})$
 &  
(considered only FUV radiation)
 & -2, -1, 0, 1 \\
\hline
\end{tabular}
\end{table*}

We estimate metal line luminosities using the spectral synthesis code CLOUDY \citep[v23.02]{Ferland:2017, Chatzikos:2023}. For each simulation cell, we have the temperature, density, metallicity, and local radiation field. Ideally, we would run CLOUDY for every cell by specifying an input SED based on the four radiation bins, the absolute radiation field strength, gas density, gas temperature, gas metallicity, and slab size. However, each snapshot contains over $10^5$ cells, and the total number of cells over $\sim200$ snapshots (from $z\sim20$ to $z\sim9$) exceeds $\sim2\times10^7$. It is therefore impractical to compute emission lines for each cell individually. We, therefore, construct a lookup table for two types of cells: PDR cells and \HII cells by classifying each cell according to whether its ionization fraction is below or above $y_{\rm HII, thres}$.

Figure~\ref{table:cloudy_parameter}(a) shows the histogram of the ionization fraction $y_{\rm HII}$ for snapshot~208 ($z=10.5\,,  t_{\rm Univ} = 462\, {\rm Myr}$), where the strongest burst of star formation begins ($z=10.5$). We find that the distribution is clearly bimodal: ionized cells are concentrated at $y_{\rm HII}\gtrsim 0.9$, whereas non-ionized cells are predominantly located at $y_{\rm HII}\lesssim 0.2$. We observe a similar trend in other snapshots and thus adopt a threshold fraction of $y_{\rm HII, thres}=0.8$. The overall distribution does not change significantly when we choose a threshold in the range $y_{\rm HII, thres}=0.5-0.9$.

Figure~\ref{fig:parameter_histgram}(b)(c) presents the histograms of gas number density and temperature, respectively. From these plots, we determine the parameter ranges of $n_{\rm H}$ and $T$ as shown in Table~\ref{table:cloudy_parameter}. Note that the critical density for star formation is $\sim10^4\,\mathrm{cm}^{-3}$. Even though snapshot~208 does not contain cells at such high densities, we extend our parameter space up to $10^4\,\mathrm{cm}^{-3}$ for completeness.

Figure~\ref{fig:parameter_histgram}(d) shows the metallicity histogram. We find that both \HII and PDR cells span similar metallicity ranges, so we adopt the same parameter range for both in Table~\ref{table:cloudy_parameter}. Figure~\ref{fig:parameter_histgram}(e) shows the histogram of gas column density of each cell, revealing that both \HII and PDR cells lie within $\log_{10}(N_{\rm H}/\mathrm{cm}^{-2})=17\text{-}22$. The upper limit is consistent with the star formation criterion described in Section~\ref{subsec:zoom_in_simulation}, $\sim0.1\,\mathrm{pc}\times10^4\,\mathrm{cm}^{-3}=3\times10^{21}\,\mathrm{cm}^{-2}$. Figure~\ref{fig:parameter_histgram}(f) shows the histogram of FUV radiation fields, where each field is normalized to the Milky Way value $G_0=3.024\times10^{-4}\,\mathrm{erg\,s^{-1}\,cm^{-2}}$, calculated over 11.2-13.6\,eV using the fitting function of \citet{Mathis:1983}. The value $G_0$ corresponds to $J_{21}=8$, where $J_{21}$ is the specific FUV intensity in units of $10^{-21}\, {\rm erg\,s^{-1}cm^{-2}sr^{-1}Hz^{-1}}$. The figure indicates that \HII and PDR cells exhibit different FUV distributions, and we set the parameter ranges accordingly in Table~\ref{table:cloudy_parameter}.

To run CLOUDY, we must also determine the SED shape. For PDR cells, we consider only radiation in the FUV bin because the number density of ionizing photons is smaller by $\sim20$ orders of magnitude compared to FUV photons in these cells. We therefore adopt a power-law SED of $\nu^{-1}$ to approximate a constant flux over 11.2-13.6\,eV (FUV). In general, the FUV radiation field can vary from place to place. For example, cells close to stars may follow stellar SED slopes (e.g., $\beta_{\rm UV} \sim -2.5$; \citealt{Yanagisawa:2024}), whereas cells farther from stars often follow a background FUV slope, such as $\nu^{-4.41}$ \citep{Mathis:1983}. However, since the energy range is narrow, 
the precise slope does not significantly affect our results. For \HII cells, we parametrize the SED with two flux ratios: ${\rm FE1}\equiv{\rm FUV}/{\rm EUV1}$ and ${\rm E12}\equiv{\rm EUV1}/{\rm EUV2}$. As an input SED, we adopt a power-law of $\nu^{-1}$ for the three energy bins (FUV, EUV1, EUV2). By varying the spectral slope of the input SED from the fiducial value to -2 (-0.44), the integrated energy fluxes in the FUV, EUV1, and EUV2 bins change by only 0.3(-0.18)\%, 2.9 (-1.6)\%, and 5.0 (-2.9)\%, respectively. Figure~\ref{fig:radiation_histgram} shows the histogram of the flux ratios in each bin. We calculate the flux in each bin from the photon number density by using the median energy of that bin. Based on these distributions, we set the parameter values in Table~\ref{table:cloudy_parameter}. In this study, we neglect EUV3 radiation ($h\nu > 54.4\, {\rm eV}$). Our focus is on the \OIII and \CII emission lines, whose ionization energy are 35.12 eV and 11.26 eV, respectively. We then expect the EUV3 radiation does not affect their emissivities. However, when we target lines from ions with ionization potentials above 54.4 eV, such as HeII 1640 \AA, we need to include the EUV3 radiation. We leave this parameter expansion to the future work.

To summarize, we construct a lookup table using seven parameters: gas density $n_{\rm H}$, gas metallicity $Z_{\rm gas}$, FUV radiation field $G$, gas temperature $T$, gas column density $N_{\rm H}$, and (for \HII cells) the two flux ratios (FE1, E12), as listed in Table~\ref{table:cloudy_parameter}.

\begin{figure*}[t]
    \centering
    \includegraphics[width =\linewidth, clip]{\figdir/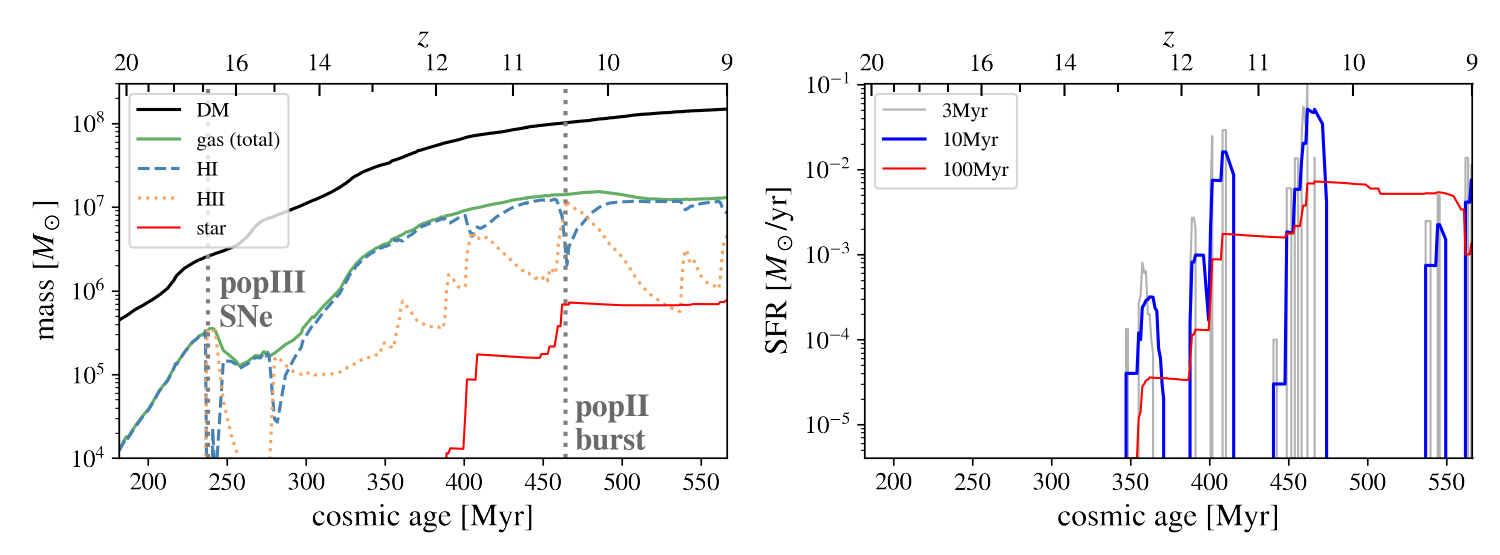}
    \caption{ Mass (left panel) and SFR (right panel) as a function of the cosmic age. 
In the left panel, the black, green, blue dashed, orange dotted, and thin red lines are masses of dark matter (DM), total gas, neutral hydrogen, ionized gas, and stars, respectively. The vertical gray dotted lines at $t_{\rm Univ} \sim 240 $ Myr ($\sim$ 460 Myr) represent the timing of popIII hypernovae  (popII bursty star formation). In the right panel, thin gray, blue, and red lines show SFRs averaged over 3, 10, and 100 Myrs. They are computed by summing the stellar mass formed over the past
X Myr and dividing by X Myr, where $X = 3, 10, \text{or}\,  100$.
}
    \label{fig:snap_evolution}
\end{figure*}
In our CLOUDY calculations, we adopt a gas slab in an open geometry, assuming uniform density, temperature, metallicity, FUV flux, and SED shape. Accordingly, we create input SEDs parametrized by FE1 and E12, and use the ``grid'' command to vary $G, Z, n_{\rm H}, T$ and $N_{\rm H}$. We include an isotropic background from the cosmic microwave background (CMB) at $z=10$ and assume that the abundance ratios of individual elements follow solar values \citep[\textsc{GASS} command;][]{Grevesse:2010}. In this work, we adopt the solar O/C ratio as a fiducial baseline, which is a common assumption in both photoionization models and galaxy simulations in the absence of direct observational constraints at $z > 6$. Observations indicate that O/C may vary from nearly solar \citep{D'Eugenio:2024} to 2.5–5 $\times$ solar \citep{Arellano-Cordova:2022, Cameron:2023}, but for most ALMA-detected \OIII/\CII emitters at $z > 6$ the O/C abundance remains unconstrained. \OIII/\CII scales nearly linearly with O/C \citep{Katz:2022, Nyhagen:2024}.

The calculations are iterated until convergence and then stopped when the slab reaches the column density $N_{\rm H}$. The flux of each line is then extracted as the emergent emissivity.

We compute the line emissivity in each cell via interpolation from the resulting emissivity table. For \HII\ (PDR) cells, the interpolation is performed in the seven-(five-)dimensional parameter space, using $2^7$ ($2^5$) data points that surround the cell's parameter values. We have additionally performed a spot-check validation of the nebular line luminosities by computing the H$\alpha$ and H$\beta$ fluxes in two independent ways:
 (1) directly evaluating each cell's emissivity from the simulation outputs ($y_{\rm HII}, n_{\rm e}, n_{\rm p}$) using the case-B recombination and collisional coefficients from \citet{Katz:2022_MgII}, and
(2) interpolating the corresponding emissivities from our CLOUDY-based lookup table. The resulting H$\alpha$ and H$\beta$ luminosities agree within a factor of 1.7, indicating that the lookup-table method introduces only minor uncertainties.

Our emission calculations are conducted within the virial radius of the halo, which ranges from $R_{\rm vir}\sim0.2\,\mathrm{kpc}$ at $z=20$ to $R_{\rm vir}\sim1.5\,\mathrm{kpc}$ at $z=9$. We include all gases within this region because strong SN feedback can expand \HII regions out to the halo scale \citepalias[see Figure~2 in][]{Sugimura:2024}.

\section{Results} \label{sec:result}
\subsection{Galaxy formation histories} \label{subsec: galaxy formation histories}
The formation history of our simulated galaxy is shown in Figure~\ref{fig:snap_evolution}. In the left panel, we plot the gas, dark matter halo, and stellar masses, while in the right panel we plot the star formation rate (SFR) averaged over 3, 10, and 100~Myr intervals, both as functions of cosmic age.

Pop~III star formation begins at $z=24$. At $z \sim 17$, the EUV feedback and subsequent Pop~III hypernova explosion sweep out gas to halo scales, reducing the gas mass by 0.3~dex. After Pop~III SNe, metal-enriched gas re-accumulates, and Pop~II star formation begins at $z \sim 13$.

Although the newly formed PopII stars undergo SN explosions stochastically over timescales of 4–40Myr, the halo potential is sufficiently deep that these SNe do not trigger halo-wide gas blowouts.
However, FUV radiation and SN feedback continue to suppress star formation by photodissociation of ${\rm H_2}$ and heating of the gas to $T \sim 10^4\,\mathrm{K}$. This hot gas has high pressure and prevents gravitational collapse, leading to a delay in star formation. During this suppression phase, the cloud grows by accreting surrounding gas and cools predominantly via Lyman-alpha emission. 
Once the cloud becomes gravitationally unstable at $z\sim10.5$, the subsequent star formation starts rapidly with relatively high star formation efficiencies ($f_*$ defined in eq.~(\ref{eq:SFE})) of 10\% \citepalias{Sugimura:2024}.

The right–hand panel of Figure~\ref{fig:snap_evolution} shows the star-formation rate (SFR) averaged over three different time windows. The 3 Myr– and 10 Myr–averaged SFRs track each other closely, although the 3 Myr profile resolves the underlying, highly stochastic bursts that occur on $\sim1 \,{\rm Myr}$ scales.
The 100 Myr–averaged SFR (red) instead shows a stepwise increase up to \(z\simeq9.1\). Although three {\it mini-quench} episodes are apparent at \(z\sim12,\;11,\;10\), each lasts less than 100 Myrs, and is therefore smoothed out in the 100 Myr‐averaged SFR, giving the appearance of a nearly constant SFR during those intervals.

In observational work, star-formation histories (SFHs) are usually inferred from SED fitting of the stellar continuum, and thus the shape of SFH tends to be similar to our 100 Myr–averaged behavior. Short-timescale burstiness can still be probed observationally through indicators such as the H$\alpha$-to-UV luminosity ratio \citep[e.g.][]{Asada:2024,Endsley:2024}, which is sensitive to SFR variations on $\sim$10\,Myr scales.

From the next section, we focus mainly on emission lines during the Pop~II formation phase at $z=9-13$, since the SFR during the Pop~III formation phase at $z=24-14$ is too small and highly stochastic \citepalias[also see Figure 4 in][]{Sugimura:2024}. Emission lines from that earlier phase may still be important when studying Pop~III stellar signatures; however, we leave that for future work. During the Pop~II formation phase, there are 137 snapshots, and we will present the corresponding results.

\begin{figure*}
    \centering
    \includegraphics[width =\linewidth, clip]{\figdir/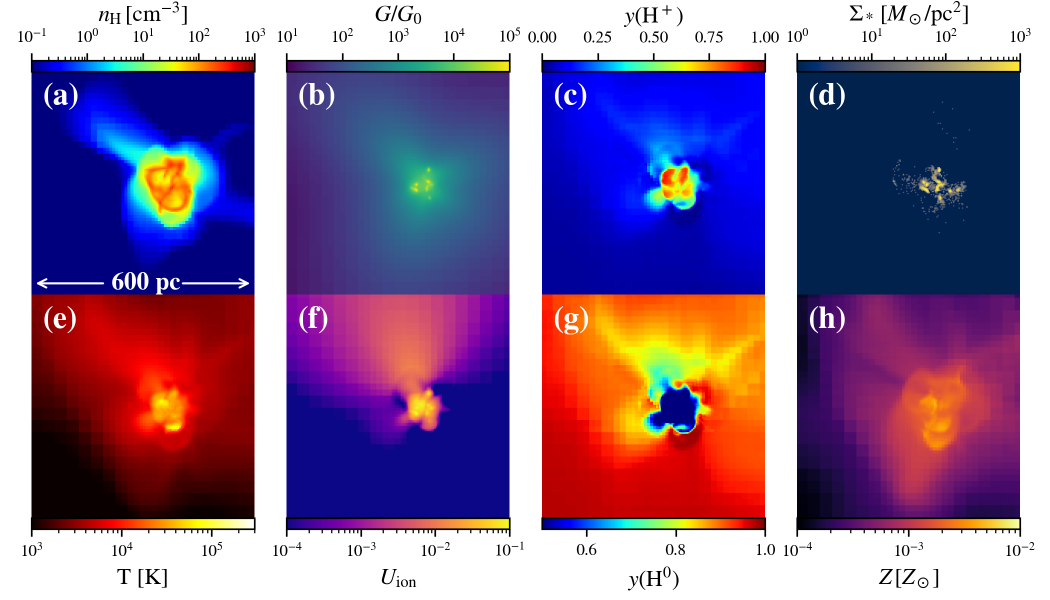}
    \caption{ Projected distributions of our galaxy at $z = 10.45$ ($t_{\rm Univ}=462\, {\rm Myr}$), where the largest bursty star formation proceeds with ${\rm sSFR} = 74.5\, {\rm Gyr^{-1}}$. 
In the upper row, we plot gas number density ($n_{\rm H}$), FUV radiation field ($G$), ionization fraction ($y({\rm H^+})$), and surface stellar mass density ($\Sigma_*$). In the bottom row, we show gas temperature ($T$), ionization parameter ($U_{\rm ion}$), neutral hydrogen fraction ($y({\rm H^0})$), and gas-phase metallicity ($Z$). Note that all $n_{\rm H}, T, y({\rm H^+}), y({\rm H^0}), U_{\rm ion}$, and $Z$ are gas density-weighted averaged along the selected line of sight. 
The FUV radiation field $G$ is a photon number density-weighted average.
One side length of the panel and the projected depth of each panel is 600 pc. Note that this snapshot (snapshot 208) has a maximum spatial resolution of 2 pc. In the gas collapsing phase ($n_{\rm H} \gtrsim 10^4 \,{\rm cm^{-3}}$), the spatial resolution reaches the order of 0.1 pc.}
    \label{fig:projection}
\end{figure*}

\subsection{Structual properties at Pop~II bursty phase} \label{subsec:structual_properties}
Figure~\ref{fig:projection} shows the projected maps of physical properties at $z=10.45$, where the SFR (specific SFR; sSFR) reaches a maximum value of $0.051 \, M_\odot{\rm yr^{-1}}$ (74.5 ${\rm Gyr^{-1}}$). The corresponding stellar and gas masses are \(M_* = 6.9\times 10^5~M_{\odot}\), \(M_{\mathrm{gas}(\mathrm{total})} = 1.39\times 10^7~M_{\odot}\), \(M_{\mathrm{HI}} = 7.72\times 10^6~M_{\odot}\), \(M_{\mathrm{HII}} = 6.22\times10^6~M_{\odot}\), and \(M_{\mathrm{H_2}} = 0.19~M_{\odot}\), respectively. Here, $M_{\rm HI}, M_{\rm HII}$ and $M_{\rm H_2}$ are gas masses of neutral atomic hydrogen gas, hydrogen ionized gas, and hydrogen molecular gas.
Each panel of Figure~\ref{fig:projection} shows the distributions of gas density, FUV field $G$, ionization fraction, stellar mass surface density, gas temperature, ionization parameter $U_{\rm ion}$, neutral gas fraction, and gas metallicity. The ionization parameter is defined as the ratio
of ionizing photon number density to gas number density.
The corresponding sliced maps are shown in Appendix~\ref{sec:slice} (Figure~\ref{fig:slice}). Note that snapshot~208 shown in Figure \ref{fig:projection} has an effective maximum resolution of 2 pc due to the extended ionization bubbles. The gas collapsing phase with molecular gas ($n_{\rm H} \gtrsim 10^4 \,{\rm cm^{-3}}$) reaches a spatial resolution on the order of 0.1 pc (see also Figures 9 and 12 in \citet{Sugimura:2024}.)

Within the 100 pc scale star-forming cloud, we identify two central cavities enclosed by dense gas with 
$n_{\mathrm{H}}\simeq500\;\mathrm{cm^{-3}}$ (panel (a)).
The cavities are fully ionized (panel (a)) and correlated with metallicity enhancement (panel (d)), implying that they are driven by SNe.
Cells with an apparent ionized fraction of $y(\mathrm{H}^+)\simeq0.5$ arise from projection effects. In the slice maps presented in Appendix \ref{sec:slice}, $y(\mathrm{H}^+)$ is almost exclusively either 1 or 0. Panel (g) 
confirms that the gas surrounding the cavities is neutral. This dense cloud is gravitationally unstable and is 
collapsing to form stars.

While SNe occur in the center, some stellar clusters remain bound \citep{Garcia:2023,Garcia:2025}. Panel~(d) presents the surface stellar mass density, which peaks at $\Sigma_{\ast} \sim 3\times 10^3~M_{\odot}~\mathrm{pc}^{-2}$. The clumpy distribution of stellar clusters and their high surface density are interestingly consistent with the recently observed stellar clusters at $z \gtrsim 6$ \citep{Vanzella:2023_sunrise_arc,Welch:2023, Bradley:2024, Mowla:2024, Adamo:2024}, and other recent simulation results of stellar cluster formation at $z > 6$ \citep{Calura:2025, Pascale:2025}.

The temperature map in panel (e) shows that the galaxy central regions have a temperature of $2\times 10^4\, {\rm K}$ due to the photoionization heating by EUV photons from massive stars. We observe some regions with higher temperatures of $T \gtrsim 5 \times 10^5 \, {\rm K}$ as a result of SN explosions, where the temperature can initially reach $T\sim 10^8\, {\rm K}$. Both photoionized regions and collisionally-ionized SN bubbles can contribute to ionization, but the recombination time is usually longer than the cooling time, making long-lived ionized regions \citep{Hartley_Ricotti:2016}. Outside of the ionizing bubbles, the neutral gas maintains a warm temperature of $T\sim (5-8)\times10^3\, {\rm K}$. 
This is because the strong FUV radiation penetrating beyond the ionized bubbles, with $G/G_0 \sim 10^3$ (corresponding to $J_{21} \sim 10^4$), almost completely suppresses ${\rm H_2}$ cooling in the low-metallicity environments (a few $\times 10^{-3} Z_\odot$; panel (h)), making Lyman-alpha cooling the dominant process (\citealp{Omukai:2008}; \citetalias{Sugimura:2024}, also see Figure \ref{fig:rho_T}).
Such high-FUV radiation is also inferred from the latest high-z ($z >6$) observations \citep{Fudamoto:2025}.

The UV radiation emitted by these stars creates ionized regions that are highly anisotropic and inhomogeneous, as shown by the maps of the ionization fraction (panel~(c)) and the ionization parameter (panel~(f)). There are ionizing bubbles in the center with $\sim$~100~pc scale and the ionization parameter peaks in the bubbles with \(\log U_{\rm ion} \sim -1\), which are co-located with recent star-formation events.

Interestingly, the FUV distribution differs from the \(U_{\mathrm{ion}}\) map. 
The central brightest regions, with \(G/G_0 \sim 10^4\), coincide spatially with the ionized regions. However, the EUV radiation drops rapidly outside the \HII regions while the FUV radiation extends over 600~pc, maintaining a high value of \(G/G_0 \sim 10^2\).

These EUV and FUV radiation distributions largely affect the \OIII and \CII distributions as explained in Section \ref{subsec:FIR_line_properties}.

\begin{figure*}
    \centering
    \includegraphics[width =0.85\linewidth, clip]{\figdir/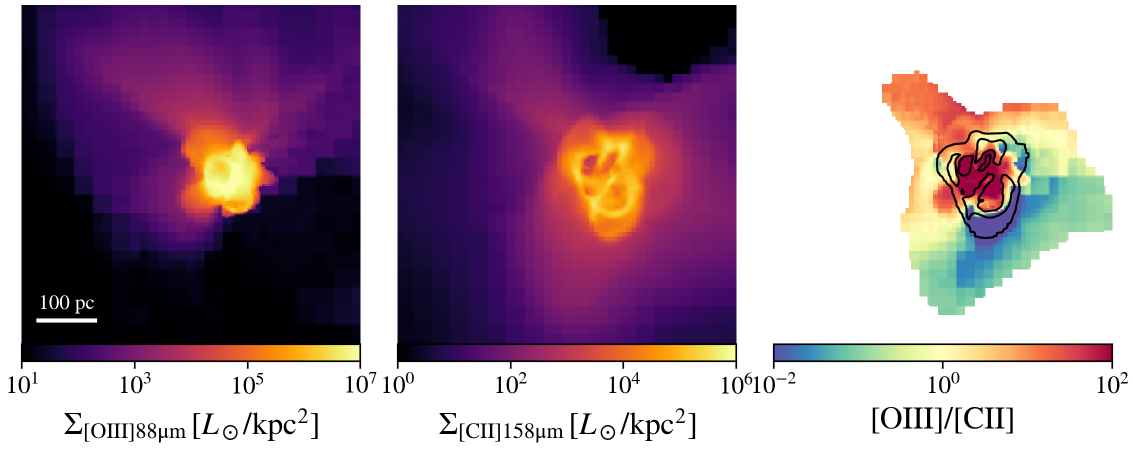}
    \caption{FIR emission line distributions of our galaxy at $z = 10.45$, at the same time as in Figure~\ref{fig:projection}. The left and middle panels show the surface brightness of \OIII 88 \mum and \CII 158 \mum, respectively. The right panel presents the line ratio of \OIII 88 \mum to \CII 158 \mum. Pixels with low surface brightness ($\Sigma_{\rm [CII]158 \mu m} > 10^2 \,{L_\odot/{\rm kpc^2}}$) are masked. The black contours represent the \CII flux $10^4, 10^5 \,{L_\odot/{\rm kpc^2}}$, respectively. The side length and projected depth of each panel are 600 pc, as in Figure \ref{fig:projection}. A movie showing the time evolution of emission map is available (\href{https://www.dropbox.com/scl/fi/fgkfeilhy4ddxeczo4ykt/line_map.mp4?rlkey=dsaakprnriwy82jxz1p31oo1z&dl=0}{link}).
}
    \label{fig:projection_line}
\end{figure*}
\subsection{FIR line properties at Pop~II bursty phase} \label{subsec:FIR_line_properties}
\subsubsection{Luminosity map}
In Figure~\ref{fig:projection_line}, we show the emission maps of \OIII 88\mum and \CII 158 \mum, as well as the emission ratio of \OIII/\CII  at $z=10.5$.
Their luminosities are \(L_{\rm [OIII]} =3.73\times 10^4 \,L_\odot\) and \(L_{\rm \CII} = 1.59\times 10^3\,L_\odot\), respectively. Since both \OIII 88 \mum and \CII 158 \mum are optically thin and self-absorption can be negligible for these lines\footnote{We verify that \CII emission is optically thin in our models by computing the ${\rm C^{+}}$ column density and optical depth. Several local environments, including starbursts, ULIRGs, and dense PDRs, exhibit moderately optically thick \CII emission with $\tau \sim 1-5$ \citep{Neri:2014, Guevara:2020, Jackson:2020, Keilmann:2025}. However, at the low metallicity adopted in this work ($Z=0.005 \, Z_\odot$), we find $N({\rm C^{+}})\sim 10^{16-17}\, {\rm cm^{-2}}$ even during the peak starburst phase at $z = 10.5$. The corresponding optical depth is $\tau_{\rm [CII]}\sim 10^{-3}$. Varying the excitation temperature between 50 K and 250 K \citep{Goldsmith:2012} changes $\tau$ by at most a factor of $\sim 4$ but does not alter this conclusion. We therefore adopt the optically thin approximation for \CII throughout this analysis.}, their luminosities and the resulting \OIII/\CII ratios are expected to be the same for different viewing angles.

As shown by \citet{Nakazato:2023}, the \OIII emission traces the \(U_{\mathrm{ion}}\) map with $U_{\rm ion} \gtrsim 10^{-2}$. It is concentrated within a compact area of $\sim$ 100 pc, while the extended component in the upper region is 100 times fainter.

However, the \CII emission map is clearly different from the \OIII map, instead resembling the distributions of gas density in Figure \ref{fig:projection}. The bright \CII emission arises from dense gas with $\sim 500\, {\rm cm^{-3}}$, which surrounds SN cavities. Diffuse \CII with $10^3\, L_\odot{\rm kpc^{-2}}$ traces the diffuse gas of $n_{\rm H}\sim 1\, {\rm cm^{-3}}$. The \CII emitting structure is asymmetric and extends out to $\sim$ 250 pc. 

For quantitative analysis, we compute the half light radii of \OIII and \CII \footnote{We take the center as the brightest point of \OIII.}, and obtain $r_{\rm e}\OIII = 75\, {\rm pc}$, $r_{\rm e}\CII= 90\, {\rm pc}$, and the ratio between the radii is $r_{\rm e}\CII/r_{\rm e}\OIII = 1.2$. These values should be interpreted with caution because \CII is predominantly emitted from shell-like regions offset from the \OIII region and both the \CII and \OIII distributions are anisotropic and extended (Figure \ref{fig:projection_line} and the accompanying \href{https://www.dropbox.com/scl/fi/4zh89k5huqp4fxxcutp9l/line_map.mp4?rlkey=loxnxwqxewolv9yndmogrh0l3&dl=0}{movie}). Nevertheless, we track the size ratio throughout the bursty phase at $z \approx 10.5$ and find that it fluctuates between $r_{\rm e}\CII/r_{\rm e}\OIII = 1-2$.

Emission line ratios are a common tool for exploring galaxy properties.
In particular, the line ratio of \OIII88 \mum and \CII158 \mum is especially interesting, as it involves two prominent FIR emission lines that arise from gas in different phases and thus can be used to constrain the physical conditions of the high-\(z\) ISM. The right panel in Figure \ref{fig:projection_line} shows the line ratio map of the \OIII/\CII emission lines, where each 2D pixel is calculated as the ratio of the \OIII to \CII surface brightness. We see that the central ionized (\OIII-emitting) regions have a very large ratio of 100, whereas the ratio suddenly decreases to 0.1-0.01 in the dense-neutral regions around the ionized ones. Taking the global ratio within $R_{\rm vir}$, we obtain \OIII/\CII $= 23.4$. 

Observations of \OIII/\CII  have been made in several galaxies at \(z \geq 6\), as introduced in Section~\ref{sec:intro}, but most of these ratios are measured as a whole-galaxy average. Currently, only the \(z=7.13\) galaxy A1689-zD1 shows the spatial variation of \OIII/\CII, as reported in \citet{Wong:2022}. They found that the central region has a ratio \(\sim3.5\) times higher than the galaxy-wide average, which is consistent with our result. Other theoretical studies \citep{Schimek:2024, Nyhagen:2024} have also presented \OIII/\CII ~maps for \(z \sim6\) galaxies. They reproduced the line ratio of \OIII/\CII $\sim$ 1-10 partially, but most regions have \OIII/\CII $\sim$ 0.1-0.01, resulting in the low total ratio of \OIII/\CII $= 0.17$. This might be because they trace large-scale structures ($\sim$10 kpc), including CGM and gas bridge during a merger phase, where a large amount of neutral gas can produce bright \CII emission. 

Overall, our simulations with a high-spatial resolution of 0.1 pc can precisely capture the difference between \OIII and \CII distributions and yield high \OIII/\CII ratios.

\begin{figure*}
    \centering
    \includegraphics[width =\linewidth, clip]{\figdir/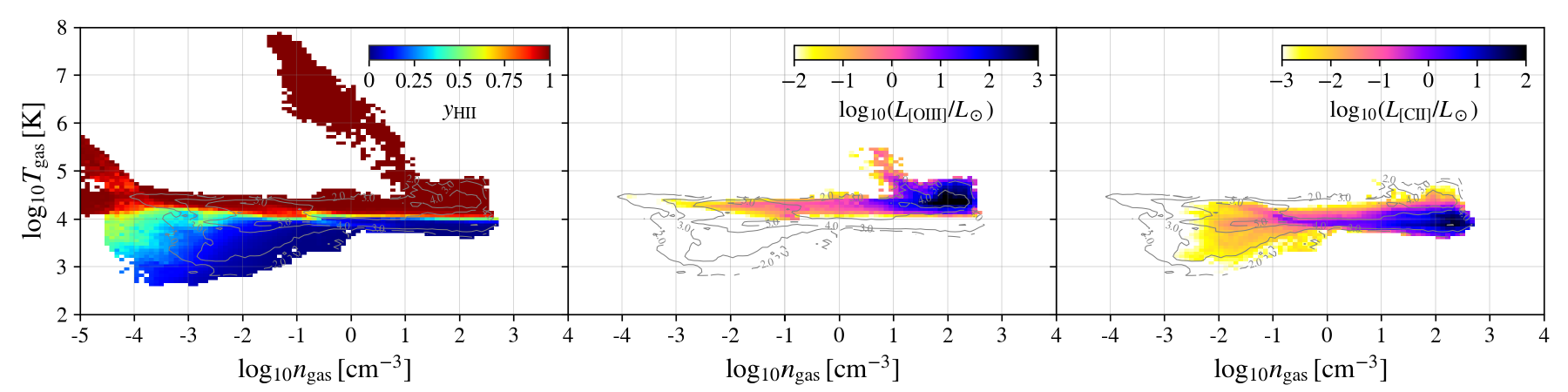}
    \caption{ 
    Phase-space diagrams of gas number density versus temperature for all cells within $2R_{\rm vir}(=3.0 \,{\rm kpc})$ at $z=10.45$. The left panel shows the ionization fraction in the phase space. The middle (right) panel presents the luminosities of \OIII 88\mum (\CII 158 \mum) in each bin. In the middle panel, we omit the hot gas with $T > 3\times 10^5\, {\rm K}$ produced by SN heating. The gray contours indicate four levels of gas mass in each bin: $\log M_{\rm gas}\, [M_\odot]=2, 3, 4, 5$. The bin sizes are $\Delta \log n_{\rm H} \, [\mathrm{cm^{-3}}]= 0.09$ and $\Delta \log T  \, [\mathrm{K}]= 0.065$ in all panels. 
}
    \label{fig:rho_T}
\end{figure*}
\subsubsection{Physical conditions of gas emitting \OIII 88\mum and \CII158\mum lines}\label{subsubsec:origin_gas}
Figure~\ref{fig:rho_T} shows the phase diagram of gas at \(z=10.5\), with colorbars corresponding to the ionization fraction, \OIII, and \CII  luminosities. 
The gray contours indicate the gas mass in each bin.
The hot gas with \(T \gtrsim 3\times10^5\)\,K is due to SN feedback, while radiation feedback from stars heats the gas up to \(\sim2\times10^4\)\,K. 
From the left panel, we see that such heated gas is completely ionized.

In the middle panel, we show the physical state of the gas that emits the \OIII line, which originates entirely from ionized cells, as the ionizing energy of ${\rm O^+}$ is 35.1 eV. 
We find that the ionized gas spans a wide range of densities, from \(0.1\,\mathrm{cm}^{-3}\) to \(500\,\mathrm{cm}^{-3}\), and contributes to the \OIII emission. Such diffused ionized gas originates from \HII bubbles in the upper part in panel (f) of Figure \ref{fig:projection}. We also compute the ionized-gas density weighted by mass and by the \OIII 88\mum luminosity, obtaining \(\langle n_{\rm HII} \rangle = 37\) and \(120\;\mathrm{cm^{-3}}\), respectively. Intriguingly, the latter value matches the very recent measurements of
\citet{Harikane:2025}, who derive electron densities for EoR galaxies from the \OIII 88\mum line and find values lower than those inferred from rest-UV/optical line ratios.
The luminosity-weighted average ionized gas density $\langle n_{\rm HII} \rangle$ is discussed further in Section~\ref{subsubsec:redshift_evolution_of_OIII_CII}.

The hot SNe gas with $T = 5\times 10^5-10^8\, {\rm K}$ is out the range of our cloudy grid ($T_{\rm max} = 10^5 \, {\rm K}$) and \OIII emission from such gas is artificial since the SNe gas ionize oxygen to OIV or OV by collisional excitation \citep{Osterbrock:2006}. However, we have checked that such artificial \OIII emission from SNe cells contributes only 0.2 \% of the total emission, which can be negligible.

The right panel shows the same type of diagram but for the \CII line, indicating that 80\% of the emission comes from neutral gas. 
In particular, gas with \(n_{\mathrm{H}} \gtrsim 100\,\mathrm{cm}^{-3}\) is the main contributor (70\% of the total luminosity: 14\% from \HII and 56\% from PDR) to the \CII emission. We derive neutral‐gas densities weighted by mass and by the PDR-\CII 158\mum luminosity, in the same manner as for the $\langle n_{\rm HII} \rangle$. The resulting values are $\langle n_{\rm HI} \rangle = 36$ and $168\;\mathrm{cm^{-3}}$, respectively. Notice that even though our simulation follows higher gas densities up to $n_{\rm H} \simeq 10^4\, {\rm cm^{-3}}$, these gases are converted into stars immediately, and are not seen in most snapshots (see Fig. 9 of \citetalias{Sugimura:2024} for a snapshot where gases reaches $n_{\rm H}\gtrsim 10^4\, {\rm cm^{-3}}$).

Observationally, the ionization state of the \CII-producing gas is investigated by taking the line ratio of \CII158\mum and \NII205\mum. This is because the ionizing potentials of neutral carbon (11.3 eV) and nitrogen (14.5 eV) are just below and above that of hydrogen (13.6 eV). Therefore, \NII lines come only from \HII regions, while the \CII can be attributed to both \HII regions and PDR. Also, a similar ionizing potential makes the line ratio independent of the ionization parameter \citep[e.g.][]{Nagao:2012}. Using \NII/\CII, \citet{Cormier:2015} studied the contribution from neutral gas for local metal-poor dwarf galaxies and found that typically 85 percent of the emission comes from neutral gas. \citet{Witstok:2022} studied the \CII emitting gas of $z > 6$ observed galaxies and reported observed \CII emissions were primarily from PDR. 

PDR with large metallicity ($\sim Z_\odot$) usually consists of two-phase medium: warm neutral medium (WNM) and cold neutral medium (CNM) \citep{Wolfire:1995}. However, our $z \gtrsim 10$ galaxy has almost only WNM due to the strong FUV radiation of$G/G_0 \sim 10^{3-5}$ from massive stars and the low abundance of metals and dust that work as coolants ($\sim 10^{-3} Z_\odot$). A decreasing mass in CNM in the high-z Universe has also been argued in previous studies \citep[e.g.,][]{Vallini:2013}. 
\begin{figure*}
    \centering
    \includegraphics[width =0.8\linewidth, clip]{\figdir/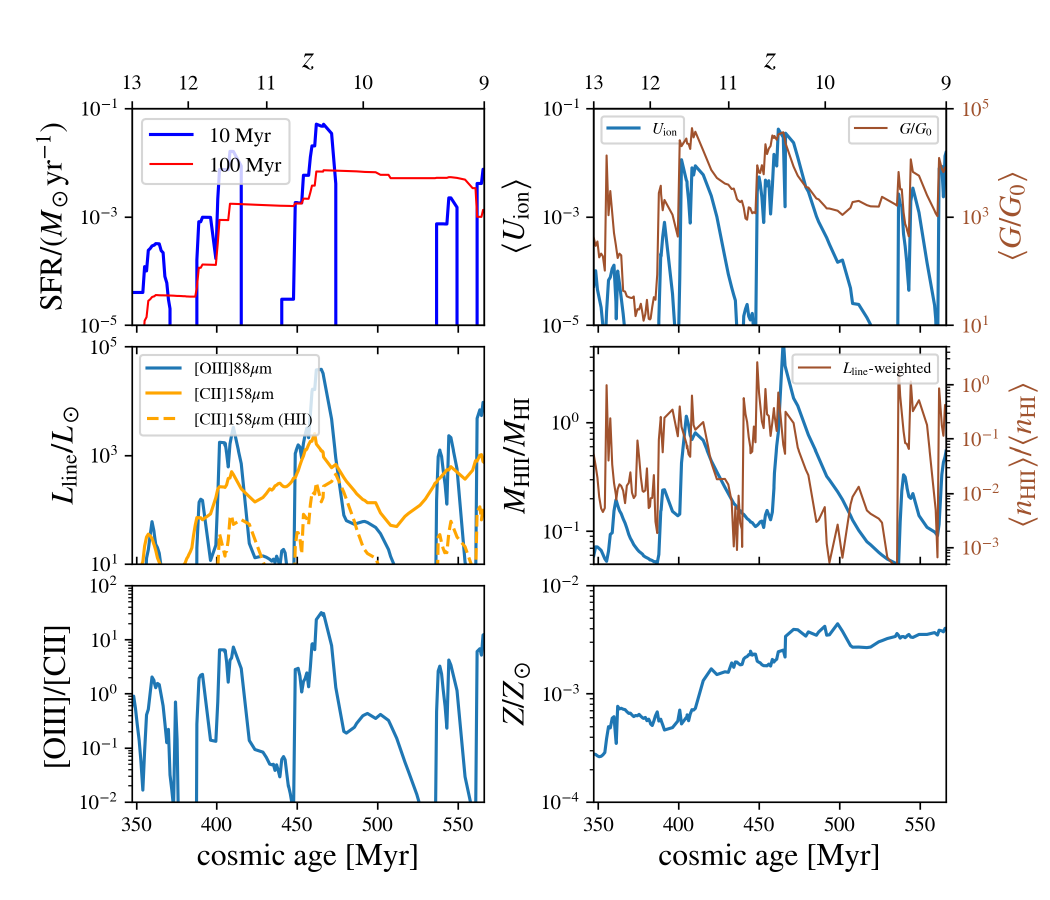}
    \caption{
    Redshift evolution of physical properties within $2R_{\rm vir}$. In the left column, the top panel shows the SFR averaged over 10 Myr (blue) and 100 Myr (red), as in Figure \ref{fig:snap_evolution} (right). The middle panel presents line luminosities of \OIII 88 \mum (blue) and \CII 158 \mum (orange). The orange dashed line shows the contribution of \CII luminosity from \HII regions. The bottom panel is the line ratio of \OIII/\CII. In the right column, the ionization parameter and FUV radiation field (top), the ratios of gas mass and averaged-gas densities of \HII and \HI gas (middle), and gas metallicity (bottom) are plotted. The ionization parameter and FUV radiation field are calculated as averages weighted by gas number densities and FUV photon numbers, respectively. The averaged-gas density for \HII and \HI gas is weighted by \OIII and \CII luminosity, respectively.
}
    \label{fig:redshift_evolution}
\end{figure*}
\subsection{Time evolution of global emission properties} \label{subsec:global_emission_properties}
Figure~\ref{fig:redshift_evolution} provides a global view of how the galaxy's emission properties evolve with time from $z = 13$ to $z=9$. Each panel shows the redshift evolution of the SFR, FUV field $G$ and ionization parameter \(U_{\mathrm{ion}}\), emission line luminosities of \OIII and \CII, mass ratio of ionized and neutral gas ($M_{\rm HII}/M_{\rm HI}$), the line ratio \OIII/\CII, and metallicity. We calculate the global $G$ and \(U_{\mathrm{ion}}\) as averages weighted by FUV photon numbers and gas number densities, respectively.

In the left-middle panel, we show the \OIII and \CII luminosities. The \OIII \,evolution follows that of the SFR averaged over 10\,Myr, because O\(^{2+}\) ions exist in \HII regions produced by young (\(\lesssim10\)\,Myr), massive stars. The ionizing photon production rate \(Q(\mathrm{H})\) \([\mathrm{s}^{-1}]\) decreases rapidly when stellar ages exceed 10\,Myr. \citet{Xiao:2018} show that \(Q(\mathrm{H})\) at 3, 10, 30, and 100\,Myr for a single stellar population with \(M_* = 10^6\,M_{\odot}\) and \(Z=10^{-5}\) are $\log\,Q(\mathrm{H})\,[\mathrm{s}^{-1}] = 52.5, 51.2, 49.1$, and $47.1$, respectively (also see Figure \ref{fig:bpass_age} in Appendix \ref{sec:bpass_age}). The upper-right panel shows the evolution of \(U_{\mathrm{ion}}\) and $G$. The $U_{\rm ion}$ peaks at the $z\sim 10.5$ bursty phase with $\log U_{\rm ion} \simeq -1.5$  suddenly decreases to $\log U_{\rm ion} = -5$ within 50 Myrs. The time fluctuation is consistent with that of 10\,Myr-averaged SFR and \OIII. 

Meanwhile, \CII emission persists even during the ``mini-quenched'' phase, and its evolution follows that of the FUV field. This is because the \CII line predominantly arises from neutral gas irradiated by FUV photons (right-middle panel). In Appendix \ref{sec:bpass_age}, we have calculated FUV luminosity with $M_* = 10^6\,M_{\odot}$, $Z=10^{-5}$, and age of 3, 10, 30, and 100\,Myr using BPASS {\it single} SEDs \citep{Stanway:2018, Byrne:2022}\footnote{The RT calculations in our simulation adopt BC03. According to \citet{Eldridge:2017}, BPASS single SEDs are similar to other ones such as Starburst 99 \citep{Leitherer:1999}, GALAXEV \citep{Bruzual_Charlot:2003}, and \citet{Maraston:2005} at the age of $<$ 1 Gyrs.}, and obtained $\log L_{\rm FUV}\, [L_\odot] = 8.1, 7.7, 7.1$, and 6.3. The FUV luminosity decreases only 1 dex from 3 Myr to 30 Myr, while the ionizing photon number rate decreases by 3.5 dex. The decrease in \CII luminosity during the mini-quench phase is consistent with the decline in FUV flux in the upper-right panel. 
Since we show the time evolution with a very short snapshot interval of 1 Myr in comparison to other galaxy simulations \citep[e.g., ][]{Arata:2020, Katz:2019}, we can examine this trend in detail for the first time. 

The orange dotted line indicates the \CII  emission originating from \HII regions. 
Its time evolution also tracks the 10\,Myr-averaged SFR, but it contributes less than 1\% most of the time as shown in section \ref{subsubsec:origin_gas}. Only at the cosmic age of 472-479\,Myr does \CII from \HII regions contribute about 56-66\%.

The left-bottom panel illustrates the time evolution of the \OIII/\CII  ratio. 
As discussed above, the relatively flat \CII evolution implies that the fluctuations in the line ratio largely trace those of \OIII, $\log U_{\rm ion}$ and 10 Myr-averaged SFR, which is a similar behaviour only recently reported by \citet{Kohandel:2025}. We see that the most intense starburst phases reach \OIII/\CII $\sim$ 50 at the starting time of bursty star formation ($t_{\rm Univ}\sim 460\, {\rm Myr}$). The ratio rapidly decreases to 0.2 at the beginning of the mini-quench but increases to 0.4 at $t_{\rm Univ} \sim 500 \, {\rm Myr}$, due to the different decreasing timescales of \OIII and \CII luminosities. Interestingly, the \OIII/\CII fluctuation is similar to that of the mass ratio of $M_{\rm HII}/M_{\rm HI}$ ratio even the latter amplitude is 1-dex smaller than that of \OIII/\CII as shown in the right-middle panel of Figure \ref{fig:redshift_evolution}. In the same panel, we also plot the time evolution of the gas density ratio between ionized and neutral components, $\langle n_{\rm HII}\rangle/\langle n_{\rm HI}\rangle$, where each density is weighted by the luminosity of the \OIII 88\mum and \CII158 \mum lines, respectively. The density ratio also shows a similar trend. We will explain the correlation between \OIII/\CII and $M_{\rm HII}/M_{\rm HI}$ ($\langle n_{\rm HII}\rangle /\langle n_{\rm HI}\rangle$) in Section \ref{subsec:physical_origin}.

Finally, we plot the time evolution of the metallicity in the right-bottom panel because some theoretical simulations have argued that \OIII/\CII correlates with metallicity \citep[e.g.][]{Arata:2020}. We find that the metallicity fluctuation is not very large compared to radiation fluctuations. Since accelerated gas by SNe remains bound within the halo at $z \lesssim 13$, the metallicity gradually increases with time even during the mini-quenched phase. Therefore, the redshift evolution of FIR luminosity does not trace gas metallicity very much in comparison to other properties (SFR, EUV/FUV radiation fields). However, the luminosity would be correlated with the metallicity when we compare galaxies with the same SFR \citep[e.g.,][]{Inoue:2014}.

\subsection{Comparison to observations} \label{subsec:comparison_to_observations}
\begin{figure*}[t]
  \centering
  \begin{minipage}[t]{0.49\linewidth}
    \centering
    \includegraphics[width=\linewidth]{\figdir/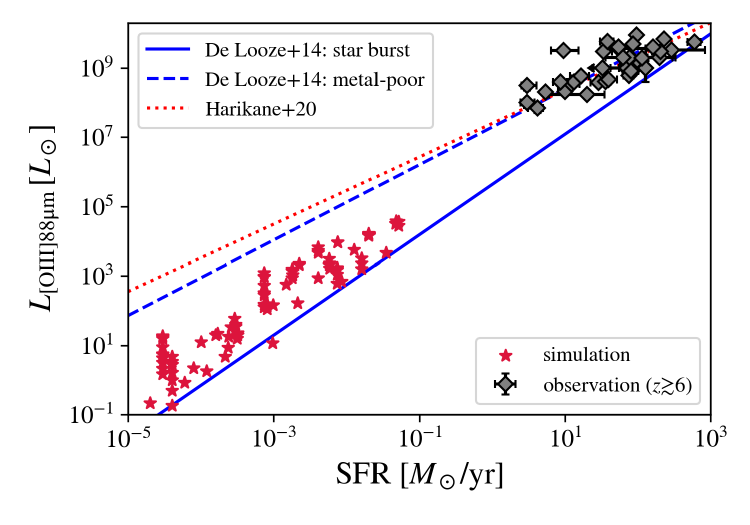}
  \end{minipage}
  \hfill
  \begin{minipage}[t]{0.49\linewidth}
    \centering
    \includegraphics[width=\linewidth]{\figdir/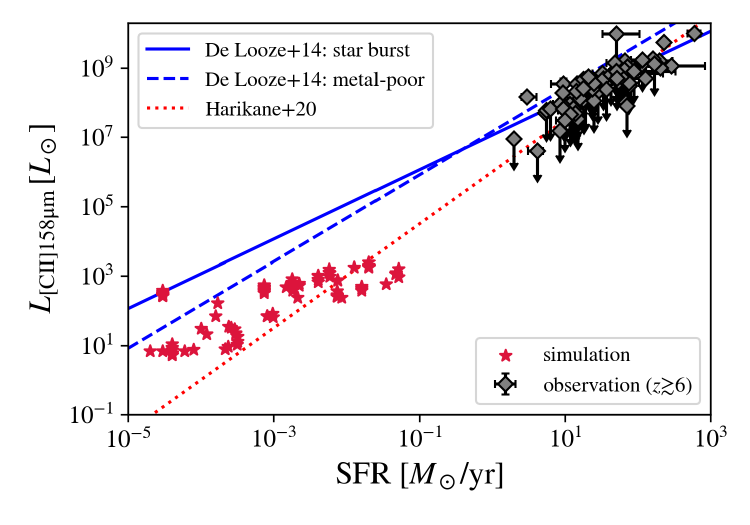}
  \end{minipage}
  \caption{(Left) The \OIII 88 $\mu\mathrm{m}$ luminosity versus SFR. We adopt 10 Myr-averaged SFR. The red stars represent our simulation snapshots ($z=9-13$). For comparison, we show the \OIII-SFR relations derived from observations of local galaxies by \citet{De_Looze:2014} and $z\gtrsim 6$ galaxies by \citet{Harikane:2020}. Gray points are the observational results of high-z $(z > 6)$ galaxies from \citet{Hashimoto:2018, Tamura:2019, Inoue:2016, Hashimoto:2019, Carniani:2017, Harikane:2020, Wong:2022, Witstok:2022, Zavala:2024, Algera:2024, Schouws:2024, Bakx:2024, Fujimoto:2024, Algera:2025, Witstok:2025}.
  (Right) The \CII 158 $\mu\mathrm{m}$ luminosity versus SFR. The red stars again represent our simulation snapshots ($z=9-13$). For comparison, we show the \CII-SFR relations derived from observations of local galaxies \citep{De_Looze:2014}, ALPINE survey for $z\sim 5$ galaxies \citep{Schaerer:2020}, and $z\gtrsim 6$ galaxies \citep{Harikane:2020}. Gray points are the observational results of high-z $(z > 6)$ galaxies from \citet{Laporte:2019, Bakx:2020, Hashimoto:2019, Maiolino:2015, Harikane:2020, Wong:2022, Witstok:2022, Algera:2024, Schouws:2025, Bakx:2024, Schaerer:2015, Watson:2015, Pentericci:2016, Ota:2014, Smit:2018, Bradac:2017, Matthee:2017, Carniani:2017, Carniani:2018, Kanekar:2013, Matthee:2019, Knudsen:2016, Fujimoto:2019, Capak:2015, Sommovigo:2022, Fudamoto:2024, Algera:2025}.}
  \label{fig:Line_SFR_obs}
\end{figure*}

\begin{figure*}
  \centering
  \begin{minipage}[t]{0.49\linewidth}
    \centering
    \includegraphics[width=\linewidth]{\figdir/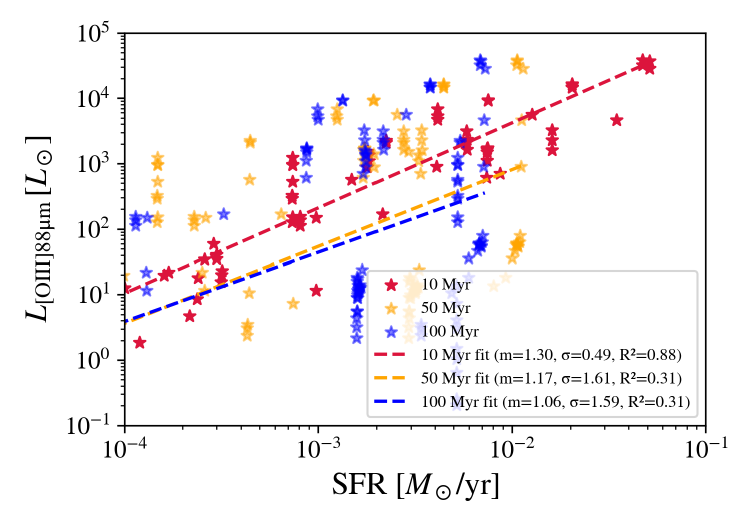}
  \end{minipage}
  \hfill
  \begin{minipage}[t]{0.49\linewidth}
    \centering
    \includegraphics[width=\linewidth]{\figdir/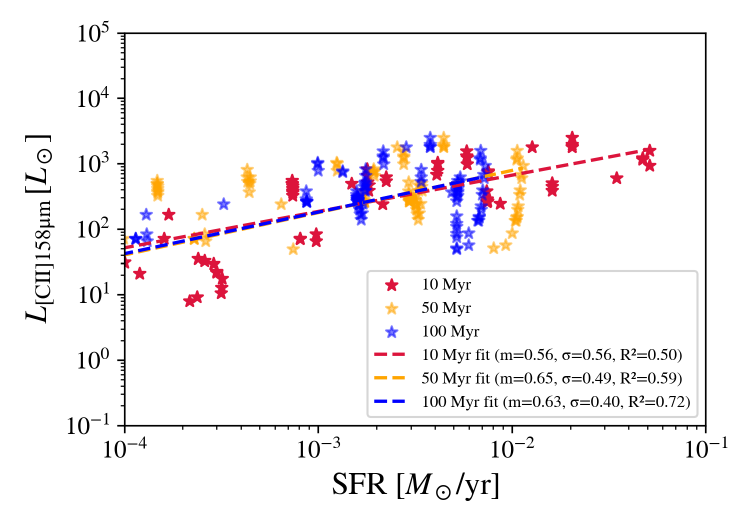}
  \end{minipage}
  \caption{(Left) The \OIII 88\mum versus SFR for our simulation snapshots ($z = 9-13$) using different SFR averaging timescales: 10 Myr (red), 50 Myr (yellow), and 100 Myr (blue). (Right) Same as the left panel, but for the \CII 158 \mum luminosity versus SFR. In both panels, the relations are fitted with the function $\log_{10}(L_{\rm line}) = m \times \log_{10}({\rm SFR}) + C$, where $m$ is the slope and $C$ is the intercept. The scatter ($\sigma$) and the coefficient of determination ($R^2$) for each fit are shown in the white boxes.}
  \label{fig:Line_SFR_with_diff_averaged_SFR}
\end{figure*}

\begin{figure*}
  \centering
  \begin{minipage}{0.48\linewidth}
    \centering
    \includegraphics[width=\linewidth]{\figdir/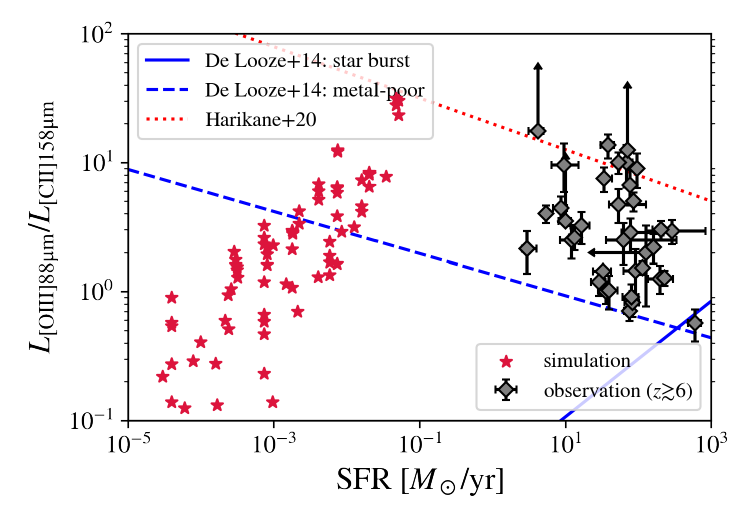}
  \end{minipage}
  \hfill
  \begin{minipage}{0.48\linewidth}
    \centering
    \includegraphics[width=\linewidth]{\figdir/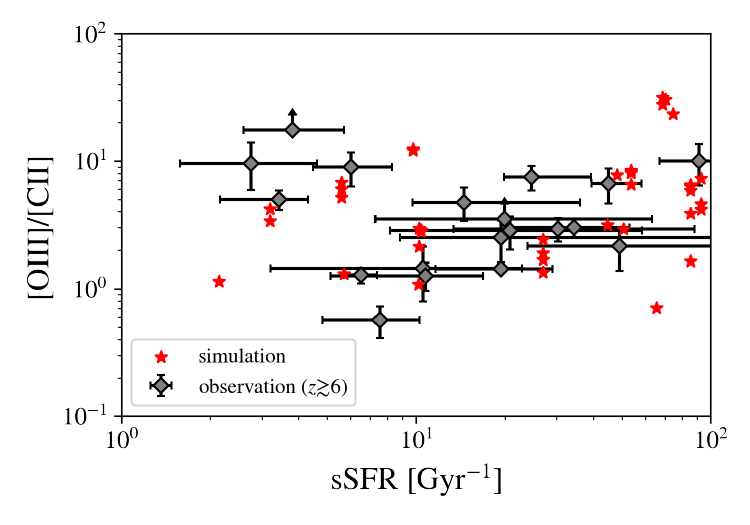}
  \end{minipage}
  \caption{(Left) Total \OIII/\CII line luminosity ratio as a function of SFR for our simulated galaxy at $z = 9-13$ (red stars). The gray points show the observations of $z\gtrsim 6$ galaxies where both \OIII and \CII lines are targeted. Some \CII lines are upper-limits, and the ratio is plotted accordingly. The blue and red lines are relationships for local galaxies \citep{De_Looze:2014} and $z\gtrsim 6$ galaxies \citep{Harikane:2020}, respectively, as in Figure \ref{fig:Line_SFR_obs}. (Right) Total \OIII/\CII line luminosity ratio as a function of sSFR. The red stars show our simulation results with ${\rm SFR} > 10^{-3},M_\odot,{\rm yr^{-1}}$, and the gray points correspond to $z \gtrsim 6$ galaxies with stellar masses reported by JWST \citep{Zavala:2024, Witstok:2025, Stiavelli:2023, Fujimoto:2024, Harshan:2024, Rowland:2025, Ren:2025, Jones:2024, Mawatari:2025}.}
  \label{fig:OIII_CII_SFR_obs}
\end{figure*}
There are empirical correlations between FIR lines and SFR in galaxies, which have been extensively investigated through both observations and simulations. These are widely used to validate theoretical predictions against observational constraints.

In Figure~\ref{fig:Line_SFR_obs}, we show the \OIII-SFR (left) and \CII-SFR (right) relationships for our simulated galaxy at \(z=9-13\). We adopt 10 Myr-averaged SFRs defined in Figure \ref{fig:snap_evolution}, following previous theoretical works \citep[e.g., ][]{Pallottini:2019, Katz:2019, Nakazato:2023}.
For comparison, we plot the local relations for starburst galaxies and metal-poor galaxies from \citet{De_Looze:2014}, and that for high-\(z\) observed galaxies at \(z \gtrsim 6\) from \citet{Harikane:2020}. Note that the \citet{Harikane:2020} relation is obtained over SFR $\simeq 4-220\, M_\odot {\rm yr^{-1}}$; the local metal-poor-dwarf and starburst galaxies of \citet{De_Looze:2014} cover SFR $\simeq 10^{-3} -30 \, M_\odot {\rm yr^{-1}}$ and $ (1-1000)\,  M_\odot {\rm yr^{-1}}$, respectively. We extrapolate those relationships to the lower SFR regime of our simulations.
The gray points with error bars indicate observed galaxies at \(z \gtrsim 6\)  (references are in the caption of Figure \ref{fig:Line_SFR_obs}). 

For the \OIII-SFR relation, nearly all of our simulated points lie 1-2 dex above the local starburst galaxies with the same SFR due to high-ionization states with $\log U_{\rm ion} \sim [-1.5, -2]$, whereas the local galaxies typically have $\log U_{\rm ion} \sim -3$ \citep{Harikane:2020}. 
However, our \OIII luminosities are about 1-2 dex below the local metal-poor or \(z \gtrsim 6\) observed relations. The offset is readily explained by metallicity: the simulations have gas metallicities of only $(0.5-5)\times10^{-3}\,Z_{\odot}$, roughly two orders of magnitude lower than the $\sim 0.25\,Z_{\odot}$ typical of local metal-poor systems.  Because the \OIII emissivity scales nearly linearly with metallicity in this regime \citep[see Figure 1 of][]{Inoue:2014}, the predicted \OIII luminosities are correspondingly reduced by a factor of $\sim$ 100, yielding the observed 1–2 dex deficit.

For the \CII-SFR relation, our points are 1-2 dex below both local starburst and metal-poor galaxies and lie closer to the \(z \gtrsim 6\) relations. We propose the following two main reasons: (i) little contribution from \HII regions with very high-ionizing state, where carbon can be ionized to ${\rm C^{++}}$\citep[e.g.,][]{Liang:2024}, and (ii) very low-metallicity. \citet{Liang:2024} derived that $L_{\rm [CII]}/{\rm SFR}$ is roughly proportional to gas-phase metallicity, and our plots are consistent with this trend. Interestingly, the slope is shallower than that of the \OIII-SFR relation because 
the \OIII luminosity traces the 10 Myr-averaged SFR that is used in the \OIII-SFR and \CII-SFR relations, whereas the \CII luminosity shows a weak correlation with that same 10Myr-SFR as shown in Section~\ref{subsec:global_emission_properties}. 
In both panels, we see a scatter of the line intensity at a given SFR. The \OIII–SFR relation shows a scatter of 1–2 dex. Roughly one dex of this dispersion is attributable to differences in metallicities, while additional scatter arises from spatial and temporal fluctuations in the gas density.
The \CII-SFR scatter can also be explained in the same way.

In Figure \ref{fig:Line_SFR_obs}, we discuss the relation between emission line luminosity and the star formation rate (SFR) using the 10 Myr averaged SFR. However, observationally derived SFRs, based on UV-to-FIR continuum emission, typically probe longer timescales of $\gtrsim 100\,{\rm Myr}$. We therefore further investigate the line–SFR relation by adopting different SFR averaging timescales. The left and right panels of Figure \ref{fig:Line_SFR_with_diff_averaged_SFR} correspond to the $L_{\mathrm{[OIII]88\mu{\rm m}}}$–SFR and $L_{\mathrm{[CII]158\mu{\rm m}}}$–SFR relations, respectively. We fit the relations with the function $\log_{10}(L_{\rm line}) = m \times \log_{10}({\rm SFR}) + C$, where $m$ is the slope and $C$ is the intercept. We also derive the scatter ($\sigma$) and the coefficient of determination ($R^2$) for each fit.

We find that the $L_{\rm [OIII]}$-SFR relation based on the 10 Myr averaged SFR exhibits the smallest scatter ($\sigma = 0.49$), whereas the 50 Myr and 100 Myr averages yield larger $\sigma$ values ($\sigma = 1.61, 1.59$). This trend arises because the ionizing photon rate declines rapidly within $\lesssim 10$ Myr, corresponding to the lifetimes of massive O- and B-type stars. Figure \ref{fig:bpass_age} shows the time evolution of the ionizing photon production rate ($Q_{\rm H}$) and the FUV luminosity as a function of stellar age. These quantities are calculated from the BPASS single-star SED model, assuming $M_* = 10^6\,M_\odot$ and instantaneous star formation. We find that $Q_{\rm H}$ decreases by about four orders of magnitude from 10 Myr to 100 Myr.

In contrast, the scatter in the $L_{\rm [CII]}$–SFR relation slightly decreases with longer averaging timescales.
The overall scatter remains small ($\sigma = 0.4$–$0.56$), and the variation is modest (less than a factor of 1.4) compared to the $L_{\rm [OIII]}$–SFR relation, where the difference exceeds a factor of three. This behavior arises because the \CII emission primarily traces the FUV radiation field, which is dominated by stars younger than $\lesssim 100$ Myr. The right panel of Figure \ref{fig:bpass_age} shows that the FUV radiation decreases by only $\sim$1.5 dex from 10 Myr to 100 Myr.

Therefore, we conclude that $L_{\mathrm{[OIII]88\mu{\rm m}}}$ serves as a tracer of bursty star formation on $\sim$10 Myr timescales, whereas $L_{\mathrm{[CII]158\mu{\rm m}}}$ traces star formation activity averaged over longer ($\sim$100 Myr) periods. The \CII luminosity correlates well with both short- and long-term SFRs, making it a robust tracer of star formation over a wide range of timescales.
\begin{figure*}[t]
  \centering
  \begin{minipage}{0.48\linewidth}
    \centering
    \includegraphics[width=\linewidth]{\figdir/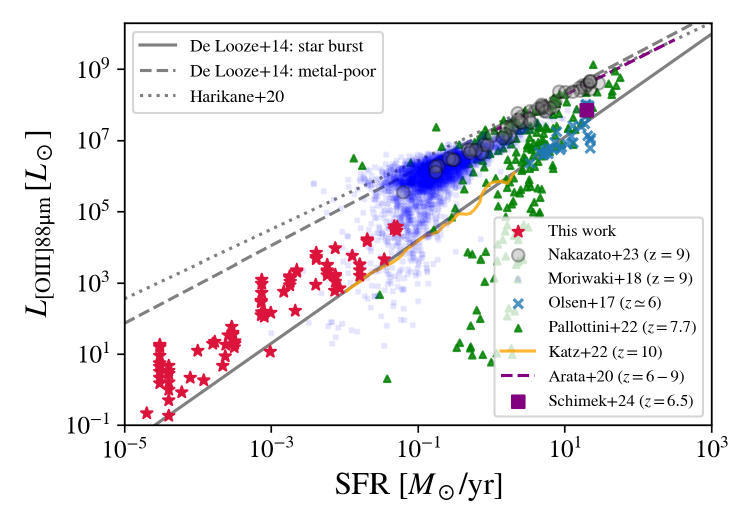}
  \end{minipage}
  \hfill
  \begin{minipage}{0.48\linewidth}
    \centering
    \includegraphics[width=\linewidth]{\figdir/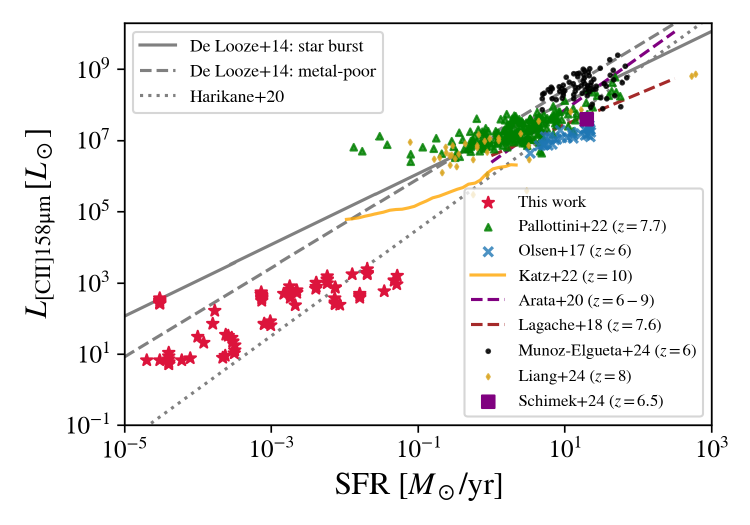}
  \end{minipage}
  \caption{Same as Figure \ref{fig:Line_SFR_obs}
  but comparison with $z\gtrsim6$ galaxy simulations. We compare our results
  with FirstLight \citep{Nakazato:2023}, \citet{Moriwaki:2018}, SERRA \citep{Pallottini:2022}, SIGAME \citep{Olsen:2017}, \citet{Katz:2022}, \citet{Arata:2020}, TNG\citep{Munoz-Elgueta:2024}. FIRE-box\citep{Liang:2024}, and PONOS\citep{Schimek:2024}. We also plot the semi-analytical model of \citet{Lagache:2018} in the right panel.}
  \label{fig:Line_SFR_sim}
\end{figure*}

\citet{Hashimoto:2019} and \citet{Harikane:2020} have pointed out that there exists a negative correlation between the ratio of \OIII/\CII  and the bolometric luminosity of the host. The bolometric luminosity is expected to scale with the SFR of the galaxy (i.e., $L_{\rm bol} \propto L_{\rm UV}\propto {\rm SFR}$). In the left panel of Figure \ref{fig:OIII_CII_SFR_obs}, we plot \OIII/\CII-SFR relation for our galaxy at $z = 9-13$ as red stars. We see that there is a roughly positive correlation between \OIII/\CII, but with the large scatter of the ratio with 1-2 dex at a fixed SFR, which reflects stochastic bursty star formation. We will discuss this correlation in Section \ref{subsec:comparison_to_simulation}.
Our galaxy has a variety of \OIII/\CII $= 0.1 -50$ at $z=9-13$. Recently, \citet{Schouws:2025} targeted $z=14.2$ galaxy JADES-GS-z14-0 and reported non-detection of \CII, which constrained a lower limit \OIII/\CII $>$ 3.5. Our galaxy successfully reproduces such a high line ratio during the bursty star formation phase (Figure \ref{fig:redshift_evolution}). 

The right panel of Figure~\ref{fig:OIII_CII_SFR_obs} shows the relation between \OIII/\CII and the specific SFR (sSFR). Since stellar mass estimates for high-$z$ galaxies were largely unavailable before JWST, this relation can now be investigated observationally for the first time. We plot the observed galaxies for which stellar masses have been reported (references are listed in the caption of Figure \ref{fig:OIII_CII_SFR_obs}). Because sSFR is normalized by stellar mass, our simulation results naturally cover the observed range.

Note that our simulated galaxies are significantly less massive than the observed high-$z$ systems, with stellar masses of $M_* = 10^{5-6}\,M_\odot$, comparable to proto-globular-cluster scales (see also caveats in Section~\ref{subsec:caveates}). Nevertheless, the key physical processes that determine the \OIII and \CII emission, such as ionization state, gas density, and metallicity, are expected to operate in a similar manner within individual star-forming regions of more massive galaxies. The consistency between the simulated and observed \OIII/\CII–sSFR relations supports this interpretation. In this framework, the resulting line luminosities would scale roughly linearly with the total mass, metallicity, and the amount of star-forming gas, implying that our results can be extrapolated to represent sub-regions of massive galaxies ($M_* \gtrsim 10^8\,M_\odot$). Very recently, \citet{Knudsen:2025} measured \OIII/\CII ratios for individual star-forming clumps within A1689-zD1 and found that the ratios vary by less than a factor of 2, further supporting our scaling expectation. Therefore, the \OIII/\CII ratios obtained in our low-mass systems likely capture the fundamental physical origin of the high ratios observed in massive galaxies at $z > 6$.

\begin{figure}
    \centering
    \includegraphics[width =\linewidth, clip]{\figdir/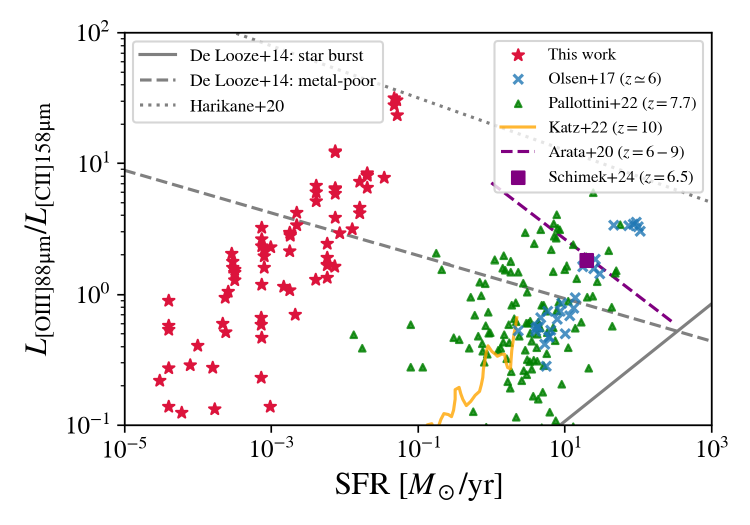}
    \caption{Same as Figure \ref{fig:OIII_CII_SFR_obs} but comparison with previous theoretical works at $z\gtrsim 6$ that investigated both \OIII and \CII lines. References are in the caption of Figure \ref{fig:Line_SFR_sim}. 
}
    \label{fig:OIII_CII_SFR_sim}
\end{figure}

\subsection{Comparison to simulations} \label{subsec:comparison_to_simulation}
In addition to the comparison to observations in the previous subsection, we compare our results with other cosmological simulations at \(z \gtrsim 6\) in Figure \ref{fig:Line_SFR_sim} (references are in the caption). Notice that the other simulations (except PONOS simulation, \citet{Schimek:2024}) treat several different galaxies at the same redshift, whereas our simulation focuses on only one galaxy. The scatter in the plots represents sample‐to‐sample variation in the other simulations, whereas in our simulation it reflects temporal variation. Additionally, our simulation tracks a very low-mass (low-SFR) regime, making direct comparison to previous studies not straightforward. Nevertheless, it provides a valuable opportunity to understand the FIR line\textendash SFR relationship in high-\(z\) galaxies across a broad mass (SFR) range.

For \OIII-SFR, some simulations of \cite{Moriwaki:2018, Arata:2020, Pallottini:2022, Nakazato:2023} are located on the relationships of local metal-poor and $z>6$ galaxies and have more luminous \OIII 88\mum than the other ones \citep{Olsen:2017, Schimek:2024, Katz:2022}, which are around the local star-burst relations. As mentioned in Section~\ref{subsec:comparison_to_observations}, our simulated galaxy is consistent with the relation of local metal-poor galaxies rather than that of local star-burst because of its very low-metallicity of $\log(Z/Z_\odot) = [-3.5, -2.7]$. 

For \CII-SFR, the slope of our data is shallower than that of \OIII-SFR, again as mentioned in Section \ref{subsec:comparison_to_observations}. This trend is similar to the other simulations with on-the-fly RT \citep{Pallottini:2022, Katz:2022}.

Figure \ref{fig:OIII_CII_SFR_sim} shows the \OIII/\CII- SFR relation in comparison to other theoretical studies at $z\gtrsim 6$. We find that our positive correlation is also seen in \citet{Olsen:2017, Pallottini:2022}, and \citet{Katz:2022}, although \citet{Arata:2020} finds an anti-correlation. The simulation trends and the lack of a clear correlation in the current observational samples can be understood in terms of galaxy mass and the associated stochastic bursty star formation. Low–mass systems possess shallow potential wells, and stellar feedback can efficiently suppress star formation, producing highly stochastic SFRs. The \OIII/\CII then tracks the 10 Myr-averaged SFR and \(U_{\rm ion}\) (Section~\ref{subsec:global_emission_properties}; see also \citealt{Kohandel:2025}).
As galaxies grow and their gravitational potentials deepen, the amplitude of SFR fluctuations diminishes and the correlation weakens. Most of the $z\gtrsim6$ galaxies currently detected in both \OIII\ and \CII\ are already fairly massive ($M_*\gtrsim10^{9}\,M_\odot$), and their comparatively smooth star-formation histories may obscure the underlying relation seen in the simulations.

The main difference with respect to previous studies are the values of the line ratio. Our sample has large values, up to 50, while other simulations show lower values of 0.1-3, which failed to reproduce the observed high ratios in Figure \ref{fig:OIII_CII_SFR_obs}. It is difficult to explain the main reason for the discrepancy since star formation recipes and feedback models differ from simulation to simulation. One possibility may be our exceptionally short snapshot output interval. Especially, during the strongest starburst at $z \approx 10.5$, we save snapshots every 1 Myr. This fine cadence captures the brief peak of the burst and, therefore, the moment when the \OIII/\CII ratio reaches its maximum ($\OIII/\CII \sim 30$). By contrast, many other simulations report results at only a single redshift or at coarser snapshot spacing, so they miss such short-lived extremes.
Another possibility is that the underestimate of \OIII/\CII ratio is caused by RT with 10 pc-scale or worse resolution in other simulations.
The 10 pc-scale resolution is not sufficient to trace the ionization structure precisely, and \citet{Pallottini:2019} mentioned that their simulation with 28~pc resolution could not resolve high-density ionized gas with $n_{\rm HII} \gtrsim 1\, {\rm cm^{-3}}$, which can lead to an underestimate of \OIII luminosity and overestimate of \CII luminosity.
Our simulations with 0.1 pc scale resolution can resolve high-density ionized gas ($\sim 500\, {\rm cm^{-3}}$) and obtain luminous \OIII ($\propto n_{\rm HII}^2$). We will discuss the effects of resolution on RT simulations in Section \ref{subsubsec:OIII_CII_simulation}.

\section{Discussion} \label{sec:discussion}
\subsection{Physical origins of high \OIII/\CII} \label{subsec:physical_origin}
There are several possibilities proposed to explain high-\OIII/\CII ratio at high-redshift galaxies. \citet{Harikane:2020} and \citet{Katz:2022} examined the following eight effects: (A) higher ionization parameter, (B) lower gas metallicities, (C) higher (electron) gas density, (D) higher O/C ratio, (E) lower PDR covering fraction, (F) CMB attenuation effect, (G) spatially extended halo, and (H) inclination effect. However, these eight effects interact with each other, making it difficult to identify the primary mechanism of high-\OIII/\CII ratio. We then get back to basics and discuss the following three fundamental factors: gas density, temperature, and ion fractions. Here we derive emission line luminosities from a simple model assuming constant temperature and density in each gas phase. The intrinsic luminosities of forbidden lines (i.e., collisionally excited lines) can be written as 
\begin{align}
L_{\rm int} &= \Lambda(n_{\rm gas}, T) n_{\rm gas} n_{\rm ion} V_{\rm gas} \notag \\
&= \Lambda(T) n_{\rm gas}^2 X_{\rm ion} V_{\rm gas}\,\,\,\, (\text{for}\, n_{\rm gas} < n_{\rm crit}),
\end{align}
where $\Lambda, V_{\rm gas}$, and $X_{\rm ion}$ are cooling rate [${\rm erg\,s^{-1} cm^{3}}$], gas volume, and the fraction of a target ion, respectively. Here, $n_{\rm gas}$ denotes the number density of each gas phase and is written as $n_{\rm gas} = n_{\rm HII}$ in \HII regions and $n_{\rm gas} = n_{\rm HI}$ in PDRs. The main collision partners in \HII regions and PDR are electrons and neutral hydrogen atoms, and their densities are $n_{\rm c}=n_{\rm e}\simeq n_{\rm HII}$ for \HII regions and $n_{\rm c}=n_{\rm HI}$ for PDR. The observed luminosity is decreased to some extent via several effects such as aperture and inclination effects: 
\begin{equation}
L_{\rm obs} = f_{\rm obs}\,L_{\rm int}  \,\,\, (0 < f_{\rm obs} < 1)\,,
\label{eq:lum_obs_int}
\end{equation}
where $f_{\rm obs}$ is the suppression factor related to observations. This factor is introduced as an ad-hoc parameterization to encapsulate various observational effects into a single term. Here, however, we focus on the physical origins of the intrinsic \OIII/\CII ratios and do not apply this suppression factor. As discussed in Section \ref{subsubsec:OIII_CII_simulation}, its impact is small and cannot account for the observed high \OIII/\CII ratios at $z \gtrsim 6$.
The intrinsic \OIII and \CII luminosities and their ratio are given as \footnote{The critical densities of \OIII 88\mum and \CII 158\mum in \HII regions with $T_{\rm HII} = 10^4\, {\rm K}$ are 510 and 50 ${\rm cm^{-3}}$, respectively \citep{Osterbrock:2006} The critical density of \CII 158\mum in PDR with $T_{\rm HI} = 5000 (100)\, {\rm K}$ is $1.5(2.7)\times 10^3\, {\rm cm^{-3}}$ \citep{Draine:2011}.}
\begin{align}
L_{\rm [OIII], int} &=  \Lambda_{\rm [OIII]}(T_{\rm e}) n_{\rm HII}^2 X_{\rm O^{++}} V_{\rm HII}\,,\\
L_{\rm [CII], int} &=  \Lambda_{\rm [CII]}(T_{\rm e}) n_{\rm HII}^2 X_{\rm C^{+}}^{\rm (HII)} V_{\rm HII} \notag  \\
 &+ \Lambda_{\rm [CII]}(T_{\rm HI}) n_{\rm HI}^2 X_{\rm C^{+}}^{\rm (HI)} V_{\rm HI} \,, \\
\therefore \frac{L_{\rm [CII], int}}{L_{\rm [OIII], int}} &= \frac{\Lambda_{\rm [CII]}(T_{\rm e})}{\Lambda_{\rm [OIII]}(T_{\rm e})}\cdot \frac{X^{({\rm HII})}_{\rm C^+}}{X_{\rm O^{++}}}  \notag \\
&+ \frac{\Lambda_{\rm [CII]}(T_{\rm HI})}{\Lambda_{\rm [OIII]}(T_{\rm e})}\frac{n_{\rm HI}^2}{n_{\rm HII}^2}\frac{X^{({\rm HI})}_{\rm C^+}}{X_{\rm O^{++}}}\frac{V_{\rm HI}}{V_{\rm HII}}\,, 
\label{eq:line_ratio}
\end{align}
where $T_{\rm HI}$ is the gas temperature in PDRs and $T_{\rm e} (= T_{\rm HII})$ are
the electron and gas temperature in \HII regions, respectively.

Here, we neglect the \CII emission from molecular gas since the molecular gas mass fraction is less than 1\% even at the bursty star formation phase in our simulation. The dominance of atomic hydrogen gas at high-z galaxies is also reported in other simulations \citep{Casavecchia:2025}. Also, the FIRE simulations \citep{Liang:2024} show that \CII emission from molecular gas contributes less than 10\% at $z = 0-8$.
From Appendix \ref{sec:cooling_rate}, we obtain $\Lambda_{[\rm CII]}(T_{\rm e})/\Lambda_{[\rm OIII]}(T_{\rm e}) \simeq 1$ in \HII regions.  
Stellar radiation in high-redshift galaxies tends to be dominated by young and low-metallicity populations and to have hard SEDs ($=\text{high}\, U_{\rm ion}$). Consequently, $X_{\rm C+}^{\rm (HII)}$ is much smaller than that of local galaxies: ${X^{({\rm HII})}_{\rm C^+}}\ll{X_{\rm O^{++}}}$, and the first term in Eq.~(\ref{eq:line_ratio}) can be negligible. The cooling rate of \CII (\OIII) is calculated considering collisions with atomic hydrogen (electrons) in PDR (\HII regions). They are approximated as follows: 
\begin{align}
\Lambda_{[\rm CII]}(T_{\rm HI}) &= 10^{-24}\exp\left(-\frac{91.2}{T_{\rm HI}}\right) \notag  \\
&\times (16+0.344\sqrt{T_{\rm HI}} - 47.7/T_{\rm HI}) \,  {\rm erg\, cm^3 s^{-1}} \label{eq:lambda_CII}\,,\\
\Lambda_{[\rm OIII]}(T_{\rm e}) &= 10^{-19} T_{\rm e}^{-1/2} \,  {\rm erg\, cm^3 s^{-1}}. \label{eq:lambda_OIII}
\end{align}
Eq. (\ref{eq:lambda_CII}) is taken from \citet{Barinovs:2005} and Eq. (\ref{eq:lambda_OIII}) is obtained from Appendix \ref{sec:cooling_rate} based on \citet{Osterbrock:2006}.  The temperature $T_{\rm HI}$ in Eq. (\ref{eq:lambda_CII}) and 
$T_{\rm e}$ in Eq. (\ref{eq:lambda_OIII}) are expressed in Kelvin. Note that Eq.~(\ref{eq:lambda_CII}) 
does not include the effect of CMB attenuation. In Appendix~\ref{sec:CMB_attenuation} we derive the attenuation factor and show that the CMB at $z=10$ ($z=7$) suppresses $\Lambda_{\CII}$ by 10–20 \% ($\sim5$ \%).

From Figures \ref{fig:parameter_histgram} and \ref{fig:rho_T}, the typical temperatures of PDR and HII regions are $T_{\rm HI} \sim 5000\, {\rm K}$ and $T_{\rm e} \sim 2\times 10^4\, {\rm K}$, and the corresponding cooling rates are $\Lambda_{\rm [CII]}= 3.96\times 10^{-23} \,  {\rm erg\, cm^3 s^{-1}}$ and $\Lambda_{\rm [OIII]}= 7.58 \times 10^{-22} \,  {\rm erg\, cm^3 s^{-1}}$, respectively. For ${X^{({\rm HI})}_{\rm C^+}}/{X_{\rm O^{++}}}$, we can approximate as the element abundance ratio ${A{\rm (C)}}/{A{\rm (O)}}$ because 
$X_{\rm C^{+}}^{\rm (HI)}/X_{\rm C,{tot}}^{\rm (HI)}\sim 1$ in a neural gas and $X_{\rm O^{++}}/X_{\rm O,{tot}}^{\rm (HII)}\sim 1$ in \HII regions are satisfied. Here $X_{\rm C,{tot}}^{\rm (HI)}$ and $X_{\rm O,{tot}}^{\rm (HII)}$ 
denote the total fraction of carbon and oxygen, summed over all ionization states in each region. We then rewrite Eq. (\ref{eq:line_ratio}) as 
\begin{align}
 \frac{L_{\rm [OIII], int}}{L_{\rm [CII], int}} &= 
\frac{\Lambda_{\rm [OIII]}(T_{\rm e})}{\Lambda_{\rm [CII]}(T_{\rm HI})} \frac{M_{\rm HII}}{M_{\rm HI}}\frac{n_{\rm HII}}{n_{\rm HI}}\frac{A({\rm O})}{A({\rm C})} \notag \\
& = 5.7  \left(\frac{M_{\rm HII}/M_{\rm HI}}{0.3}\right)\left(\frac{n_{\rm HII}/n_{\rm HI}}{0.5}\right)\frac{\rm (O/C)}{\rm (O/C)_\odot}, \label{eq:line_ratio_ver2}
\end{align}
where $M_{\rm HII} \, (M_{\rm HI})$ is ionized (neutral hydrogen) gas mass and ${\rm O/C}\equiv A({\rm O})/A({\rm C})$ is the abundance ratio.
We adopted typical values for the gas mass ratio from the middle-right panel in Figure \ref{fig:redshift_evolution} and for the density ratio from Figure \ref{fig:rho_T}. Since the above derivation assumes constant temperature and density in each phase, the density ratio that appears in Eq.~(\ref{eq:line_ratio_ver2}), $n_{\rm HII}/n_{\rm HI}$, should be considered a global luminosity-weighted ratio: $n_{\rm HII}$ is the average weighted by the \OIII 88\mum line luminosity and $n_{\rm HI}$ is the average weighted by the \CII 158\mum line luminosity. This ratio is time dependent and varies between 0.3-1 during bursts of star formation, to $10^{-2}$-$10^{-3}$ during quiescence periods between starbursts. Eq.~(\ref{eq:line_ratio_ver2}) shows that line ratios of about 5-6 can be obtained assuming typical values of 
$M_{\rm HII}/M_{\rm HI} \sim 0.3$ and $n_{\rm HII}/n_{\rm HI}\sim 0.5$. During star-formation bursts, the ionized gas mass can be larger than the neutral gas mass, as shown in Figure \ref{fig:snap_evolution}, leading to even more extreme values of the line ratio ($>10$).

\subsubsection{Redshift Evolution of \OIII/\CII}\label{subsubsec:redshift_evolution_of_OIII_CII}
Here, we discuss how the \OIII/\CII  ratio evolves with redshift based on Eq.~(\ref{eq:line_ratio_ver2}).
Observations suggest that \(\OIII/\CII\) ratios are higher at \(z \gtrsim 6\) than in local star-forming galaxies (see Sections~\ref{sec:intro} and~\ref{subsec:comparison_to_observations}). To unveil the origin of this redshift evolution, we examine the redshift dependence of four key factors that influence the line ratio: cooling rates, gas masses, gas densities, and elemental abundances.

\paragraph{Cooling Rates.}
As metallicity rises, both PDRs and \HII regions cool to lower temperatures via metal and dust cooling \citep[e.g.,][]{Omukai:2005}.
PDRs can cool to \(\sim 100\)\,K, while \HII regions cool to \(\sim 8000\)\,K \citep[e.g.,][]{Osterbrock:2006}.
Correspondingly, the cooling rates change to \(\Lambda_{\rm [CII]} = 7.62\times10^{-24}\,\mathrm{erg\,s^{-1}\,cm^{3}}\)\footnote{Here we assume that \CII emission from CNM dominates at lower redshifts since the gas density is over 10 times higher than that of WNM.} and \(\Lambda_{\rm [OIII]} = 1.2\times10^{-21}\,\mathrm{erg\,s^{-1}\,cm^{3}}\).
Interestingly, the \OIII 88\,\(\mu\mathrm{m}\) cooling rate increases slightly, whereas that of \CII 158\,\(\mu\mathrm{m}\) decreases.\footnote{However, rest-optical \OIII cooling rates rise with temperature \citep[see Figure~3.2 of][]{Osterbrock:2006}.}
Consequently, the ratio \(\Lambda_{\rm [OIII]}/\Lambda_{\rm [CII]}\) increases from about 19 to 157 as one moves from high-\(z\) to lower-\(z\). However, the impact of cooling rates is reversed
by the three other effects discussed below, and the \OIII/\CII ratio turns out to be higher in high-$z$ galaxies.

\paragraph{Gas Mass.}
From Figure~\ref{fig:redshift_evolution}, our simulated galaxies typically show a high ionized-to-neutral gas mass ratio, \(M_{\rm HII} / M_{\rm HI} > 0.3\), when the line ratio is large, \(\OIII/\CII > 3\).  
Such high mass ratios are also found in other high-redshift galaxies, as we will discuss in Section~\ref{subsec:synergy_of_JWST_ALMA}.  
In contrast, local galaxies, including the Milky Way, typically show much lower values, \(M_{\rm HII}/M_{\rm HI} \approx 0.01\) \citep{Tielens:2005},
suggesting a significantly lower ionized gas fraction relative to neutral gas at low-z. This effect is therefore expected to reduce the line ratio at low-z. 

\paragraph{Density.}
In our simulations, we have calculated luminosity(\OIII88\mum and \CII 158\mum)-weighted gas density for \HII regions ($\langle n_{\rm HII} \rangle$) and PDR ($\langle n_{\rm HI} \rangle$). During the bursty star formation phases, our galaxy exhibits $\langle n_{\rm HII} \rangle = 20-200\, {\rm cm^{-3}}$ and $\langle n_{\rm HI} \rangle = 70-600 \, {\rm cm^{-3}}$, with the corersponding gas density ratio fluctuating in the range $\langle n_{\rm HII}\rangle/\langle n_{\rm HI} \rangle=0.3-1.0$. Very recently, \citet{Harikane:2025} estimate electron densities of EoR galaxies from FIR \OIII88\mum, finding that their five target galaxies have $n_{\rm e} \lesssim 500\, {\rm cm^{-3}}$, with some galaxies showing densities even below $100\, {\rm cm^{-3}}$. These values are systematically lower than the electron densities estimated by optical (\OII-based) and UV (\textsc{Ciii]}-based) emission lines, but are interestingly consistent with our \OIII88\mum-weighted ionized gas density.

For local galaxies, \citet{Nakazato:2023} and \citet{Harikane:2025} estimated the FIR\OIII-weighted ionized gas density using \OIII 88\mum/52\mum line ratio. These two FIR-\OIII lines were observed by \citet{Fernandez-Ontiveros:2016} and \citet{Brauher:2008}, and the obtained electron densities are $\sim 100\, {\rm cm^{-3}}$. The neutral gas densities of local star formation galaxies are estimated by \citet{Malhotra:2001}, and their typical values are $n_{\rm HI} \sim 10^3\, {\rm cm^{-3}}$. Therefore, the density ratio \(n_{\rm HII}/n_{\rm HI}\) is expected to decrease, typically from \(0.5\) at high-z to \(0.1\) in local galaxies.

\paragraph{Elemental Abundances.}
Because we adopt solar abundance ratios, O/C remains constant in our current model.
In reality, however, high-redshift galaxies may be enriched primarily by Type\,II SNe, possibly elevating O/C above the solar value \citep{Nomoto:2006,Stark:2017}.
For example, Type\,II SNe yield at \(1/20\,Z_{\odot}\) can have an O/C ratio up to eight times higher than the solar value.
Some observations of \(z>6\) galaxies also suggest about twice the solar O/C ratio \citep{Stark:2017}, which would naturally boost \(\OIII/\CII\). 
Several studies have shown that the \OIII/\CII ratio scales approximately linearly with O/C. \citet{Katz:2022} enhanced the O/C abundance by a factor of 8 relative to solar and found a corresponding $\sim 8\times$ increase in \OIII/\CII (see their Figure 7), while \citet{Nyhagen:2024} tested four O/C values and likewise reported nearly linear scaling. 

In lower-redshift galaxies with solar metallicities, O/C would converge closer to the solar ratio.\footnote{We note that a fraction of both carbon and oxygen is locked up in solid grains. Because carbon is more heavily depleted onto dust than oxygen \citep{Draine:2011}, the gas-phase O/C ratio can be up to $\sim$ 2.5 times higher than the solar value.}

Collectively, the above four factors and Eq.~(\ref{eq:line_ratio_ver2}) imply that \(\OIII/\CII\) could decline from approximately 5.7 (or 11.2, if O/C ratio is twice the solar value) at high-\(z\) to around 0.3 at lower-\(z\), which is consistent with local star burst galaxies (\OIII/\CII $\sim 0.04-1$) reported by \cite{De_Looze:2014}.

\subsubsection{Role of numerical resolution and $f_{\rm obs}$ in reproducing high \OIII/\CII Ratios at $z>6$} \label{subsubsec:OIII_CII_simulation}

In Figure~\ref{fig:OIII_CII_SFR_obs}, our simulated galaxies achieve the high \(\OIII/\CII\) ratios (\(\sim 3\text{--}30\)), reproducing the values observed at \(z>6\).
In contrast, Figure~\ref{fig:OIII_CII_SFR_sim} shows that other simulations yield lower \(\OIII/\CII\) values, falling short of the high observed ratios at \(z>6\).
A likely reason for this discrepancy is our exceptionally high spatial resolution, which is sufficient to resolve dense \HII regions whose typical size is estimated by the Str\"omgren radius with 1 pc order (Eq. (\ref{eq:stromgren_raius})).
Previous on-the-fly RT simulations \citep[e.g.,][]{Pallottini:2019,Katz:2019} employed cell sizes an order of magnitude larger than these \HII regions\footnote{Strictly speaking, the size of an individual \HII region around each star particle depends on $Q(\mathrm{H})$, which scales with the star-particle mass.
For reference, \citet{Katz:2022} employ stellar particles of $400\,M_\odot$, while \citet{Pallottini:2019} use $1.2\times10^{4}\,M_\odot$.
The corresponding Str\"omgren radii are $\sim 3$ pc and $\sim9$ pc, respectively—both are below the spatial resolution of their simulations.
}.
Consequently, their simulations can produce partially ionized cells where ionizing photons are absorbed within under-resolved volumes, whereas our higher resolution allows cells to be either predominantly neutral \((x_{\rm HII}\approx 0)\) or ionized \((x_{\rm HII}\approx 1)\) (see Figure~\ref{fig:parameter_histgram}).
Insufficient resolution can lead to an underestimate of both the ionized gas mass and its density, suppressing \(\OIII/\CII\) to below unity according to Equation~(\ref{eq:line_ratio_ver2}). We note that we have not performed artificial coarse-graining of our simulation grids to directly test the effect of unresolved \HII regions on the global \OIII/\CII ratios. A dedicated convergence study using consistent lower-resolution radiation-hydrodynamic simulations would be needed to fully assess this systematic uncertainty.

In the preceding discussion, we focused on the physical factors that affect the intrinsic line ratios. However, observational factors also influence the observed line luminosities through $f_{\rm obs}$ in Eq.~(\ref{eq:lum_obs_int}).
Extended \CII\ emission can be missed if a small aperture is adopted, and a large velocity width (FWHM) can drive the peak flux below the detection threshold.  
\citet{Harikane:2020} used relatively large apertures ($\sim2''$) to capture halo–scale \CII\ flux, yet still reported high \OIII/\CII\ ratios of 3.4–8.8.  
They demonstrated that even if the extended \CII\ profile \citep{Fujimoto:2019} is applied on smaller–aperture data \citep{Inoue:2016, Laporte:2019}, the recovered \CII\ flux increases by at most $\sim0.6$ dex, and the \OIII/\CII\ ratio remains high.  
Large FWHM values can arise, for example, when a galaxy is observed edge-on \citep{Kohandel:2019, Kohandel:2020, Rizzo:2022}. However, \citet{Katz:2022} analysed simulated galaxies from different viewing angles and found that the \CII\ line is not always broader than \OIII.  
We therefore conclude that aperture losses and velocity–width effects can lower $f_{\rm obs}$ for \CII luminosities, thereby increasing \OIII/\CII\ ratios in certain sources, but are unlikely to be the primary cause of the observed high ratios.

\subsection{Synergy of JWST detected lines} \label{subsec:synergy_of_JWST_ALMA}
\begin{figure*}
    \centering
    \includegraphics[width =0.8\linewidth, clip]{\figdir/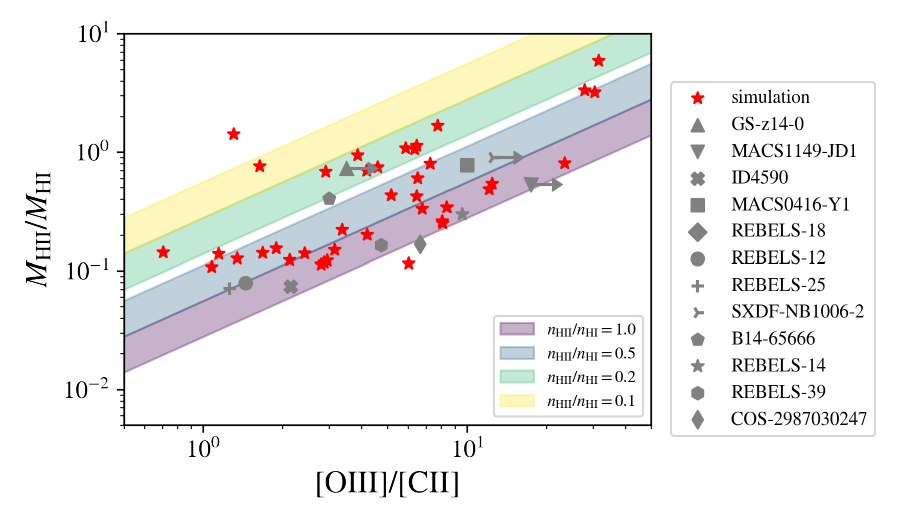}
    \vspace{-6mm}
    \caption{The relationship between \OIII/\CII and $M_{\rm HII}/M_{\rm HI}$. The red stars are our simulation results with ${\rm SFR} > 10^{-3}\, M_\odot {\rm yr^{-1}}$ and gray plots are observational results at $z \gtrsim 6$. The information of $L_{\OIII 88\mu m}, L_{\CII 158\mu m}, L_{\rm H\beta}$, and $12 + \log({\rm O/H})$ are from the following references; GS-z14-0 \citep{Schouws:2024, Schouws:2025, Helton:2025}, MACS1149-JD1 \citep{Hashimoto:2018, Laporte:2019, Stiavelli:2023}, ID4590 \citep{Fujimoto:2024, Heintz:2023}, MACS0416-Y1 \citep{Tamura:2019, Bakx:2020, Hagimoto:2025}, REBELS-12, REBELS-25, REBELS-14, and REBELS-40 \citep{Algera:2024, Rowland:2025}, SXDF-NB1006 \citep{Inoue:2016, Ren:2025}, B14-65666 \citep{Hashimoto:2019, Jones:2024}, COS2987030247 \citep{Witstok:2022, Mawatari:2025}. Dust-corrected $L_{\rm H\beta}$ are adopted. We calculate $M_{\rm HII}$ assuming $n_{\rm HII} = 100\, {\rm cm^{-3}}$. The colorband shows the different density ratio cases with $n_{\rm HII}/n_{\rm HI} = 0.1-1.0$, with different neutral gas temperatures of $T_{\rm HI}= 500-5000\, {\rm K}$.
}
    \label{fig:OIII_CII_MHII_MHI}
\end{figure*}
In Section \ref{subsec:physical_origin}, we have shown $\OIII/\CII \propto (M_{\rm HII}/M_{\rm HI}) (n_{\rm HII}/n_{\rm HI})$. To examine the validity of this relation, we derive the mass ratio $M_{\rm HII}/M_{\rm HI}$ observationally for the first time by combining JWST and ALMA observations. 

The ionized gas mass $M_{\rm HII}$ can be calculated using ${\rm H}\beta$ luminosity \citep{Finkelman:2010}:
\begin{align}
M_{\rm HII} &= \frac{L_{\rm H\beta}m_{\rm H}/n_{\rm e}}{4\pi j_{\rm H\beta}/(n_{\rm e}n_{\rm p})} \notag \\
&= 1.01 \times 10^6\, M_\odot \left(\frac{L_{\rm H\beta}}{10^{40}\, {\rm erg/s}}\right) \left(\frac{n_{\rm e}}{100 \, {\rm cm^{-3}}}\right)^{-1}, 
\end{align}
where $m_{\rm H}$ and $j_{\rm H\beta}$ are the proton mass and H$\beta$ emission coefficient, respectively. The value of $j_{\rm H\beta}$ can be changed by electron temperatures, but its temperature dependency is insignificant \citep{Osterbrock:2006, Draine:2011}.
For $M_{\rm HI}$, we use the conversion relation from \CII luminosity to gas mass derived by \citet{Vallini:2025}\footnote{We do not adopt the conversion factor derived from our own simulations, because we track only a single galaxy whose bursty star–formation history at $z>10$ produces a much larger scatter in the metallicity–$\alpha_{\rm[CII]}$ relation than that reported by \citet{Vallini:2025}.  Once galaxies grow sufficiently massive, their SFR fluctuations should subside and the scatter is expected to converge.  We therefore apply the calibration of \citet{Vallini:2025} to the observed systems, assuming that the inferred gas mass is dominated by neutral atomic hydrogen, $M_{\rm HI}$.
A conversion between $L_{\rm[CII]}$ and the neutral–gas mass ($M_{\rm HI}$) has also been investigated by \citet{Heintz:2021,Heintz:2025} using GRB afterglows and damped–Ly$\alpha$ spectrum. Because those methods can include large amounts of circum-galactic gas, we do not employ their relations here and instead focus on the multi-phase ISM in star-forming regions.
}: 
\begin{align}
M_{\rm HI}= \alpha_{\CII} L_{\CII}\,,
\end{align}
where
\begin{align}
& \log (\alpha_{\CII}/M_\odot L_\odot^{-1}) = -0.39 \log(Z/Z_\odot) + 0.67.
\end{align}

Figure \ref{fig:OIII_CII_MHII_MHI} shows the relation between \OIII/\CII and $M_{\rm HII}/M_{\rm HI}$. We plot every $z>6$ object for which \OIII 88 $\mu$m, \CII 158 $\mu$m, H$\beta$, and gas-phase metallicity are currently available (references are in the caption of Figure \ref{fig:OIII_CII_MHII_MHI}). We successfully obtain the positive correlation between \OIII/\CII and $M_{\rm HII}/M_{\rm HI}$, as expected from Eq.~(\ref{eq:line_ratio_ver2}), albeit with large scatter. The theoretical relation given by Eq.~(\ref{eq:line_ratio_ver2}) is plotted in Figure \ref{fig:OIII_CII_MHII_MHI}) as colored bands corresponding to different density ratios. The width of the bands, corresponding to a factor of $\lesssim 1.5$ difference, indicates the impact of varying the cooling-rate ratio by changing neutral-gas temperatures in $T_{\rm HI} = 500$–$5000$ K. We also plots our simulation results with SFR $ > 10^{-3} \, M_\odot {\rm yr^{-1}}$, which exhibit a clear positive correlation between \OIII/\CII and $M_{\rm HII}/M_{\rm HI}$. The physical conditions of large $M_{\rm HII}/M_{\rm HI}$ environments are similar to density-bounded ISM, implying low PDR covering fractions and large escape fractions. Our results show GS-z14-0, MACS1149-JD1, SXDF-NB1006-2, MACS0416-Y1 and B14-65666 have relatively high mass ratios of $M_{\rm HII}/M_{\rm HI} \gtrsim 0.3$, and might have large escape fractions.

\begin{figure}
    \centering
    \includegraphics[width =\linewidth, clip]{\figdir/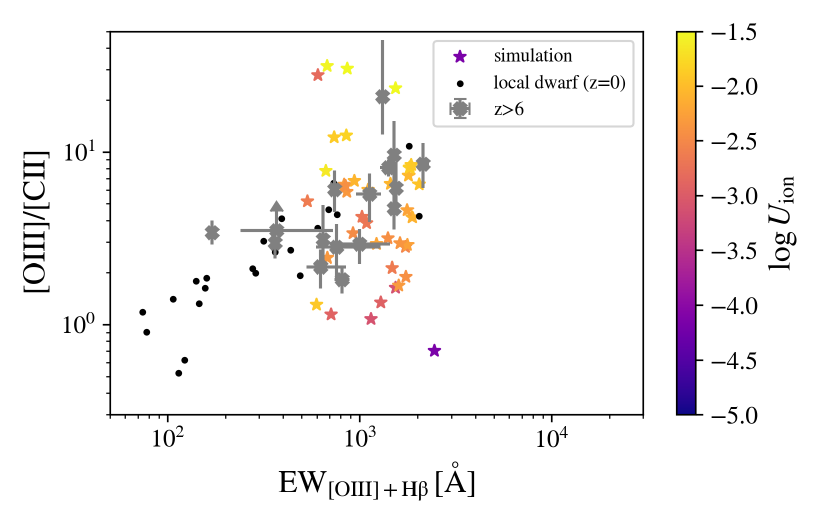}
    \vspace{-4mm}
    \caption{ \OIII/\CII  ratio as a function of $\OIII_{4959, 5007}+{\rm H}\beta$ equivalent width. Star plots are our simulation results at $z=9-13$ with SFR $ > 10^{-3} \, M_\odot {\rm yr^{-1}}$, and a colorbar represents the averaged ionization parameters. Black plots are $z>6$ galaxies \citep{Witstok:2022, Schouws:2025}, and the tabulated data from \citet{Algera:2025}. The gray plots are local dwarf galaxies from the Dwarf Galaxy Survey \citep{Kumari:2024}.
}
    \label{fig:OIII_CII_EW}
\end{figure}
We also investigate another synergetic diagnostic enabled by the combination of
JWST and ALMA, the relation between the \OIII/\CII ratio and equivalent width of $\OIII_{4959, 5007}+{\rm H}\beta$ (EW(\OIII$+$H$\beta$)). Recent observations suggest that the \OIII/\CII\ line ratio correlates with the EW(\OIII$+$H$\beta$),
because bursty star formation not only elevates the \OIII/\CII\ line ratio (see Section~\ref{subsec:global_emission_properties}) but also
enhances the the emission lines (\OIII$+$H$\beta$) relative to the underlying stellar continuum. Figure \ref{fig:OIII_CII_EW} presents the relation between the \OIII/\CII ratio and EW(\OIII$+$H$\beta$) for our simulation results with SFR $ > 10^{-3} \, M_\odot {\rm yr^{-1}}$. We have calculated the stellar continuum by applying BPASS {\it single} SED to each stellar particle as a function of age, stellar mass, and metallicity. The EW of an emission line is defined as \citep{Inoue:2011} 
\begin{equation}
{\rm EW}_{\rm line} \equiv \frac{L_{\rm line}}{L_{*, \lambda {\rm line}}},
\end{equation}
where $L_{\rm line}$ is the line luminosity and $L_{*, \lambda {\rm line}}$ is the stellar luminosity density at the line wavelength $\lambda_{\rm line}$. 
Although we neglect the nebular continuum contribution to the continuum, which would reduce the EW, the effect is expected to be modest, at most a factor of 3 \citep{Inoue:2011}.
The $\OIII_{4959, 5007}+{\rm H}\beta$ EW is obtained as 
\begin{align}
{\rm EW}(\OIII_{4959, 5007}&+{\rm H}\beta) = {\rm EW}(\OIII_{4959}) \notag \\
&+ {\rm EW}(\OIII_{5007}) + {\rm EW}({\rm H}\beta). 
\end{align}
Our simulated galaxy exhibits large equivalent widths of 500-2000 \AA, reflecting the very young stellar populations with ages less than 10 Myrs. 
The distribution of our simulation results also encompasses recent high-redshift observations, which report ${\rm EW}(\OIII_{4959, 5007}+{\rm H}\beta) = 500 -1300 \, {\rm \AA}$ and \OIII/\CII $\simeq$ 3-10 \citep{Witstok:2022, Schouws:2025}. In this regime, our simulation results exhibit high ionization parameters, 
with $\log U_{\rm ion}\gtrsim -2$.

\subsection{Caveats} \label{subsec:caveates}
There are some caveats for this study. 
First of all, our analysis focuses on a single low-mass system followed across different redshifts. It is therefore unclear whether the results from such a small sample can be generalized to the broader galaxy population.
Our simulation achieves 0.1 pc resolution and includes on-the-fly radiation transfer, making it computationally expensive to treat larger samples. 
The primary aim of this study is to investigate the physical origins of emission-line ratios by post-processing an existing high-resolution RHD simulation \citep{Sugimura:2024}. Therefore, while it is crucial to increase the sample size for a statistical discussion, performing new simulations for other halos is left for future work.

Second, our galaxy is less massive than the observed high-z galaxies with $M_* \gtrsim 10^8\, M_\odot$. However, it is currently computationally infeasible for zoom-in simulations to simultaneously follow the formation and evolution of such massive galaxies while maintaining 0.1 pc resolution and on-the-fly RT. 
Nonetheless, if the cluster-scale phenomena identified here also take place in star-forming regions of more massive and higher-metallicity galaxies at $z \gtrsim 6$, our explanation should remain valid and naturally account for the large \OIII/\CII ratios observed in these systems. To test this hypothesis, we performed additional CLOUDY calculations across a range of metallicities ($0.005$ to 0.3 $Z_\odot$) using BPASS SEDs with ages of 3 Myr. The \OIII/\CII ratios remain nearly constant with metallicity at fixed $U$ and $N_{\rm H}$. Since simulations of more massive EoR galaxies find similar $U$ and $N_{\rm H}$ values in star-forming regions \citep[$\log U\sim -1.5$ to $-2$, $\log N_{\rm H} \sim 20–22\, {\rm cm^{-2}}$;][]{Nakazato:2023, Kohandel:2025, Gelli:2025}, our predictions should be applicable to higher-mass and/or higher-metallicity systems. We caution, however, that this extrapolation neglects possible additional effects such as large-scale (kpc-scale) geometry and CGM \CII emission.

Our simulated galaxy will evolve into a dwarf at $z = 0$, and at $z > 10$ its stellar mass is only $M_* \sim 10^{5-6}\, M_\odot$, comparable to stellar clusters. Such low-mass galaxies can nevertheless be detected through gravitational lensing. Indeed, several proto-stellar clusters with $M_* \sim 10^{5-6}\, M_\odot$ are gravitationally magnified and observed at $z=6-10$ \citep[e.g., ][]{Vanzella:2023_sunrise_arc, Adamo:2024}. Since several new JWST observations targeting low-mass systems at high-$z$ Universe will be conducted in lensing fields (e.g., GO5119, GO6882, GO7049), the ISM states of these galaxies can be compared directly. 

In summary, this study lacks statistics and massive samples. However, our sub-parsec resolution and on-the-fly RT simulations can reveal the most detailed ISM structures and ionizing states at high-z for the first time, and explain the physical origins of the observed emission line ratios. 

\section{Conclusion and Summary} \label{sec:summary}
We present emission-line calculations applied to a zoom-in simulation with 0.1\,pc resolution and on-the-fly radiative transfer.
By combining \textsc{cloudy} with seven parameters, we accurately compute multi-wavelength emission lines ranging from rest-UV to rest-IR.
We then investigate the radiative properties of a galaxy at $z=9-13$ with a snapshot span of 1 Myrs (total 137 snapshots), focusing on the \OIII~88\,\(\mu\mathrm{m}\) and \CII~158\,\(\mu\mathrm{m}\) lines.
Our findings are as follows:

\noindent(i) The \OIII emission mainly originates from central \HII regions with \(\log(U_{\mathrm{ion}})\approx -1\), whereas \CII is predominantly emitted from the surrounding dense PDRs, tracing the overall gas distribution.
Consequently, the spatial distribution of \OIII/\CII is high (\(\gtrsim100\)) within the central \(\sim100\)\,pc but abruptly drops to \(\sim0.01\) just outside the ionizing bubble.

\noindent(ii) The time evolution of \OIII follows that of the SFR averaged over 10\,Myr and the ionization parameter \(U_{\mathrm{ion}}\).
Meanwhile, \CII  traces the FUV radiation field, which varies more smoothly due to continuous FUV emission from older (10--100\,Myr) stellar populations.
As a result, \OIII/\CII largely reflects the evolution of \OIII, and only the most starburst phase at \(z\sim10\) reaches \(\OIII/\CII \ge 10\).

\noindent(iii) The relationship between each emission line and the SFR shows that \OIII--SFR exhibits a steeper slope than local correlations, consistent with recent ALMA observations and other simulations, whereas \CII--SFR is flatter than local analogs.
We successfully reproduce the high \(\OIII/\CII\) ratio seen in observations without introducing exotic abundance ratios of O/C.
The ratio \(\OIII/\CII\) is proportional to the ratios of gas mass \(M_{\mathrm{HII}} / M_{\mathrm{HI}}\), density \(n_{\mathrm{HII}} / n_{\mathrm{HI}}\), and element abundance \(A({\rm O})/A({\rm C})\). 

\noindent(iv) To examine the validity of the derived scaling relation, we derived \(M_{\mathrm{HII}} / M_{\mathrm{HI}}\) for observed $z > 6$ galaxies using ${\rm H\beta}$ and \CII. Both simulations and observations show a correlation between \OIII/\CII and \(M_{\mathrm{HII}} / M_{\mathrm{HI}}\) with the scatter arising from different ratios of gas densities and cooling rates.
We also explore the positive correlation between ${\rm EW}(\OIII_{4959, 5007}+{\rm H}\beta)$ and \OIII/\CII. These quantities also trace high-ionization parameters. Galaxies with high line ratios of \OIII/\CII $ \gtrsim 3$ also have large equivalent widths of 500--2000\,\AA.

In the near future, both ALMA and JWST will probe these detailed ISM states, such as electron densities and ionized and neutral gas mass, enabling more comprehensive investigations of high-redshift galaxies. 

\section*{Acknowledgements}
We are grateful to an anonymous referee for valuable comments that have greatly improved the paper. We thank Takashi Hosokawa and Masato Hagimoto for providing us with valuable comments. YN thanks Hiddo Algera for kindly sharing his results.
 This work made use of v2.3 of the Binary Population and Spectral Synthesis (BPASS) models as described in \citet{Byrne:2022} and \citet{Stanway:2018}. Numerical analyses were carried out on the analysis servers at Center for Computational Astrophysics, National Astronomical Observatory of Japan. The Flatiron Institute is a division of the Simons Foundation.
YN acknowledges funding from JSPS KAKENHI Grant Number 23KJ0728 and a JSR fellowship. 
AKI was supported by JSPS KAKENHI Grant Number 23H00131.

\appendix
\counterwithin{figure}{section}
\section{Sliced maps} \label{sec:slice}
\begin{figure*}
    \centering
    \includegraphics[width =\linewidth, clip]{\figdir/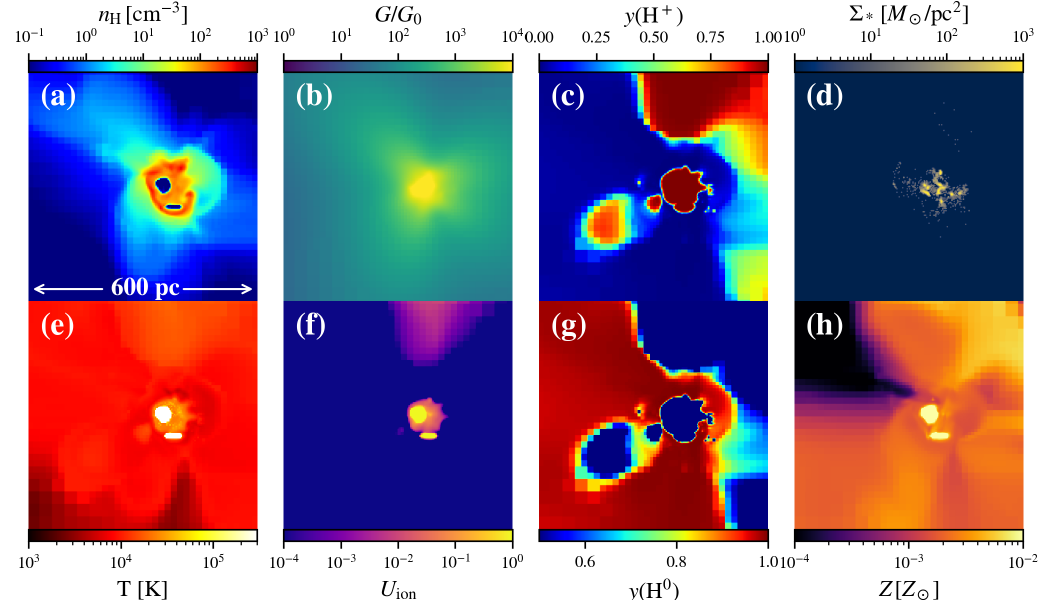}
    \caption{ Same as Figure \ref{fig:projection} but sliced maps at $z = 10.45$. 
    Only $\Sigma_*$ is a projected map.
}
    \label{fig:slice}
\end{figure*}

To complement the projected maps in Figure~\ref{fig:projection}, we present sliced maps in Figure~\ref{fig:slice}, which show the same physical quantities.
We can see that the two central cavities have very high temperatures ($T > 5\times10^5\,\mathrm{K}$) and are metal-enriched ($Z \sim 10^{-2}\,Z_{\odot}$), which are clear signatures of SN explosions. Panel (c) reveals ionized bubbles in the upper and left regions, in addition to the central ones. However, these regions are less apparent in the projected distributions.

\section{Time Evolution of Ionizing Photon Production and FUV Luminosity} \label{sec:bpass_age}
\begin{figure*}
    \centering
    \includegraphics[width =\linewidth, clip]{\figdir/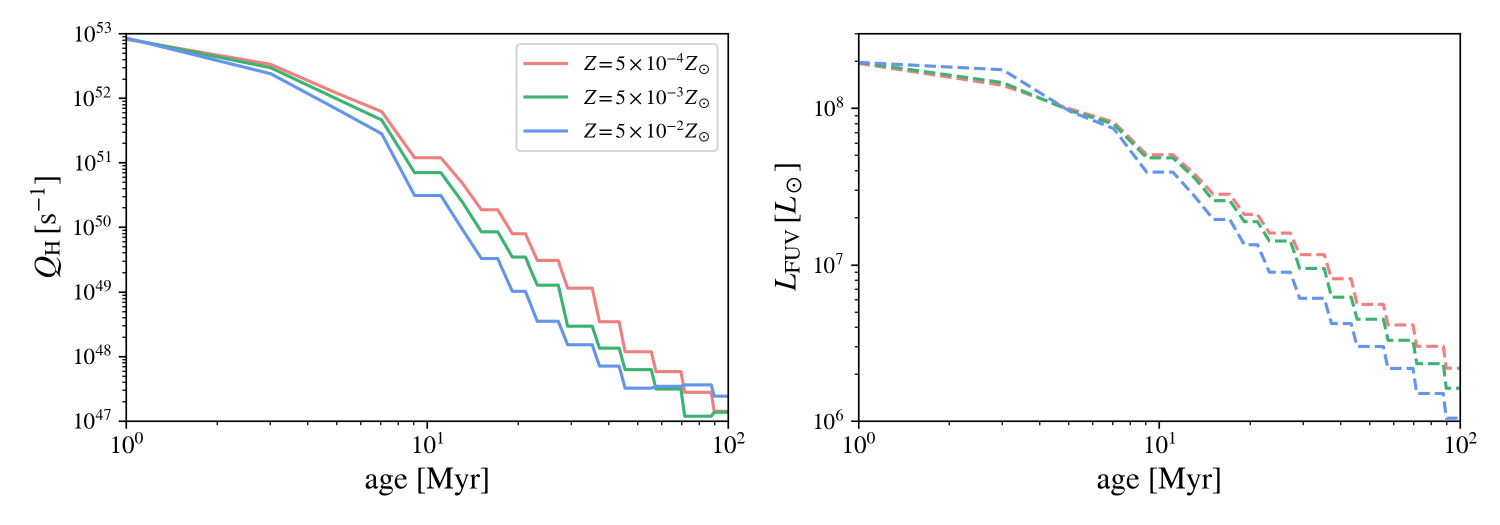}
    \caption{Time evolution of the ionizing photon production rate (left) and FUV luminosity (right) as a function of stellar age, calculated from the BPASS single-star SED model assuming $M_* = 10^6\,M_\odot$ and instantaneous star formation.
We also show models with different metallicities: $Z = 5 \times 10^{-4} Z_\odot$ (red), $Z = 5 \times 10^{-3} Z_\odot$ (green), and $Z = 5 \times 10^{-2} Z_\odot$ (blue).}
    \label{fig:bpass_age}
\end{figure*}

Figure~\ref{fig:bpass_age} shows the time evolution of the ionizing photon production rate ($Q_{\rm H}$) and the FUV luminosity as a function of stellar age. These quantities are calculated using the BPASS single-star SED model \citep{Eldridge:2017}, assuming an instantaneous starburst with a total stellar mass of $M_* = 10^6\,M_\odot$. We find that $Q_{\rm H}$ decreases by about four orders of magnitude from 10 Myr to 100 Myr, while the FUV luminosity decreases by about two orders of magnitude.

\section{Cooling rate} \label{sec:cooling_rate}
A cooling rate for a collisionally excited line in the case of low-density and no stimulated emission is expressed as:
\begin{equation}
\Lambda_{21} = h \nu_{21} q_{12}, \label{eq:cooling_rate_CEL}
\end{equation}
where $\nu$ is the frequency of the emission line and $q_{12}$ is the collisional excitation rate per unit time per unit volume $[{\rm cm^{3} s^{-1}}]$. The cooling rate is defined to calculate volume emissivity as 
\begin{equation}
\epsilon_{21} = n_{\rm c} n_1 \Lambda_{21}, 
\end{equation}
where $n_1$ is the number density of lower levels. Since \OIII and \CII transitions are $^3 P_{1} \rightarrow ^3 P_{0}$ and $^2 P_{3/2} \rightarrow ^2 P_{1/2}$, $n_1$ correspond to the base levels of their ions. In the case of collisions with electrons,  the collision de-excitation $q_{21}$ is written as \citep{Osterbrock:2006}
\begin{align}
&q_{21} = \frac{8.629\times 10^{-6}}{T^{-1/2}} \frac{\Upsilon_{(1,2)}}{g_2}\, [{\rm cm^3 s^{-1}}], \\
&\Upsilon_{(1,2)[{\rm OIII]}} = 0.55 , \\
&\Upsilon_{(1,2)[{\rm CII]}} = 2.15.
\end{align}
Here $\Upsilon_{(1,2)}$ is the collision strength (dimensionless) and the temperature is expressed in kelvin [${\rm K}$].
From eq. (\ref{eq:detailed_balance}), we obtain $q_{12} = q_{21} g_2/g_1 \exp(-T_*/T)$, and the equivalent temperature for \OIII 88\mum and \CII 158\mum are 163 K and 91.2 K, respectively. Finally, eq. (\ref{eq:cooling_rate_CEL}) can be written as 
\begin{align}
\Lambda_{21} &= h\nu_{21} \frac{8.629\times 10^{-6}}{T^{-1/2}} \frac{\Upsilon_{(1,2)}}{g_1}\exp\left(-\frac{T_*}{T}\right) \\
& \simeq h\nu_{21} \frac{8.629\times 10^{-6}}{T^{-1/2}} \frac{\Upsilon_{(1,2)}}{g_1}.\label{eq:cooling_rate_approx}
\end{align}
The approximated eq. (\ref{eq:cooling_rate_approx}) can be satisfied when we consider FIR emission lines from \HII regions, i.e., $T_* \ll T_{\rm e}$ and $\exp\left(-{T_*}/{T}\right)\rightarrow 1$. The cooling rates for \OIII 88\mum and \CII 158 \mum are 
\begin{align}
\Lambda_{[\rm OIII]}(T_{\rm e}) =& 1.07\times 10^{-19} {T_{\rm e}}^{-1/2} \,\, {\rm erg\,s^{-1} cm^3} \\
\Lambda_{[\rm CII]}(T_{\rm e}) =& 1.17\times 10^{-19} {T_{\rm e}}^{-1/2} \,\, {\rm erg\,s^{-1} cm^3}.
\end{align}
We see that $\Lambda_{[\rm OIII]}(T_{\rm e})/\Lambda_{[\rm CII]}(T_{\rm e}) \approx 1$. 

\section{CMB attenuation} \label{sec:CMB_attenuation}
In our \textsc{Cloudy} setup, we include attenuation by the cosmic microwave background (CMB), as detailed in Section~\ref{sec:method}. Below, we quantify its impact, following 
\citet{Goldsmith:2012}, \citet{Arata:2020}, and \citet{Liang:2024}.

For \CII emission, we consider two-level populations and define level 1 (2) as the lower (upper) level. 
The volume emissivity of \CII emission $\epsilon_{\rm [CII]}\, [{\rm erg\,s^{-1}cm^{3}}]$is 
\begin{equation}
    \epsilon_{\rm [CII]} = \left[A_{21}n_2 + B_{21}n_2J_{\nu}(T_{\rm b})-B_{12}n_1 J_\nu(T_{\rm b})\right]h\nu_{\rm[CII]}, \label{eq:cooling_rate_CII}
\end{equation}
where $A_{21}, B_{21}, B_{12}$ are Einstein's coefficients, and they have the following relationship:
\begin{align}
B_{12} &= \frac{g_2}{g_1}B_{21} \\
\frac{A_{21}}{B_{21}} &= \frac{8\pi h\nu_{\rm[CII]}^3}{c^2}.  
\end{align}
The value of $A_{21}$ is $A_{21}=2.29\times 10^{-6}\, {\rm s^{-1}}$ \citep{Nussbaumer_Storey:1981}.
The statistical weight for level 1 (2) is $g_1=2 (g_2=4)$.
Here $J_{\nu}(T_{\rm b})$ is the background radiation at temperature of $T_{\rm b}$, i.e., Blackbody radiation at CMB temperature; 
\begin{equation}
    J_\nu (T_{\rm b}) = \frac{4\pi}{c} B_{\rm bb}(T_{\rm CMB}) = \frac{8\pi h\nu^3}{c^2}\frac{1}{\exp(T_*/T_{\rm CMB})-1},  \label{eq;J_nu_Tb}
\end{equation}
where $T_*$ is an equivalent temperature, $T_* \equiv h\nu_{\rm[CII]}/k_{\rm B} = 91.2 \, {\rm K}$. The excitation temperature of the transition is defined to satisfy the following relationship; 
\begin{equation}
    \frac{n_2}{n_1} = \frac{g_2}{g_1} \exp \left(- \frac{T_*}{T_{\rm ex}}\right). \label{eq:def_Tex}
\end{equation}
By substituting eq.(\ref{eq:def_Tex}) into eq.(\ref{eq:cooling_rate_CII}), we obtain 
\begin{equation}
    \epsilon_{\rm [CII]} = n_2 A_{21}h \nu \left[1 - \frac{\exp(T_*/T_{\rm ex})-1}{\exp(T_*/T_{\rm b})-1}\right] \equiv n_2 A_{21}h \nu \eta
\end{equation}
The $\eta$ is the correction term for background radiation. To calculate this term, we derive $\exp(T_*/T_{\rm ex})$ from the rate equation; 
\begin{equation}
    n_2 \left[A_{21}+B_{21}J_\nu + C_{21}\right] = n_1 \left[B_{12}J_\nu + C_{12}\right]. \label{eq:rate_eq}
\end{equation}
From eq. (\ref{eq:rate_eq}), we get 
\begin{align}
   & \frac{n_2}{n_1} = \frac{B_{12}J_\nu + C_{12}}{A_{21} + B_{21}J_\nu + C_{21}} \\
   \therefore\, &  \frac{g_2}{g_1} \exp\left(- \frac{T_*}{T_{\rm ex}}\right) = \frac{B_{12}J_\nu + n_{\rm c}q_{12}^{\rm c}}{A_{21}+B_{21}J_\nu + n_{\rm c}q_{21}^{\rm c}}, \label{eq:n2_n1}
\end{align}
where $n_{\rm c}$ and $q_{12}^{\rm c}(q_{21}^{\rm c})$ are the number density of a collision partner and collision rate efficiency for the partner. The partners for \HII region, PDR, and molecular gas are electrons, neutral hydrogen atoms, and hydrogen molecules, respectively. The corresponding collision rate efficiencies are \citep{Goldsmith:2012}
\begin{align}
    q_{21}^{e} &= 8.7\times 10^{-8} \, {\rm cm^3s^{-1}}(T/2000)^{-0.37} \\
    q_{21}^{\rm HI} &= 4.0\times 10^{-11} \, {\rm cm^3s^{-1}}(16 + 0.35T^{0.5} + 48T^{-1}) \\
    q_{21}^{\rm H_2} &= 3.8\times 10^{-10}  \, {\rm cm^3s^{-1}} (T/100)^{0.14}, 
\end{align}
where the temperature $T$ is expressed in kelvin [${\rm K}$].
We define $G$ as the same as \citet{Goldsmith:2012, Liang:2024}, $G\equiv 1/[\exp(T_*/T_{\rm b})-1]$, and rewrite eq. (\ref{eq;J_nu_Tb}) as 
\begin{equation}
    B_{21}J_\nu = A_{21} G. \label{eq:def_G}
\end{equation}
From the detailed balance, we obtain 
\begin{equation}
    \frac{q_{12}^{\rm c}}{q_{21}^{\rm c}} = \frac{g_2}{g_1}\exp\left(- \frac{T_*}{T_{\rm kin}}\right). \label{eq:detailed_balance}
\end{equation}
Inserting eq. (\ref{eq:def_G}) and eq. (\ref{eq:detailed_balance}) into eq.(\ref{eq:n2_n1}), we get
\begin{equation}
    \exp\left(\frac{T_*}{T_{\rm ex}}\right) = \frac{A_{21}(1+G) + n_{\rm c}q_{21}^{\rm c}}{GA_{21} + n_{\rm c}q_{21}^{\rm c}\exp\left(- \frac{T_*}{T_{\rm kin}}\right)}. \label{eq:exp_T*_Tex}
\end{equation}
In the case without CMB radiation, $G$ reaches zero and $T_{\rm ex}\rightarrow T_{\rm kin}$. When the second term in the numerator of eq.(\ref{eq:exp_T*_Tex}) is weak, i.e., diffuse gas $n_{\rm c} \ll A_{21}/q_{21}^{\rm c}$, the CMB effect ($G > 0$) becomes large to the excitation temperature $T_{\rm ex}$, decreasing $\eta$. 

To calculate eq. (\ref{eq:exp_T*_Tex}) for \CII emission, we need to consider multi-phase gas; \HII, \HI, and molecular gas. For \HII region, the typical gas temperature is $\sim 10^4\, {\rm K}$, and we can approximate $\exp \left(-T_*/T_{\rm kin}\right)\rightarrow 0$, obtaining 
\begin{equation}
   \exp\left(\frac{T_*}{T_{\rm ex}}\right) = \frac{(1+G)A_{21}+n_{\rm e}q_{21}^{\rm c}}{G A_{21}}. \label{eq:exp_T*_Tex_HII}
\end{equation}
We will obtain the cooling rate from \HII regions by inserting eq.(\ref{eq:exp_T*_Tex_HII}) into eq.(\ref{eq:cooling_rate_CII}). 
For PDR (molecular gas), we replace $n_{\rm c}$ and $q_{21}^{\rm c}$ as $n_{\rm HI}(n_{\rm H_2})$ and $q_{21}^{\rm HI}(q_{21}^{\rm H_2})$, respectively. We then obtain the cooling rate $\Lambda_{\rm HI}(\Lambda_{\rm H_2})$. 
The total luminosity is the sum of the luminosity from each region ($L_{\rm [CII]}=L_{\rm [CII]}^{\rm (HII)} + L_{\rm [CII]}^{\rm (HI)} + L_{\rm [CII]}^{\rm (H_2)}$), but we can approximate $L_{\rm [CII]}\approx L_{\rm [CII]}^{\rm (HI)}$ since the most \CII emission comes from PDR. 
Then we can evaluate the CMB effect as 
\begin{align}
     \frac{L_{\rm [CII]}(T_{\rm CMB})}{L_{\rm [CII]}(T_{\rm b}=0)} &\sim  \frac{L_{\rm [CII]}^{{\rm (HI)}}(T_{\rm CMB})}{L_{\rm [CII]}^{{\rm (HI)}}(T_{\rm b}=0)} \notag \\
     &= \frac{\epsilon_{\rm [CII]}^{\rm (HI)}(T_{\rm CMB})}{\epsilon_{\rm [CII]}^{\rm (HI)}(T_{\rm b}=0)}= \eta^{\rm HI} \notag \\
    &= 1 - \frac{\exp(T_*/T_{\rm ex})-1}{\exp(T_*/T_{\rm b})-1}. 
\end{align}
\begin{figure}
    \centering
    \includegraphics[width =\linewidth, clip]{\figdir/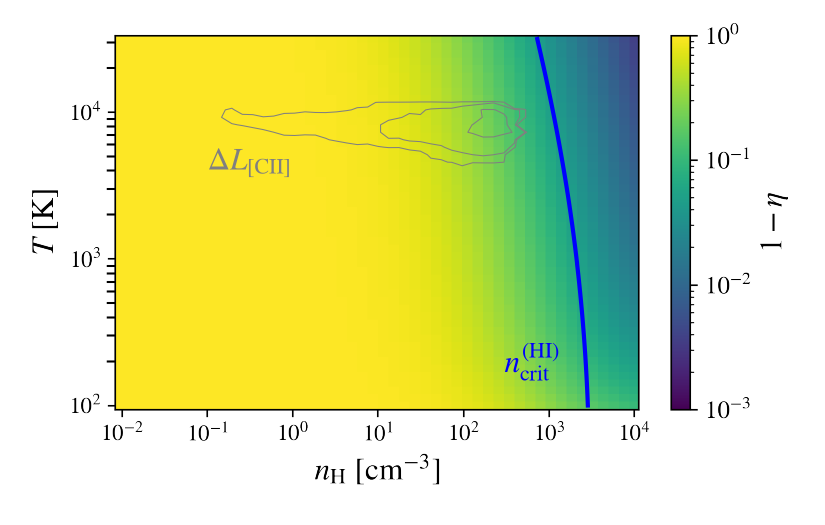}
    \caption{ The CMB effect to \CII 158\mum luminosity at $z=10$ as a function of hydrogen nuclei density and temperature. The color scale indicates the reduction strength $1-\eta$, where $\eta$ is the luminosity ratio of \CII with (without) the CMB stimulated emission and absorption. The blue solid line represents the critical density of hydrogen atoms. The gray contours show the \CII luminosity in the phase-diagram with $\log L_{\rm [CII]}/[L_\odot/\Delta\log n_{\rm H}/\Delta \log T])=-0.5, 0.5, 1.5$, where $\Delta\log n_{\rm H}=0.12$ and $\Delta \log T=0.05$. Note that the contoured \CII luminosities are calculated by considering the CMB effect in CLOUDY.
}
    \label{fig:CMB_effect}
\end{figure}
Figure \ref{fig:CMB_effect} shows how much \CII emission is suppressed by CMB at $z=10.0$. Here, we only consider collisions of neutral hydrogen atoms. For diffused gas with $n_{\rm H} \lesssim 30\, {\rm cm^{-3}}$, the CMB effect becomes significant, resulting in a reduction of luminosity over 80\%. The threshold diffuse gas number density is lower at lower-redshift (e.g., $n_{\rm H} \lesssim 10(1)\, {\rm cm^{-3}}$ at $z=7(4)$). If the gas number density is higher than the critical density $n_{\rm crit}$, the rate equation is dominated by the collision terms, and the CMB effect becomes negligible. 
We also plot the contours for \CII emission as the same as the right panel of Figure \ref{fig:rho_T} and \CII mainly comes from dens and warm gas with $n_{\rm H}=10^{2-3}\, {\rm cm^{-3}}$ and $T= 5000-10^4\, {\rm K}$. In this region, the CMB at $z=10(7)$ affects 10-20 ($\sim$ 5)\%. 

Furthermore, the cosmic microwave background (CMB) at $z=0$ can hinder detectability by adding background noise that obscures the line signal \citep{daCunha:2013,Kohandel:2019}. We do not account for this effect here because our focus is on the intrinsic luminosity of high-redshift galaxies. However, it could become significant when making detailed comparisons of line profiles between simulations and observations.
\bibliography{RAMSES_RT}{}

\begin{thebibliography}{}
\expandafter\ifx\csname natexlab\endcsname\relax\def\natexlab#1{#1}\fi
\providecommand{\url}[1]{\href{#1}{#1}}
\providecommand{\dodoi}[1]{doi:~\href{http://doi.org/#1}{\nolinkurl{#1}}}
\providecommand{\doeprint}[1]{\href{http://ascl.net/#1}{\nolinkurl{http://ascl.net/#1}}}
\providecommand{\doarXiv}[1]{\href{https://arxiv.org/abs/#1}{\nolinkurl{https://arxiv.org/abs/#1}}}

\bibitem[{{Adamo} {et~al.}(2024){Adamo}, {Bradley}, {Vanzella}, {Claeyssens},
  {Welch}, {Diego}, {Mahler}, {Oguri}, {Sharon}, {Abdurro'uf}, {Hsiao}, {Xu},
  {Messa}, {Lassen}, {Zackrisson}, {Brammer}, {Coe}, {Kokorev}, {Ricotti},
  {Zitrin}, {Fujimoto}, {Inoue}, {Resseguier}, {Rigby}, {Jim{\'e}nez-Teja},
  {Windhorst}, {Hashimoto}, \& {Tamura}}]{Adamo:2024}
{Adamo}, A., {Bradley}, L.~D., {Vanzella}, E., {et~al.} 2024, \nat, 632, 513,
  \dodoi{10.1038/s41586-024-07703-7}

\bibitem[{{Algera} {et~al.}(2025){Algera}, {Rowland}, {Smit}, {Fisher},
  {Ramambason}, {Kumari}, {Vallini}, {Inami}, {Nanayakkara}, {Stefanon},
  {Aravena}, {Bakx}, {Bouwens}, {Bowler}, {Cescon}, {Chen}, {Dayal}, {De
  Looze}, {Ferrara}, {Fudamoto}, {Komarova}, {van Leeuwen}, {Ormerod},
  {Schouws}, {Sommovigo}, {Vijayan}, {Wang}, {van der Werf}, \&
  {Witstok}}]{Algera:2025}
{Algera}, H., {Rowland}, L., {Smit}, R., {et~al.} 2025, arXiv e-prints,
  arXiv:2509.16071, \dodoi{10.48550/arXiv.2509.16071}

\bibitem[{{Algera} {et~al.}(2024){Algera}, {Inami}, {Sommovigo}, {Fudamoto},
  {Schneider}, {Graziani}, {Dayal}, {Bouwens}, {Aravena}, {da Cunha},
  {Ferrara}, {Hygate}, {van Leeuwen}, {De Looze}, {Palla}, {Pallottini},
  {Smit}, {Stefanon}, {Topping}, \& {van der Werf}}]{Algera:2024}
{Algera}, H. S.~B., {Inami}, H., {Sommovigo}, L., {et~al.} 2024, \mnras, 527,
  6867, \dodoi{10.1093/mnras/stad3111}

\bibitem[{{Arata} {et~al.}(2020){Arata}, {Yajima}, {Nagamine}, {Abe}, \&
  {Khochfar}}]{Arata:2020}
{Arata}, S., {Yajima}, H., {Nagamine}, K., {Abe}, M., \& {Khochfar}, S. 2020,
  \mnras, 498, 5541, \dodoi{10.1093/mnras/staa2809}

\bibitem[{{Arellano-C{\'o}rdova} {et~al.}(2022){Arellano-C{\'o}rdova}, {Berg},
  {Chisholm}, {Arrabal Haro}, {Dickinson}, {Finkelstein}, {Leclercq}, {Rogers},
  {Simons}, {Skillman}, {Trump}, \& {Kartaltepe}}]{Arellano-Cordova:2022}
{Arellano-C{\'o}rdova}, K.~Z., {Berg}, D.~A., {Chisholm}, J., {et~al.} 2022,
  \apjl, 940, L23, \dodoi{10.3847/2041-8213/ac9ab2}

\bibitem[{{Asada} {et~al.}(2024){Asada}, {Sawicki}, {Abraham}, {Brada{\v{c}}},
  {Brammer}, {Desprez}, {Estrada-Carpenter}, {Iyer}, {Martis}, {Matharu},
  {Mowla}, {Muzzin}, {Noirot}, {Sarrouh}, {Strait}, {Willott}, \&
  {Harshan}}]{Asada:2024}
{Asada}, Y., {Sawicki}, M., {Abraham}, R., {et~al.} 2024, \mnras, 527, 11372,
  \dodoi{10.1093/mnras/stad3902}

\bibitem[{{Bakx} {et~al.}(2020){Bakx}, {Tamura}, {Hashimoto}, {Inoue}, {Lee},
  {Mawatari}, {Ota}, {Umehata}, {Zackrisson}, {Hatsukade}, {Kohno}, {Matsuda},
  {Matsuo}, {Okamoto}, {Shibuya}, {Shimizu}, {Taniguchi}, \&
  {Yoshida}}]{Bakx:2020}
{Bakx}, T. J.~L.~C., {Tamura}, Y., {Hashimoto}, T., {et~al.} 2020, \mnras, 493,
  4294, \dodoi{10.1093/mnras/staa509}

\bibitem[{{Bakx} {et~al.}(2024){Bakx}, {Algera}, {Venemans}, {Sommovigo},
  {Fujimoto}, {Carniani}, {Hagimoto}, {Hashimoto}, {Inoue}, {Salak},
  {Serjeant}, {Vallini}, {Eales}, {Ferrara}, {Fudamoto}, {Imamura}, {Inoue},
  {Knudsen}, {Matsuo}, {Sugahara}, {Tamura}, {Taniguchi}, \&
  {Yamanaka}}]{Bakx:2024}
{Bakx}, T. J.~L.~C., {Algera}, H. S.~B., {Venemans}, B., {et~al.} 2024, \mnras,
  532, 2270, \dodoi{10.1093/mnras/stae1613}

\bibitem[{{Barinovs} {et~al.}(2005){Barinovs}, {van Hemert}, {Krems}, \&
  {Dalgarno}}]{Barinovs:2005}
{Barinovs}, {\u{G}}., {van Hemert}, M.~C., {Krems}, R., \& {Dalgarno}, A. 2005,
  \apj, 620, 537, \dodoi{10.1086/426860}

\bibitem[{{Baumschlager} {et~al.}(2024){Baumschlager}, {Shen}, \&
  {Wadsley}}]{Baumschlager:2024}
{Baumschlager}, B., {Shen}, S., \& {Wadsley}, J.~W. 2024, \aap, 691, A219,
  \dodoi{10.1051/0004-6361/202348164}

\bibitem[{{Behroozi} {et~al.}(2013){Behroozi}, {Wechsler}, \&
  {Wu}}]{Behroozi:2013}
{Behroozi}, P.~S., {Wechsler}, R.~H., \& {Wu}, H.-Y. 2013, \apj, 762, 109,
  \dodoi{10.1088/0004-637X/762/2/109}

\bibitem[{{Brada{\v{c}}} {et~al.}(2017){Brada{\v{c}}}, {Garcia-Appadoo},
  {Huang}, {Vallini}, {Quinn Finney}, {Hoag}, {Lemaux}, {Borello Schmidt},
  {Treu}, {Carilli}, {Dijkstra}, {Ferrara}, {Fontana}, {Jones}, {Ryan}, {Wagg},
  \& {Gonzalez}}]{Bradac:2017}
{Brada{\v{c}}}, M., {Garcia-Appadoo}, D., {Huang}, K.-H., {et~al.} 2017, \apjl,
  836, L2, \dodoi{10.3847/2041-8213/836/1/L2}

\bibitem[{{Bradley} {et~al.}(2024){Bradley}, {Adamo}, {Vanzella}, {Sharon},
  {Brammer}, {Coe}, {Diego}, {Kokorev}, {Mahler}, {Oguri}, {Abdurro'uf},
  {Bhatawdekar}, {Christensen}, {Fujimoto}, {Hashimoto}, {Hsiao}, {Inoue},
  {Jim{\'e}nez-Teja}, {Messa}, {Norman}, {Ricotti}, {Tamura}, {Windhorst},
  {Xu}, \& {Zitrin}}]{Bradley:2024}
{Bradley}, L.~D., {Adamo}, A., {Vanzella}, E., {et~al.} 2024, arXiv e-prints,
  arXiv:2404.10770, \dodoi{10.48550/arXiv.2404.10770}

\bibitem[{{Brauher} {et~al.}(2008){Brauher}, {Dale}, \& {Helou}}]{Brauher:2008}
{Brauher}, J.~R., {Dale}, D.~A., \& {Helou}, G. 2008, \apjs, 178, 280,
  \dodoi{10.1086/590249}

\bibitem[{{Bruzual} \& {Charlot}(2003)}]{Bruzual_Charlot:2003}
{Bruzual}, G., \& {Charlot}, S. 2003, \mnras, 344, 1000,
  \dodoi{10.1046/j.1365-8711.2003.06897.x}

\bibitem[{{Byrne} {et~al.}(2022){Byrne}, {Stanway}, {Eldridge}, {McSwiney}, \&
  {Townsend}}]{Byrne:2022}
{Byrne}, C.~M., {Stanway}, E.~R., {Eldridge}, J.~J., {McSwiney}, L., \&
  {Townsend}, O.~T. 2022, \mnras, 512, 5329, \dodoi{10.1093/mnras/stac807}

\bibitem[{{Calura} {et~al.}(2025){Calura}, {Pascale}, {Agertz}, {Andersson},
  {Lacchin}, {Lupi}, {Meneghetti}, {Nipoti}, {Ragagnin}, {Rosdahl}, {Vanzella},
  {Vesperini}, \& {Zanella}}]{Calura:2025}
{Calura}, F., {Pascale}, R., {Agertz}, O., {et~al.} 2025, \aap, 698, A207,
  \dodoi{10.1051/0004-6361/202452876}

\bibitem[{{Cameron} {et~al.}(2023){Cameron}, {Katz}, {Rey}, \&
  {Saxena}}]{Cameron:2023}
{Cameron}, A.~J., {Katz}, H., {Rey}, M.~P., \& {Saxena}, A. 2023, \mnras, 523,
  3516, \dodoi{10.1093/mnras/stad1579}

\bibitem[{{Capak} {et~al.}(2015){Capak}, {Carilli}, {Jones}, {Casey},
  {Riechers}, {Sheth}, {Carollo}, {Ilbert}, {Karim}, {Lefevre}, {Lilly},
  {Scoville}, {Smolcic}, \& {Yan}}]{Capak:2015}
{Capak}, P.~L., {Carilli}, C., {Jones}, G., {et~al.} 2015, \nat, 522, 455,
  \dodoi{10.1038/nature14500}

\bibitem[{{Carniani} {et~al.}(2018){Carniani}, {Maiolino}, {Smit}, \&
  {Amor{\'\i}n}}]{Carniani:2018}
{Carniani}, S., {Maiolino}, R., {Smit}, R., \& {Amor{\'\i}n}, R. 2018, \apjl,
  854, L7, \dodoi{10.3847/2041-8213/aaab45}

\bibitem[{{Carniani} {et~al.}(2017){Carniani}, {Maiolino}, {Pallottini},
  {Vallini}, {Pentericci}, {Ferrara}, {Castellano}, {Vanzella}, {Grazian},
  {Gallerani}, {Santini}, {Wagg}, \& {Fontana}}]{Carniani:2017}
{Carniani}, S., {Maiolino}, R., {Pallottini}, A., {et~al.} 2017, \aap, 605,
  A42, \dodoi{10.1051/0004-6361/201630366}

\bibitem[{{Carniani} {et~al.}(2024){Carniani}, {D'Eugenio}, {Ji}, {Parlanti},
  {Scholtz}, {Sun}, {Venturi}, {Bakx}, {Curti}, {Maiolino}, {Tacchella},
  {Zavala}, {Hainline}, {Witstok}, {Johnson}, {Alberts}, {Bunker}, {Charlot},
  {Eisenstein}, {Helton}, {Jakobsen}, {Kumari}, {Robertson}, {Saxena},
  {{\"U}bler}, {Williams}, {Willmer}, \& {Willott}}]{Carniani:2024}
{Carniani}, S., {D'Eugenio}, F., {Ji}, X., {et~al.} 2024, arXiv e-prints,
  arXiv:2409.20533, \dodoi{10.48550/arXiv.2409.20533}

\bibitem[{{Casavecchia} {et~al.}(2025){Casavecchia}, {Maio}, {P{\'e}roux}, \&
  {Ciardi}}]{Casavecchia:2025}
{Casavecchia}, B., {Maio}, U., {P{\'e}roux}, C., \& {Ciardi}, B. 2025, \aap,
  693, A119, \dodoi{10.1051/0004-6361/202452282}

\bibitem[{{Chatzikos} {et~al.}(2023){Chatzikos}, {Bianchi}, {Camilloni},
  {Chakraborty}, {Gunasekera}, {Guzm{\'a}n}, {Milby}, {Sarkar}, {Shaw}, {van
  Hoof}, \& {Ferland}}]{Chatzikos:2023}
{Chatzikos}, M., {Bianchi}, S., {Camilloni}, F., {et~al.} 2023, \rmxaa, 59,
  327, \dodoi{10.22201/ia.01851101p.2023.59.02.12}

\bibitem[{{Cormier} {et~al.}(2015){Cormier}, {Madden}, {Lebouteiller}, {Abel},
  {Hony}, {Galliano}, {R{\'e}my-Ruyer}, {Bigiel}, {Baes}, {Boselli},
  {Chevance}, {Cooray}, {De Looze}, {Doublier}, {Galametz}, {Hughes},
  {Karczewski}, {Lee}, {Lu}, \& {Spinoglio}}]{Cormier:2015}
{Cormier}, D., {Madden}, S.~C., {Lebouteiller}, V., {et~al.} 2015, \aap, 578,
  A53, \dodoi{10.1051/0004-6361/201425207}

\bibitem[{{Cormier} {et~al.}(2019){Cormier}, {Abel}, {Hony}, {Lebouteiller},
  {Madden}, {Polles}, {Galliano}, {De Looze}, {Galametz}, \&
  {Lambert-Huyghe}}]{Cormier:2019}
{Cormier}, D., {Abel}, N.~P., {Hony}, S., {et~al.} 2019, \aap, 626, A23,
  \dodoi{10.1051/0004-6361/201834457}

\bibitem[{{da Cunha} {et~al.}(2013){da Cunha}, {Groves}, {Walter}, {Decarli},
  {Weiss}, {Bertoldi}, {Carilli}, {Daddi}, {Elbaz}, {Ivison}, {Maiolino},
  {Riechers}, {Rix}, {Sargent}, \& {Smail}}]{daCunha:2013}
{da Cunha}, E., {Groves}, B., {Walter}, F., {et~al.} 2013, \apj, 766, 13,
  \dodoi{10.1088/0004-637X/766/1/13}

\bibitem[{{Dayal} {et~al.}(2022){Dayal}, {Ferrara}, {Sommovigo}, {Bouwens},
  {Oesch}, {Smit}, {Gonzalez}, {Schouws}, {Stefanon}, {Kobayashi}, {Bremer},
  {Algera}, {Aravena}, {Bowler}, {da Cunha}, {Fudamoto}, {Graziani}, {Hodge},
  {Inami}, {De Looze}, {Pallottini}, {Riechers}, {Schneider}, {Stark}, \&
  {Endsley}}]{Dayal:2022}
{Dayal}, P., {Ferrara}, A., {Sommovigo}, L., {et~al.} 2022, \mnras, 512, 989,
  \dodoi{10.1093/mnras/stac537}

\bibitem[{{De Looze} {et~al.}(2014){De Looze}, {Cormier}, {Lebouteiller},
  {Madden}, {Baes}, {Bendo}, {Boquien}, {Boselli}, {Clements}, {Cortese},
  {Cooray}, {Galametz}, {Galliano}, {Graci{\'a}-Carpio}, {Isaak}, {Karczewski},
  {Parkin}, {Pellegrini}, {R{\'e}my-Ruyer}, {Spinoglio}, {Smith}, \&
  {Sturm}}]{De_Looze:2014}
{De Looze}, I., {Cormier}, D., {Lebouteiller}, V., {et~al.} 2014, \aap, 568,
  A62, \dodoi{10.1051/0004-6361/201322489}

\bibitem[{{D'Eugenio} {et~al.}(2024){D'Eugenio}, {Maiolino}, {Carniani},
  {Chevallard}, {Curtis-Lake}, {Witstok}, {Charlot}, {Baker}, {Arribas},
  {Boyett}, {Bunker}, {Curti}, {Eisenstein}, {Hainline}, {Ji}, {Johnson},
  {Kumari}, {Looser}, {Nakajima}, {Nelson}, {Rieke}, {Robertson}, {Scholtz},
  {Smit}, {Sun}, {Venturi}, {Tacchella}, {{\"U}bler}, {Willmer}, \&
  {Willott}}]{D'Eugenio:2024}
{D'Eugenio}, F., {Maiolino}, R., {Carniani}, S., {et~al.} 2024, \aap, 689,
  A152, \dodoi{10.1051/0004-6361/202348636}

\bibitem[{{D{\'\i}az-Santos} {et~al.}(2017){D{\'\i}az-Santos}, {Armus},
  {Charmandaris}, {Lu}, {Stierwalt}, {Stacey}, {Malhotra}, {van der Werf},
  {Howell}, {Privon}, {Mazzarella}, {Goldsmith}, {Murphy}, {Barcos-Mu{\~n}oz},
  {Linden}, {Inami}, {Larson}, {Evans}, {Appleton}, {Iwasawa}, {Lord},
  {Sanders}, \& {Surace}}]{Diaz-Santos:2017}
{D{\'\i}az-Santos}, T., {Armus}, L., {Charmandaris}, V., {et~al.} 2017, \apj,
  846, 32, \dodoi{10.3847/1538-4357/aa81d7}

\bibitem[{{Draine}(2011)}]{Draine:2011}
{Draine}, B.~T. 2011, {Physics of the Interstellar and Intergalactic Medium}

\bibitem[{{Draine} \& {Bertoldi}(1996)}]{Draine_Bertoldi:1996}
{Draine}, B.~T., \& {Bertoldi}, F. 1996, \apj, 468, 269, \dodoi{10.1086/177689}

\bibitem[{{Eldridge} {et~al.}(2017){Eldridge}, {Stanway}, {Xiao}, {McClelland},
  {Taylor}, {Ng}, {Greis}, \& {Bray}}]{Eldridge:2017}
{Eldridge}, J.~J., {Stanway}, E.~R., {Xiao}, L., {et~al.} 2017, \pasa, 34,
  e058, \dodoi{10.1017/pasa.2017.51}

\bibitem[{{Endsley} {et~al.}(2024){Endsley}, {Chisholm}, {Stark}, {Topping}, \&
  {Whitler}}]{Endsley:2024}
{Endsley}, R., {Chisholm}, J., {Stark}, D.~P., {Topping}, M.~W., \& {Whitler},
  L. 2024, arXiv e-prints, arXiv:2410.01905, \dodoi{10.48550/arXiv.2410.01905}

\bibitem[{{Ferland} {et~al.}(2017){Ferland}, {Chatzikos}, {Guzm{\'a}n},
  {Lykins}, {van Hoof}, {Williams}, {Abel}, {Badnell}, {Keenan}, {Porter}, \&
  {Stancil}}]{Ferland:2017}
{Ferland}, G.~J., {Chatzikos}, M., {Guzm{\'a}n}, F., {et~al.} 2017, \rmxaa, 53,
  385, \dodoi{10.48550/arXiv.1705.10877}

\bibitem[{{Fern{\'a}ndez-Ontiveros} {et~al.}(2016){Fern{\'a}ndez-Ontiveros},
  {Spinoglio}, {Pereira-Santaella}, {Malkan}, {Andreani}, \&
  {Dasyra}}]{Fernandez-Ontiveros:2016}
{Fern{\'a}ndez-Ontiveros}, J.~A., {Spinoglio}, L., {Pereira-Santaella}, M.,
  {et~al.} 2016, \apjs, 226, 19, \dodoi{10.3847/0067-0049/226/2/19}

\bibitem[{{Finkelman} {et~al.}(2010){Finkelman}, {Brosch}, {Funes}, {Kniazev},
  \& {V{\"a}is{\"a}nen}}]{Finkelman:2010}
{Finkelman}, I., {Brosch}, N., {Funes}, J.~G., {Kniazev}, A.~Y., \&
  {V{\"a}is{\"a}nen}, P. 2010, \mnras, 407, 2475,
  \dodoi{10.1111/j.1365-2966.2010.17070.x}

\bibitem[{{Fudamoto} {et~al.}(2024){Fudamoto}, {Inoue}, {Coe}, {Welch},
  {Acebron}, {Ricotti}, {Mandelker}, {Windhorst}, {Xu}, {Sugahara}, {Bauer},
  {Brada{\v{c}}}, {Bradley}, {Diego}, {Florian}, {Frye}, {Fujimoto},
  {Hashimoto}, {Henry}, {Mahler}, {Oesch}, {Ravindranath}, {Rigby}, {Sharon},
  {Strait}, {Tamura}, {Trenti}, {Vanzella}, {Zackrisson}, \&
  {Zitrin}}]{Fudamoto:2024}
{Fudamoto}, Y., {Inoue}, A.~K., {Coe}, D., {et~al.} 2024, \apj, 961, 71,
  \dodoi{10.3847/1538-4357/ad0f95}

\bibitem[{{Fudamoto} {et~al.}(2025){Fudamoto}, {Inoue}, {Bouwens}, {Inami},
  {Smit}, {Stark}, {Aravena}, {Pallottini}, {Hashimoto}, {Oguri}, {Bowler}, {da
  Cunha}, {Dayal}, {Ferrara}, {Fujimoto}, {Heintz}, {Hygate}, {van Leeuwen},
  {De Looze}, {Rowland}, {Stefanon}, {Sugahara}, {Witstok}, \& {van der
  Werf}}]{Fudamoto:2025}
{Fudamoto}, Y., {Inoue}, A.~K., {Bouwens}, R., {et~al.} 2025, arXiv e-prints,
  arXiv:2504.03831, \dodoi{10.48550/arXiv.2504.03831}

\bibitem[{{Fujimoto} {et~al.}(2019){Fujimoto}, {Ouchi}, {Ferrara},
  {Pallottini}, {Ivison}, {Behrens}, {Gallerani}, {Arata}, {Yajima}, \&
  {Nagamine}}]{Fujimoto:2019}
{Fujimoto}, S., {Ouchi}, M., {Ferrara}, A., {et~al.} 2019, \apj, 887, 107,
  \dodoi{10.3847/1538-4357/ab480f}

\bibitem[{{Fujimoto} {et~al.}(2024){Fujimoto}, {Ouchi}, {Nakajima}, {Harikane},
  {Isobe}, {Brammer}, {Oguri}, {Gim{\'e}nez-Arteaga}, {Heintz}, {Kokorev},
  {Bauer}, {Ferrara}, {Kojima}, {Lagos}, {Laura}, {Schaerer}, {Shimasaku},
  {Hatsukade}, {Kohno}, {Sun}, {Valentino}, {Watson}, {Fudamoto}, {Inoue},
  {Gonz{\'a}lez-L{\'o}pez}, {Koekemoer}, {Knudsen}, {Lee}, {Magdis}, {Richard},
  {Strait}, {Sugahara}, {Tamura}, {Toft}, {Umehata}, \&
  {Walth}}]{Fujimoto:2024}
{Fujimoto}, S., {Ouchi}, M., {Nakajima}, K., {et~al.} 2024, \apj, 964, 146,
  \dodoi{10.3847/1538-4357/ad235c}

\bibitem[{{Garcia} {et~al.}(2025){Garcia}, {Ricotti}, \&
  {Sugimura}}]{Garcia:2025}
{Garcia}, F. A.~B., {Ricotti}, M., \& {Sugimura}, K. 2025, arXiv e-prints,
  arXiv:2503.08779, \dodoi{10.48550/arXiv.2503.08779}

\bibitem[{{Garcia} {et~al.}(2023){Garcia}, {Ricotti}, {Sugimura}, \&
  {Park}}]{Garcia:2023}
{Garcia}, F. A.~B., {Ricotti}, M., {Sugimura}, K., \& {Park}, J. 2023, \mnras,
  522, 2495, \dodoi{10.1093/mnras/stad1092}

\bibitem[{{Gelli} {et~al.}(2025){Gelli}, {Mason}, {Pallottini}, {Heintz},
  {Chen}, {D'Odorico}, {Ferrara}, {Fynbo}, {Kohandel}, {Pollock}, {Robinson},
  \& {Salvadori}}]{Gelli:2025}
{Gelli}, V., {Mason}, C., {Pallottini}, A., {et~al.} 2025, arXiv e-prints,
  arXiv:2510.01315, \dodoi{10.48550/arXiv.2510.01315}

\bibitem[{{Goldsmith} {et~al.}(2012){Goldsmith}, {Langer}, {Pineda}, \&
  {Velusamy}}]{Goldsmith:2012}
{Goldsmith}, P.~F., {Langer}, W.~D., {Pineda}, J.~L., \& {Velusamy}, T. 2012,
  \apjs, 203, 13, \dodoi{10.1088/0067-0049/203/1/13}

\bibitem[{{Grevesse} {et~al.}(2010){Grevesse}, {Asplund}, {Sauval}, \&
  {Scott}}]{Grevesse:2010}
{Grevesse}, N., {Asplund}, M., {Sauval}, A.~J., \& {Scott}, P. 2010, \apss,
  328, 179, \dodoi{10.1007/s10509-010-0288-z}

\bibitem[{{Guevara} {et~al.}(2020){Guevara}, {Stutzki}, {Ossenkopf-Okada},
  {Simon}, {P{\'e}rez-Beaupuits}, {Beuther}, {Bihr}, {Higgins}, {Graf}, \&
  {G{\"u}sten}}]{Guevara:2020}
{Guevara}, C., {Stutzki}, J., {Ossenkopf-Okada}, V., {et~al.} 2020, \aap, 636,
  A16, \dodoi{10.1051/0004-6361/201834380}

\bibitem[{{Gullberg} {et~al.}(2015){Gullberg}, {De Breuck}, {Vieira},
  {Wei{\ss}}, {Aguirre}, {Aravena}, {B{\'e}thermin}, {Bradford}, {Bothwell},
  {Carlstrom}, {Chapman}, {Fassnacht}, {Gonzalez}, {Greve}, {Hezaveh},
  {Holzapfel}, {Husband}, {Ma}, {Malkan}, {Marrone}, {Menten}, {Murphy},
  {Reichardt}, {Spilker}, {Stark}, {Strandet}, \& {Welikala}}]{Gullberg:2015}
{Gullberg}, B., {De Breuck}, C., {Vieira}, J.~D., {et~al.} 2015, \mnras, 449,
  2883, \dodoi{10.1093/mnras/stv372}

\bibitem[{{Hagimoto} {et~al.}(2025){Hagimoto}, {Tamura}, {Inoue}, {Umehata},
  {Bakx}, {Hashimoto}, {Mawatari}, {Sugahara}, {Fudamoto}, {Harikane},
  {Matsuo}, \& {Taniguchi}}]{Hagimoto:2025}
{Hagimoto}, M., {Tamura}, Y., {Inoue}, A.~K., {et~al.} 2025, \apj, 990, 29,
  \dodoi{10.3847/1538-4357/ade87e}

\bibitem[{{Hahn} \& {Abel}(2011)}]{Hahn_Abel:2011}
{Hahn}, O., \& {Abel}, T. 2011, \mnras, 415, 2101,
  \dodoi{10.1111/j.1365-2966.2011.18820.x}

\bibitem[{{Harikane} {et~al.}(2020){Harikane}, {Ouchi}, {Inoue}, {Matsuoka},
  {Tamura}, {Bakx}, {Fujimoto}, {Moriwaki}, {Ono}, {Nagao}, {Tadaki}, {Kojima},
  {Shibuya}, {Egami}, {Ferrara}, {Gallerani}, {Hashimoto}, {Kohno}, {Matsuda},
  {Matsuo}, {Pallottini}, {Sugahara}, \& {Vallini}}]{Harikane:2020}
{Harikane}, Y., {Ouchi}, M., {Inoue}, A.~K., {et~al.} 2020, \apj, 896, 93,
  \dodoi{10.3847/1538-4357/ab94bd}

\bibitem[{{Harikane} {et~al.}(2025){Harikane}, {Sanders}, {Ellis}, {Jones},
  {Ouchi}, {Laporte}, {Roberts-Borsani}, {Katz}, {Nakajima}, {Ono}, \&
  {Gupta}}]{Harikane:2025}
{Harikane}, Y., {Sanders}, R.~L., {Ellis}, R., {et~al.} 2025, arXiv e-prints,
  arXiv:2505.09186, \dodoi{10.48550/arXiv.2505.09186}

\bibitem[{{Harshan} {et~al.}(2024){Harshan}, {Tripodi}, {Martis},
  {Rihtar{\v{s}}i{\v{c}}}, {Brada{\v{c}}}, {Asada}, {Brammer}, {Desprez},
  {Estrada-Carpenter}, {Matharu}, {Markov}, {Muzzin}, {Mowla}, {Noirot},
  {Sarrouh}, {Sawicki}, {Strait}, \& {Willott}}]{Harshan:2024}
{Harshan}, A., {Tripodi}, R., {Martis}, N.~S., {et~al.} 2024, \apjl, 977, L36,
  \dodoi{10.3847/2041-8213/ad9741}

\bibitem[{{Hartley} \& {Ricotti}(2016)}]{Hartley_Ricotti:2016}
{Hartley}, B., \& {Ricotti}, M. 2016, \mnras, 462, 1164,
  \dodoi{10.1093/mnras/stw1562}

\bibitem[{{Hashimoto} {et~al.}(2018){Hashimoto}, {Laporte}, {Mawatari},
  {Ellis}, {Inoue}, {Zackrisson}, {Roberts-Borsani}, {Zheng}, {Tamura},
  {Bauer}, {Fletcher}, {Harikane}, {Hatsukade}, {Hayatsu}, {Matsuda}, {Matsuo},
  {Okamoto}, {Ouchi}, {Pell{\'o}}, {Rydberg}, {Shimizu}, {Taniguchi},
  {Umehata}, \& {Yoshida}}]{Hashimoto:2018}
{Hashimoto}, T., {Laporte}, N., {Mawatari}, K., {et~al.} 2018, \nat, 557, 392,
  \dodoi{10.1038/s41586-018-0117-z}

\bibitem[{{Hashimoto} {et~al.}(2019){Hashimoto}, {Inoue}, {Mawatari}, {Tamura},
  {Matsuo}, {Furusawa}, {Harikane}, {Shibuya}, {Knudsen}, {Kohno}, {Ono},
  {Zackrisson}, {Okamoto}, {Kashikawa}, {Oesch}, {Ouchi}, {Ota}, {Shimizu},
  {Taniguchi}, {Umehata}, \& {Watson}}]{Hashimoto:2019}
{Hashimoto}, T., {Inoue}, A.~K., {Mawatari}, K., {et~al.} 2019, \pasj, 71, 71,
  \dodoi{10.1093/pasj/psz049}

\bibitem[{{He} {et~al.}(2019){He}, {Ricotti}, \& {Geen}}]{He:2019}
{He}, C.-C., {Ricotti}, M., \& {Geen}, S. 2019, \mnras, 489, 1880,
  \dodoi{10.1093/mnras/stz2239}

\bibitem[{{Heintz} {et~al.}(2021){Heintz}, {Watson}, {Oesch}, {Narayanan}, \&
  {Madden}}]{Heintz:2021}
{Heintz}, K.~E., {Watson}, D., {Oesch}, P.~A., {Narayanan}, D., \& {Madden},
  S.~C. 2021, \apj, 922, 147, \dodoi{10.3847/1538-4357/ac2231}

\bibitem[{{Heintz} {et~al.}(2023){Heintz}, {Gim{\'e}nez-Arteaga}, {Fujimoto},
  {Brammer}, {Espada}, {Gillman}, {Gonz{\'a}lez-L{\'o}pez}, {Greve},
  {Harikane}, {Hatsukade}, {Knudsen}, {Koekemoer}, {Kohno}, {Kokorev}, {Lee},
  {Magdis}, {Nelson}, {Rizzo}, {Sanders}, {Schaerer}, {Shapley}, {Strait},
  {Toft}, {Valentino}, {van der Wel}, {Vijayan}, {Watson}, {Bauer},
  {Christiansen}, \& {Wilson}}]{Heintz:2023}
{Heintz}, K.~E., {Gim{\'e}nez-Arteaga}, C., {Fujimoto}, S., {et~al.} 2023,
  \apjl, 944, L30, \dodoi{10.3847/2041-8213/acb2cf}

\bibitem[{{Heintz} {et~al.}(2025){Heintz}, {Pollock}, {Witstok}, {Carniani},
  {Hainline}, {D'Eugenio}, {Terp}, {Saxena}, \& {Watson}}]{Heintz:2025}
{Heintz}, K.~E., {Pollock}, C., {Witstok}, J., {et~al.} 2025, arXiv e-prints,
  arXiv:2502.06016, \dodoi{10.48550/arXiv.2502.06016}

\bibitem[{{Helton} {et~al.}(2025){Helton}, {Rieke}, {Alberts}, {Wu},
  {Eisenstein}, {Hainline}, {Carniani}, {Ji}, {Baker}, {Bhatawdekar}, {Bunker},
  {Cargile}, {Charlot}, {Chevallard}, {D'Eugenio}, {Egami}, {Johnson}, {Jones},
  {Lyu}, {Maiolino}, {P{\'e}rez-Gonz{\'a}lez}, {Rieke}, {Robertson}, {Saxena},
  {Scholtz}, {Shivaei}, {Sun}, {Tacchella}, {Whitler}, {Williams}, {Willmer},
  {Willott}, {Witstok}, \& {Zhu}}]{Helton:2025}
{Helton}, J.~M., {Rieke}, G.~H., {Alberts}, S., {et~al.} 2025, Nature
  Astronomy, \dodoi{10.1038/s41550-025-02503-z}

\bibitem[{{Howell} {et~al.}(2010){Howell}, {Armus}, {Mazzarella}, {Evans},
  {Surace}, {Sanders}, {Petric}, {Appleton}, {Bothun}, {Bridge}, {Chan},
  {Charmandaris}, {Frayer}, {Haan}, {Inami}, {Kim}, {Lord}, {Madore},
  {Melbourne}, {Schulz}, {U}, {Vavilkin}, {Veilleux}, \& {Xu}}]{Howell:2010}
{Howell}, J.~H., {Armus}, L., {Mazzarella}, J.~M., {et~al.} 2010, \apj, 715,
  572, \dodoi{10.1088/0004-637X/715/1/572}

\bibitem[{{Inoue}(2011)}]{Inoue:2011}
{Inoue}, A.~K. 2011, \mnras, 415, 2920,
  \dodoi{10.1111/j.1365-2966.2011.18906.x}

\bibitem[{{Inoue} {et~al.}(2014){Inoue}, {Shimizu}, {Tamura}, {Matsuo},
  {Okamoto}, \& {Yoshida}}]{Inoue:2014}
{Inoue}, A.~K., {Shimizu}, I., {Tamura}, Y., {et~al.} 2014, \apjl, 780, L18,
  \dodoi{10.1088/2041-8205/780/2/L18}

\bibitem[{{Inoue} {et~al.}(2016){Inoue}, {Tamura}, {Matsuo}, {Mawatari},
  {Shimizu}, {Shibuya}, {Ota}, {Yoshida}, {Zackrisson}, {Kashikawa}, {Kohno},
  {Umehata}, {Hatsukade}, {Iye}, {Matsuda}, {Okamoto}, \&
  {Yamaguchi}}]{Inoue:2016}
{Inoue}, A.~K., {Tamura}, Y., {Matsuo}, H., {et~al.} 2016, Science, 352, 1559,
  \dodoi{10.1126/science.aaf0714}

\bibitem[{{Jackson} {et~al.}(2020){Jackson}, {Allingham}, {Killerby-Smith},
  {Whitaker}, {Smith}, {Contreras}, {Guzm{\'a}n}, {Hogge}, {Sanhueza}, \&
  {Stephens}}]{Jackson:2020}
{Jackson}, J.~M., {Allingham}, D., {Killerby-Smith}, N., {et~al.} 2020, \apj,
  904, 18, \dodoi{10.3847/1538-4357/abba2e}

\bibitem[{{Jones} {et~al.}(2024){Jones}, {Bowler}, {Bunker}, {Arribas},
  {Carniani}, {Charlot}, {Perna}, {Rodr{\'\i}guez Del Pino}, {{\"U}bler},
  {Willott}, {Chevallard}, {Cresci}, {Parlanti}, {Scholtz}, \&
  {Venturi}}]{Jones:2024}
{Jones}, G.~C., {Bowler}, R., {Bunker}, A.~J., {et~al.} 2024, arXiv e-prints,
  arXiv:2412.15027, \dodoi{10.48550/arXiv.2412.15027}

\bibitem[{{Kanekar} {et~al.}(2013){Kanekar}, {Wagg}, {Chary}, \&
  {Carilli}}]{Kanekar:2013}
{Kanekar}, N., {Wagg}, J., {Chary}, R.~R., \& {Carilli}, C.~L. 2013, \apjl,
  771, L20, \dodoi{10.1088/2041-8205/771/2/L20}

\bibitem[{{Kannan} {et~al.}(2022){Kannan}, {Smith}, {Garaldi}, {Shen},
  {Vogelsberger}, {Pakmor}, {Springel}, \& {Hernquist}}]{Kannan:2022}
{Kannan}, R., {Smith}, A., {Garaldi}, E., {et~al.} 2022, \mnras, 514, 3857,
  \dodoi{10.1093/mnras/stac1557}

\bibitem[{{Katz} {et~al.}(2017){Katz}, {Kimm}, {Sijacki}, \&
  {Haehnelt}}]{Katz:2017}
{Katz}, H., {Kimm}, T., {Sijacki}, D., \& {Haehnelt}, M.~G. 2017, \mnras, 468,
  4831, \dodoi{10.1093/mnras/stx608}

\bibitem[{{Katz} {et~al.}(2024){Katz}, {Rey}, {Cadiou}, {Kimm}, \&
  {Agertz}}]{Katz:2024}
{Katz}, H., {Rey}, M.~P., {Cadiou}, C., {Kimm}, T., \& {Agertz}, O. 2024, arXiv
  e-prints, arXiv:2411.07282, \dodoi{10.48550/arXiv.2411.07282}

\bibitem[{{Katz} {et~al.}(2019){Katz}, {Galligan}, {Kimm}, {Rosdahl},
  {Haehnelt}, {Blaizot}, {Devriendt}, {Slyz}, {Laporte}, \&
  {Ellis}}]{Katz:2019}
{Katz}, H., {Galligan}, T.~P., {Kimm}, T., {et~al.} 2019, \mnras, 487, 5902,
  \dodoi{10.1093/mnras/stz1672}

\bibitem[{{Katz} {et~al.}(2022{\natexlab{a}}){Katz}, {Rosdahl}, {Kimm},
  {Garel}, {Blaizot}, {Haehnelt}, {Michel-Dansac}, {Martin-Alvarez},
  {Devriendt}, {Slyz}, {Teyssier}, {Ocvirk}, {Laporte}, \& {Ellis}}]{Katz:2022}
{Katz}, H., {Rosdahl}, J., {Kimm}, T., {et~al.} 2022{\natexlab{a}}, \mnras,
  510, 5603, \dodoi{10.1093/mnras/stac028}

\bibitem[{{Katz} {et~al.}(2022{\natexlab{b}}){Katz}, {Garel}, {Rosdahl},
  {Mauerhofer}, {Kimm}, {Blaizot}, {Michel-Dansac}, {Devriendt}, {Slyz}, \&
  {Haehnelt}}]{Katz:2022_MgII}
{Katz}, H., {Garel}, T., {Rosdahl}, J., {et~al.} 2022{\natexlab{b}}, \mnras,
  515, 4265, \dodoi{10.1093/mnras/stac1437}

\bibitem[{{Keilmann} {et~al.}(2025){Keilmann}, {Dannhauer}, {Kabanovic},
  {Schneider}, {Ossenkopf-Okada}, {Simon}, {Bonne}, {Goldsmith}, {G{\"u}sten},
  {Zavagno}, {Stutzki}, {Riechers}, {R{\"o}llig}, {Verbena}, \&
  {Tielens}}]{Keilmann:2025}
{Keilmann}, E., {Dannhauer}, S., {Kabanovic}, S., {et~al.} 2025, \aap, 697, L2,
  \dodoi{10.1051/0004-6361/202453445}

\bibitem[{{Kimm} {et~al.}(2015){Kimm}, {Cen}, {Devriendt}, {Dubois}, \&
  {Slyz}}]{Kimm:2015}
{Kimm}, T., {Cen}, R., {Devriendt}, J., {Dubois}, Y., \& {Slyz}, A. 2015,
  \mnras, 451, 2900, \dodoi{10.1093/mnras/stv1211}

\bibitem[{{Kimm} {et~al.}(2017){Kimm}, {Katz}, {Haehnelt}, {Rosdahl},
  {Devriendt}, \& {Slyz}}]{Kimm:2017}
{Kimm}, T., {Katz}, H., {Haehnelt}, M., {et~al.} 2017, \mnras, 466, 4826,
  \dodoi{10.1093/mnras/stx052}

\bibitem[{{Knudsen} {et~al.}(2016){Knudsen}, {Richard}, {Kneib}, {Jauzac},
  {Cl{\'e}ment}, {Drouart}, {Egami}, \& {Lindroos}}]{Knudsen:2016}
{Knudsen}, K.~K., {Richard}, J., {Kneib}, J.-P., {et~al.} 2016, \mnras, 462,
  L6, \dodoi{10.1093/mnrasl/slw114}

\bibitem[{{Knudsen} {et~al.}(2025){Knudsen}, {Watson}, {Richard}, {Frayer},
  {Fujimoto}, {Akins}, {Bakx}, {Bonaventura}, {Brammer}, {Christensen},
  {Hashimoto}, {Inoue}, {Matsuo}, {Micha{\l}owski}, \& {Zavala}}]{Knudsen:2025}
{Knudsen}, K.~K., {Watson}, D., {Richard}, J., {et~al.} 2025, \aap, 701, A85,
  \dodoi{10.1051/0004-6361/202453229}

\bibitem[{{Kohandel} {et~al.}(2025){Kohandel}, {Pallottini}, \&
  {Ferrara}}]{Kohandel:2025}
{Kohandel}, M., {Pallottini}, A., \& {Ferrara}, A. 2025, arXiv e-prints,
  arXiv:2505.07935, \dodoi{10.48550/arXiv.2505.07935}

\bibitem[{{Kohandel} {et~al.}(2020){Kohandel}, {Pallottini}, {Ferrara},
  {Carniani}, {Gallerani}, {Vallini}, {Zanella}, \& {Behrens}}]{Kohandel:2020}
{Kohandel}, M., {Pallottini}, A., {Ferrara}, A., {et~al.} 2020, \mnras, 499,
  1250, \dodoi{10.1093/mnras/staa2792}

\bibitem[{{Kohandel} {et~al.}(2019){Kohandel}, {Pallottini}, {Ferrara},
  {Zanella}, {Behrens}, {Carniani}, {Gallerani}, \& {Vallini}}]{Kohandel:2019}
---. 2019, \mnras, 487, 3007, \dodoi{10.1093/mnras/stz1486}

\bibitem[{{Kumari} {et~al.}(2024){Kumari}, {Smit}, {Leitherer}, {Witstok},
  {Irwin}, {Sirianni}, \& {Aloisi}}]{Kumari:2024}
{Kumari}, N., {Smit}, R., {Leitherer}, C., {et~al.} 2024, \mnras, 529, 781,
  \dodoi{10.1093/mnras/stae252}

\bibitem[{{Lagache} {et~al.}(2018){Lagache}, {Cousin}, \&
  {Chatzikos}}]{Lagache:2018}
{Lagache}, G., {Cousin}, M., \& {Chatzikos}, M. 2018, \aap, 609, A130,
  \dodoi{10.1051/0004-6361/201732019}

\bibitem[{{Laporte} {et~al.}(2019){Laporte}, {Katz}, {Ellis}, {Lagache},
  {Bauer}, {Boone}, {Inoue}, {Hashimoto}, {Matsuo}, {Mawatari}, \&
  {Tamura}}]{Laporte:2019}
{Laporte}, N., {Katz}, H., {Ellis}, R.~S., {et~al.} 2019, \mnras, 487, L81,
  \dodoi{10.1093/mnrasl/slz094}

\bibitem[{{Leitherer} {et~al.}(1999){Leitherer}, {Schaerer}, {Goldader},
  {Delgado}, {Robert}, {Kune}, {de Mello}, {Devost}, \&
  {Heckman}}]{Leitherer:1999}
{Leitherer}, C., {Schaerer}, D., {Goldader}, J.~D., {et~al.} 1999, \apjs, 123,
  3, \dodoi{10.1086/313233}

\bibitem[{{Liang} {et~al.}(2024){Liang}, {Feldmann}, {Murray}, {Narayanan},
  {Hayward}, {Angl{\'e}s-Alc{\'a}zar}, {Bassini}, {Richings},
  {Faucher-Gigu{\`e}re}, {Chung}, {Chan}, {Tolgay}, {{\c{C}}atmabacak},
  {Kere{\v{s}}}, \& {Hopkins}}]{Liang:2024}
{Liang}, L., {Feldmann}, R., {Murray}, N., {et~al.} 2024, \mnras, 528, 499,
  \dodoi{10.1093/mnras/stad3792}

\bibitem[{{Maiolino} {et~al.}(2015){Maiolino}, {Carniani}, {Fontana},
  {Vallini}, {Pentericci}, {Ferrara}, {Vanzella}, {Grazian}, {Gallerani},
  {Castellano}, {Cristiani}, {Brammer}, {Santini}, {Wagg}, \&
  {Williams}}]{Maiolino:2015}
{Maiolino}, R., {Carniani}, S., {Fontana}, A., {et~al.} 2015, \mnras, 452, 54,
  \dodoi{10.1093/mnras/stv1194}

\bibitem[{{Malhotra} {et~al.}(2001){Malhotra}, {Kaufman}, {Hollenbach},
  {Helou}, {Rubin}, {Brauher}, {Dale}, {Lu}, {Lord}, {Stacey}, {Contursi},
  {Hunter}, \& {Dinerstein}}]{Malhotra:2001}
{Malhotra}, S., {Kaufman}, M.~J., {Hollenbach}, D., {et~al.} 2001, \apj, 561,
  766, \dodoi{10.1086/323046}

\bibitem[{{Maraston}(2005)}]{Maraston:2005}
{Maraston}, C. 2005, \mnras, 362, 799, \dodoi{10.1111/j.1365-2966.2005.09270.x}

\bibitem[{{Mathis} {et~al.}(1983){Mathis}, {Mezger}, \&
  {Panagia}}]{Mathis:1983}
{Mathis}, J.~S., {Mezger}, P.~G., \& {Panagia}, N. 1983, \aap, 128, 212

\bibitem[{{Matthee} {et~al.}(2017){Matthee}, {Sobral}, {Boone},
  {R{\"o}ttgering}, {Schaerer}, {Girard}, {Pallottini}, {Vallini}, {Ferrara},
  {Darvish}, \& {Mobasher}}]{Matthee:2017}
{Matthee}, J., {Sobral}, D., {Boone}, F., {et~al.} 2017, \apj, 851, 145,
  \dodoi{10.3847/1538-4357/aa9931}

\bibitem[{{Matthee} {et~al.}(2019){Matthee}, {Sobral}, {Boogaard},
  {R{\"o}ttgering}, {Vallini}, {Ferrara}, {Paulino-Afonso}, {Boone},
  {Schaerer}, \& {Mobasher}}]{Matthee:2019}
{Matthee}, J., {Sobral}, D., {Boogaard}, L.~A., {et~al.} 2019, \apj, 881, 124,
  \dodoi{10.3847/1538-4357/ab2f81}

\bibitem[{{Mawatari} {et~al.}(2025){Mawatari}, {Costantin}, {Usui},
  {Hashimoto}, {{\'A}lvarez-M{\'a}rquez}, {Sugahara}, {Colina}, {Inoue},
  {Osone}, {Arribas}, {Marques-Chaves}, {Nakazato}, {Hagimoto}, {Hashigaya},
  {Ceverino}, {Yoshida}, {Bakx}, {Fudamoto}, {Crespo G{\'o}mez}, {Matsuo},
  {Pereira-Santaella}, {Blanco-Prieto}, {Ren}, \& {Tamura}}]{Mawatari:2025}
{Mawatari}, K., {Costantin}, L., {Usui}, M., {et~al.} 2025, arXiv e-prints,
  arXiv:2507.02053, \dodoi{10.48550/arXiv.2507.02053}

\bibitem[{{Moriwaki} {et~al.}(2018){Moriwaki}, {Yoshida}, {Shimizu},
  {Harikane}, {Matsuda}, {Matsuo}, {Hashimoto}, {Inoue}, {Tamura}, \&
  {Nagao}}]{Moriwaki:2018}
{Moriwaki}, K., {Yoshida}, N., {Shimizu}, I., {et~al.} 2018, \mnras, 481, L84,
  \dodoi{10.1093/mnrasl/sly167}

\bibitem[{{Mowla} {et~al.}(2024){Mowla}, {Iyer}, {Asada}, {Desprez}, {Tan},
  {Martis}, {Sarrouh}, {Strait}, {Abraham}, {Brada{\v{c}}}, {Brammer},
  {Muzzin}, {Pacifici}, {Ravindranath}, {Sawicki}, {Willott},
  {Estrada-Carpenter}, {Jahan}, {Noirot}, {Matharu}, {Rihtar{\v{s}}i{\v{c}}},
  \& {Zabl}}]{Mowla:2024}
{Mowla}, L., {Iyer}, K., {Asada}, Y., {et~al.} 2024, \nat, 636, 332,
  \dodoi{10.1038/s41586-024-08293-0}

\bibitem[{{Mu{\~n}oz-Elgueta} {et~al.}(2024){Mu{\~n}oz-Elgueta}, {Arrigoni
  Battaia}, {Kauffmann}, {Pakmor}, {Walch}, {Obreja}, \&
  {Buhlmann}}]{Munoz-Elgueta:2024}
{Mu{\~n}oz-Elgueta}, N., {Arrigoni Battaia}, F., {Kauffmann}, G., {et~al.}
  2024, \aap, 690, A392, \dodoi{10.1051/0004-6361/202450049}

\bibitem[{{Nagao} {et~al.}(2012){Nagao}, {Maiolino}, {De Breuck}, {Caselli},
  {Hatsukade}, \& {Saigo}}]{Nagao:2012}
{Nagao}, T., {Maiolino}, R., {De Breuck}, C., {et~al.} 2012, \aap, 542, L34,
  \dodoi{10.1051/0004-6361/201219518}

\bibitem[{{Nakazato} {et~al.}(2023){Nakazato}, {Yoshida}, \&
  {Ceverino}}]{Nakazato:2023}
{Nakazato}, Y., {Yoshida}, N., \& {Ceverino}, D. 2023, \apj, 953, 140,
  \dodoi{10.3847/1538-4357/ace25a}

\bibitem[{{Neri} {et~al.}(2014){Neri}, {Downes}, {Cox}, \&
  {Walter}}]{Neri:2014}
{Neri}, R., {Downes}, D., {Cox}, P., \& {Walter}, F. 2014, \aap, 562, A35,
  \dodoi{10.1051/0004-6361/201322528}

\bibitem[{{Nomoto} {et~al.}(2006){Nomoto}, {Tominaga}, {Umeda}, {Kobayashi}, \&
  {Maeda}}]{Nomoto:2006}
{Nomoto}, K., {Tominaga}, N., {Umeda}, H., {Kobayashi}, C., \& {Maeda}, K.
  2006, \nphysa, 777, 424, \dodoi{10.1016/j.nuclphysa.2006.05.008}

\bibitem[{{Nussbaumer} \& {Storey}(1981)}]{Nussbaumer_Storey:1981}
{Nussbaumer}, H., \& {Storey}, P.~J. 1981, \aap, 96, 91

\bibitem[{{Nyhagen} {et~al.}(2024){Nyhagen}, {Schimek}, {Cicone}, {Decataldo},
  \& {Shen}}]{Nyhagen:2024}
{Nyhagen}, C.~T., {Schimek}, A., {Cicone}, C., {Decataldo}, D., \& {Shen}, S.
  2024, arXiv e-prints, arXiv:2410.18471, \dodoi{10.48550/arXiv.2410.18471}

\bibitem[{{Olsen} {et~al.}(2017){Olsen}, {Greve}, {Narayanan}, {Thompson},
  {Dav{\'e}}, {Niebla Rios}, \& {Stawinski}}]{Olsen:2017}
{Olsen}, K., {Greve}, T.~R., {Narayanan}, D., {et~al.} 2017, \apj, 846, 105,
  \dodoi{10.3847/1538-4357/aa86b4}

\bibitem[{{Omukai} {et~al.}(2008){Omukai}, {Schneider}, \&
  {Haiman}}]{Omukai:2008}
{Omukai}, K., {Schneider}, R., \& {Haiman}, Z. 2008, \apj, 686, 801,
  \dodoi{10.1086/591636}

\bibitem[{{Omukai} {et~al.}(2005){Omukai}, {Tsuribe}, {Schneider}, \&
  {Ferrara}}]{Omukai:2005}
{Omukai}, K., {Tsuribe}, T., {Schneider}, R., \& {Ferrara}, A. 2005, \apj, 626,
  627, \dodoi{10.1086/429955}

\bibitem[{{Osterbrock} \& {Ferland}(2006)}]{Osterbrock:2006}
{Osterbrock}, D.~E., \& {Ferland}, G.~J. 2006, {Astrophysics of gaseous nebulae
  and active galactic nuclei}

\bibitem[{{Ota} {et~al.}(2014){Ota}, {Walter}, {Ohta}, {Hatsukade}, {Carilli},
  {da Cunha}, {Gonz{\'a}lez-L{\'o}pez}, {Decarli}, {Hodge}, {Nagai}, {Egami},
  {Jiang}, {Iye}, {Kashikawa}, {Riechers}, {Bertoldi}, {Cox}, {Neri}, \&
  {Weiss}}]{Ota:2014}
{Ota}, K., {Walter}, F., {Ohta}, K., {et~al.} 2014, \apj, 792, 34,
  \dodoi{10.1088/0004-637X/792/1/34}

\bibitem[{{Pallottini} {et~al.}(2019){Pallottini}, {Ferrara}, {Decataldo},
  {Gallerani}, {Vallini}, {Carniani}, {Behrens}, {Kohandel}, \&
  {Salvadori}}]{Pallottini:2019}
{Pallottini}, A., {Ferrara}, A., {Decataldo}, D., {et~al.} 2019, \mnras, 487,
  1689, \dodoi{10.1093/mnras/stz1383}

\bibitem[{{Pallottini} {et~al.}(2022){Pallottini}, {Ferrara}, {Gallerani},
  {Behrens}, {Kohandel}, {Carniani}, {Vallini}, {Salvadori}, {Gelli},
  {Sommovigo}, {D'Odorico}, {Di Mascia}, \& {Pizzati}}]{Pallottini:2022}
{Pallottini}, A., {Ferrara}, A., {Gallerani}, S., {et~al.} 2022, \mnras, 513,
  5621, \dodoi{10.1093/mnras/stac1281}

\bibitem[{{Park} {et~al.}(2021){Park}, {Ricotti}, \& {Sugimura}}]{Park:2021}
{Park}, J., {Ricotti}, M., \& {Sugimura}, K. 2021, \mnras, 508, 6176,
  \dodoi{10.1093/mnras/stab2999}

\bibitem[{{Pascale} {et~al.}(2025){Pascale}, {Calura}, {Vesperini}, {Rosdahl},
  {Nipoti}, {Giunchi}, {Lacchin}, {Lupi}, {Messa}, {Meneghetti}, {Ragagnin},
  {Vanzella}, \& {Zanella}}]{Pascale:2025}
{Pascale}, R., {Calura}, F., {Vesperini}, E., {et~al.} 2025, \aap, 699, A31,
  \dodoi{10.1051/0004-6361/202453252}

\bibitem[{{Pentericci} {et~al.}(2016){Pentericci}, {Carniani}, {Castellano},
  {Fontana}, {Maiolino}, {Guaita}, {Vanzella}, {Grazian}, {Santini}, {Yan},
  {Cristiani}, {Conselice}, {Giavalisco}, {Hathi}, \&
  {Koekemoer}}]{Pentericci:2016}
{Pentericci}, L., {Carniani}, S., {Castellano}, M., {et~al.} 2016, \apjl, 829,
  L11, \dodoi{10.3847/2041-8205/829/1/L11}

\bibitem[{{Popping} {et~al.}(2019){Popping}, {Narayanan}, {Somerville},
  {Faisst}, \& {Krumholz}}]{Popping:2019}
{Popping}, G., {Narayanan}, D., {Somerville}, R.~S., {Faisst}, A.~L., \&
  {Krumholz}, M.~R. 2019, \mnras, 482, 4906, \dodoi{10.1093/mnras/sty2969}

\bibitem[{{Ramos Padilla} {et~al.}(2023){Ramos Padilla}, {Wang}, {van der Tak},
  \& {Trager}}]{Ramod-Padilla:2023}
{Ramos Padilla}, A.~F., {Wang}, L., {van der Tak}, F.~F.~S., \& {Trager}, S.~C.
  2023, \aap, 679, A131, \dodoi{10.1051/0004-6361/202243358}

\bibitem[{{Ren} {et~al.}(2025){Ren}, {Inoue}, {{\'A}lvarez-M{\'a}rquez},
  {Hashimoto}, {Colina}, {Sugahara}, {Costantin}, {Mawatari}, {Fudamoto},
  {Arribas}, {Crespo G{\'o}mez}, {Ceverino}, {Nakazato}, {Hagimoto}, {Usui},
  {Marques-Chaves}, {Matsuo}, {Hashigaya}, {Osone}, {Blanco-Prieto}, {Tamura},
  {Yoshida}, {Bakx}, \& {Pereira-Santaella}}]{Ren:2025}
{Ren}, Y.~W., {Inoue}, A.~K., {{\'A}lvarez-M{\'a}rquez}, J., {et~al.} 2025,
  arXiv e-prints, arXiv:2510.25721, \dodoi{10.48550/arXiv.2510.25721}

\bibitem[{{Ricotti}(2016)}]{Ricotti:2016}
{Ricotti}, M. 2016, \mnras, 462, 601, \dodoi{10.1093/mnras/stw1672}

\bibitem[{{Ricotti} {et~al.}(2022){Ricotti}, {Polisensky}, \&
  {Cleland}}]{Ricotti:2022}
{Ricotti}, M., {Polisensky}, E., \& {Cleland}, E. 2022, \mnras, 515, 302,
  \dodoi{10.1093/mnras/stac1485}

\bibitem[{{Rizzo} {et~al.}(2022){Rizzo}, {Kohandel}, {Pallottini}, {Zanella},
  {Ferrara}, {Vallini}, \& {Toft}}]{Rizzo:2022}
{Rizzo}, F., {Kohandel}, M., {Pallottini}, A., {et~al.} 2022, \aap, 667, A5,
  \dodoi{10.1051/0004-6361/202243582}

\bibitem[{{Rosdahl} {et~al.}(2013){Rosdahl}, {Blaizot}, {Aubert}, {Stranex}, \&
  {Teyssier}}]{Rosdahl:2013}
{Rosdahl}, J., {Blaizot}, J., {Aubert}, D., {Stranex}, T., \& {Teyssier}, R.
  2013, \mnras, 436, 2188, \dodoi{10.1093/mnras/stt1722}

\bibitem[{{Rowland} {et~al.}(2025){Rowland}, {Stefanon}, {Bouwens}, {Hodge},
  {Algera}, {Fisher}, {Dayal}, {Pallottini}, {Stark}, {Heintz}, {Aravena},
  {Bowler}, {Cescon}, {Endsley}, {Ferrara}, {Gonzalez}, {Graziani}, {Gulis},
  {Herard-Demanche}, {Inami}, {Laza-Ramos}, {van Leeuwen}, {de Looze},
  {Nanayakkara}, {Oesch}, {Ormerod}, {Sartorio}, {Schouws}, {Smit},
  {Sommovigo}, {Toft}, {Weaver}, \& {van der Werf}}]{Rowland:2025}
{Rowland}, L.~E., {Stefanon}, M., {Bouwens}, R., {et~al.} 2025, arXiv e-prints,
  arXiv:2501.10559, \dodoi{10.48550/arXiv.2501.10559}

\bibitem[{{Salpeter}(1955)}]{Salpeter:1955}
{Salpeter}, E.~E. 1955, \apj, 121, 161, \dodoi{10.1086/145971}

\bibitem[{{Schaerer}(2002)}]{Schaerer:2002}
{Schaerer}, D. 2002, \aap, 382, 28, \dodoi{10.1051/0004-6361:20011619}

\bibitem[{{Schaerer} {et~al.}(2015){Schaerer}, {Boone}, {Zamojski}, {Staguhn},
  {Dessauges-Zavadsky}, {Finkelstein}, \& {Combes}}]{Schaerer:2015}
{Schaerer}, D., {Boone}, F., {Zamojski}, M., {et~al.} 2015, \aap, 574, A19,
  \dodoi{10.1051/0004-6361/201424649}

\bibitem[{{Schaerer} {et~al.}(2020){Schaerer}, {Ginolfi}, {B{\'e}thermin},
  {Fudamoto}, {Oesch}, {Le F{\`e}vre}, {Faisst}, {Capak}, {Cassata},
  {Silverman}, {Yan}, {Jones}, {Amorin}, {Bardelli}, {Boquien}, {Cimatti},
  {Dessauges-Zavadsky}, {Giavalisco}, {Hathi}, {Fujimoto}, {Ibar}, {Koekemoer},
  {Lagache}, {Lemaux}, {Loiacono}, {Maiolino}, {Narayanan}, {Morselli},
  {M{\'e}ndez-Hern{\`a}ndez}, {Pozzi}, {Riechers}, {Talia}, {Toft}, {Vallini},
  {Vergani}, {Zamorani}, \& {Zucca}}]{Schaerer:2020}
{Schaerer}, D., {Ginolfi}, M., {B{\'e}thermin}, M., {et~al.} 2020, \aap, 643,
  A3, \dodoi{10.1051/0004-6361/202037617}

\bibitem[{{Schimek} {et~al.}(2024){Schimek}, {Decataldo}, {Shen}, {Cicone},
  {Baumschlager}, {van Kampen}, {Klaassen}, {Madau}, {Di Mascolo}, {Mayer},
  {Montoya Arroyave}, {Mroczkowski}, \& {Warraich}}]{Schimek:2024}
{Schimek}, A., {Decataldo}, D., {Shen}, S., {et~al.} 2024, \aap, 682, A98,
  \dodoi{10.1051/0004-6361/202346945}

\bibitem[{{Schouws} {et~al.}(2024){Schouws}, {Bouwens}, {Ormerod}, {Smit},
  {Algera}, {Sommovigo}, {Hodge}, {Ferrara}, {Oesch}, {Rowland}, {van Leeuwen},
  {Stefanon}, {Herard-Demanche}, {Fudamoto}, {R{\"o}ttgering}, \& {van der
  Werf}}]{Schouws:2024}
{Schouws}, S., {Bouwens}, R.~J., {Ormerod}, K., {et~al.} 2024, arXiv e-prints,
  arXiv:2409.20549, \dodoi{10.48550/arXiv.2409.20549}

\bibitem[{{Schouws} {et~al.}(2025){Schouws}, {Bouwens}, {Algera}, {Smit},
  {Kumari}, {Rowland}, {van Leeuwen}, {Sommovigo}, {Ferrara}, {Oesch},
  {Ormerod}, {Stefanon}, {Herard-Demanche}, {Hodge}, {Fudamoto},
  {R{\"o}ttgering}, \& {van der Werf}}]{Schouws:2025}
{Schouws}, S., {Bouwens}, R.~J., {Algera}, H., {et~al.} 2025, arXiv e-prints,
  arXiv:2502.01610, \dodoi{10.48550/arXiv.2502.01610}

\bibitem[{{Smit} {et~al.}(2018){Smit}, {Bouwens}, {Carniani}, {Oesch},
  {Labb{\'e}}, {Illingworth}, {van der Werf}, {Bradley}, {Gonzalez}, {Hodge},
  {Holwerda}, {Maiolino}, \& {Zheng}}]{Smit:2018}
{Smit}, R., {Bouwens}, R.~J., {Carniani}, S., {et~al.} 2018, \nat, 553, 178,
  \dodoi{10.1038/nature24631}

\bibitem[{{Smith} {et~al.}(2017){Smith}, {Bryan}, {Glover}, {Goldbaum}, {Turk},
  {Regan}, {Wise}, {Schive}, {Abel}, {Emerick}, {O'Shea}, {Anninos}, {Hummels},
  \& {Khochfar}}]{Smith:2017}
{Smith}, B.~D., {Bryan}, G.~L., {Glover}, S. C.~O., {et~al.} 2017, \mnras, 466,
  2217, \dodoi{10.1093/mnras/stw3291}

\bibitem[{{Sommovigo} {et~al.}(2022){Sommovigo}, {Ferrara}, {Pallottini},
  {Dayal}, {Bouwens}, {Smit}, {da Cunha}, {De Looze}, {Bowler}, {Hodge},
  {Inami}, {Oesch}, {Endsley}, {Gonzalez}, {Schouws}, {Stark}, {Stefanon},
  {Aravena}, {Graziani}, {Riechers}, {Schneider}, {van der Werf}, {Algera},
  {Barrufet}, {Fudamoto}, {Hygate}, {Labb{\'e}}, {Li}, {Nanayakkara}, \&
  {Topping}}]{Sommovigo:2022}
{Sommovigo}, L., {Ferrara}, A., {Pallottini}, A., {et~al.} 2022, \mnras, 513,
  3122, \dodoi{10.1093/mnras/stac302}

\bibitem[{{Stacey} {et~al.}(2010){Stacey}, {Hailey-Dunsheath}, {Ferkinhoff},
  {Nikola}, {Parshley}, {Benford}, {Staguhn}, \& {Fiolet}}]{Stacey:2010}
{Stacey}, G.~J., {Hailey-Dunsheath}, S., {Ferkinhoff}, C., {et~al.} 2010, \apj,
  724, 957, \dodoi{10.1088/0004-637X/724/2/957}

\bibitem[{{Stanway} \& {Eldridge}(2018)}]{Stanway:2018}
{Stanway}, E.~R., \& {Eldridge}, J.~J. 2018, \mnras, 479, 75,
  \dodoi{10.1093/mnras/sty1353}

\bibitem[{{Stark} {et~al.}(2017){Stark}, {Ellis}, {Charlot}, {Chevallard},
  {Tang}, {Belli}, {Zitrin}, {Mainali}, {Gutkin}, {Vidal-Garc{\'\i}a},
  {Bouwens}, \& {Oesch}}]{Stark:2017}
{Stark}, D.~P., {Ellis}, R.~S., {Charlot}, S., {et~al.} 2017, \mnras, 464, 469,
  \dodoi{10.1093/mnras/stw2233}

\bibitem[{{Stiavelli} {et~al.}(2023){Stiavelli}, {Morishita}, {Chiaberge},
  {Grillo}, {Leethochawalit}, {Rosati}, {Schuldt}, {Trenti}, \&
  {Treu}}]{Stiavelli:2023}
{Stiavelli}, M., {Morishita}, T., {Chiaberge}, M., {et~al.} 2023, \apjl, 957,
  L18, \dodoi{10.3847/2041-8213/ad0159}

\bibitem[{{Sugimura} {et~al.}(2020){Sugimura}, {Matsumoto}, {Hosokawa},
  {Hirano}, \& {Omukai}}]{Sugimura:2020}
{Sugimura}, K., {Matsumoto}, T., {Hosokawa}, T., {Hirano}, S., \& {Omukai}, K.
  2020, \apjl, 892, L14, \dodoi{10.3847/2041-8213/ab7d37}

\bibitem[{{Sugimura} {et~al.}(2023){Sugimura}, {Matsumoto}, {Hosokawa},
  {Hirano}, \& {Omukai}}]{Sugimura:2023}
---. 2023, \apj, 959, 17, \dodoi{10.3847/1538-4357/ad02fc}

\bibitem[{{Sugimura} {et~al.}(2024){Sugimura}, {Ricotti}, {Park}, {Garcia}, \&
  {Yajima}}]{Sugimura:2024}
{Sugimura}, K., {Ricotti}, M., {Park}, J., {Garcia}, F. A.~B., \& {Yajima}, H.
  2024, arXiv e-prints, arXiv:2403.04824, \dodoi{10.48550/arXiv.2403.04824}

\bibitem[{{Tamura} {et~al.}(2019){Tamura}, {Mawatari}, {Hashimoto}, {Inoue},
  {Zackrisson}, {Christensen}, {Binggeli}, {Matsuda}, {Matsuo}, {Takeuchi},
  {Asano}, {Sunaga}, {Shimizu}, {Okamoto}, {Yoshida}, {Lee}, {Shibuya},
  {Taniguchi}, {Umehata}, {Hatsukade}, {Kohno}, \& {Ota}}]{Tamura:2019}
{Tamura}, Y., {Mawatari}, K., {Hashimoto}, T., {et~al.} 2019, \apj, 874, 27,
  \dodoi{10.3847/1538-4357/ab0374}

\bibitem[{{Teyssier}(2002)}]{Teyssier:2002}
{Teyssier}, R. 2002, \aap, 385, 337, \dodoi{10.1051/0004-6361:20011817}

\bibitem[{{Tielens}(2005)}]{Tielens:2005}
{Tielens}, A.~G.~G.~M. 2005, {The Physics and Chemistry of the Interstellar
  Medium}

\bibitem[{{Toyouchi} {et~al.}(2025){Toyouchi}, {Yajima}, {Ferrara}, \&
  {Nagamine}}]{Toyouchi:2025}
{Toyouchi}, D., {Yajima}, H., {Ferrara}, A., \& {Nagamine}, K. 2025, \mnras,
  \dodoi{10.1093/mnras/staf1182}

\bibitem[{{Truelove} {et~al.}(1997){Truelove}, {Klein}, {McKee}, {Holliman},
  {Howell}, \& {Greenough}}]{Truelove:1997}
{Truelove}, J.~K., {Klein}, R.~I., {McKee}, C.~F., {et~al.} 1997, \apjl, 489,
  L179, \dodoi{10.1086/310975}

\bibitem[{{Tsuna} {et~al.}(2023){Tsuna}, {Nakazato}, \& {Hartwig}}]{Tsuna:2023}
{Tsuna}, D., {Nakazato}, Y., \& {Hartwig}, T. 2023, \mnras, 526, 4801,
  \dodoi{10.1093/mnras/stad3043}

\bibitem[{{Vallini} {et~al.}(2017){Vallini}, {Ferrara}, {Pallottini}, \&
  {Gallerani}}]{Vallini:2017}
{Vallini}, L., {Ferrara}, A., {Pallottini}, A., \& {Gallerani}, S. 2017,
  \mnras, 467, 1300, \dodoi{10.1093/mnras/stx180}

\bibitem[{{Vallini} {et~al.}(2013){Vallini}, {Gallerani}, {Ferrara}, \&
  {Baek}}]{Vallini:2013}
{Vallini}, L., {Gallerani}, S., {Ferrara}, A., \& {Baek}, S. 2013, \mnras, 433,
  1567, \dodoi{10.1093/mnras/stt828}

\bibitem[{{Vallini} {et~al.}(2015){Vallini}, {Gallerani}, {Ferrara},
  {Pallottini}, \& {Yue}}]{Vallini:2015}
{Vallini}, L., {Gallerani}, S., {Ferrara}, A., {Pallottini}, A., \& {Yue}, B.
  2015, \apj, 813, 36, \dodoi{10.1088/0004-637X/813/1/36}

\bibitem[{{Vallini} {et~al.}(2025){Vallini}, {Pallottini}, {Kohandel},
  {Sommovigo}, {Ferrara}, {Bethermin}, {Herrera-Camus}, {Carniani}, {Faisst},
  {Zanella}, {Pozzi}, {Dessauges-Zavadsky}, {Gruppioni}, {Veraldi}, \&
  {Accard}}]{Vallini:2025}
{Vallini}, L., {Pallottini}, A., {Kohandel}, M., {et~al.} 2025, arXiv e-prints,
  arXiv:2504.14001.
\newblock \doarXiv{2504.14001}

\bibitem[{{Vanzella} {et~al.}(2023){Vanzella}, {Claeyssens}, {Welch}, {Adamo},
  {Coe}, {Diego}, {Mahler}, {Khullar}, {Kokorev}, {Oguri}, {Ravindranath},
  {Furtak}, {Hsiao}, {Abdurro'uf}, {Mandelker}, {Brammer}, {Bradley},
  {Brada{\v{c}}}, {Conselice}, {Dayal}, {Nonino}, {Andrade-Santos},
  {Windhorst}, {Pirzkal}, {Sharon}, {de Mink}, {Fujimoto}, {Zitrin},
  {Eldridge}, \& {Norman}}]{Vanzella:2023_sunrise_arc}
{Vanzella}, E., {Claeyssens}, A., {Welch}, B., {et~al.} 2023, \apj, 945, 53,
  \dodoi{10.3847/1538-4357/acb59a}

\bibitem[{{Watson} {et~al.}(2015){Watson}, {Christensen}, {Knudsen}, {Richard},
  {Gallazzi}, \& {Micha{\l}owski}}]{Watson:2015}
{Watson}, D., {Christensen}, L., {Knudsen}, K.~K., {et~al.} 2015, \nat, 519,
  327, \dodoi{10.1038/nature14164}

\bibitem[{{Welch} {et~al.}(2023){Welch}, {Coe}, {Zitrin}, {Diego}, {Windhorst},
  {Mandelker}, {Vanzella}, {Ravindranath}, {Zackrisson}, {Florian}, {Bradley},
  {Sharon}, {Brada{\v{c}}}, {Rigby}, {Frye}, \& {Fujimoto}}]{Welch:2023}
{Welch}, B., {Coe}, D., {Zitrin}, A., {et~al.} 2023, \apj, 943, 2,
  \dodoi{10.3847/1538-4357/aca8a8}

\bibitem[{{Wise} {et~al.}(2012){Wise}, {Turk}, {Norman}, \& {Abel}}]{Wise:2012}
{Wise}, J.~H., {Turk}, M.~J., {Norman}, M.~L., \& {Abel}, T. 2012, \apj, 745,
  50, \dodoi{10.1088/0004-637X/745/1/50}

\bibitem[{{Witstok} {et~al.}(2022){Witstok}, {Smit}, {Maiolino}, {Kumari},
  {Aravena}, {Boogaard}, {Bouwens}, {Carniani}, {Hodge}, {Jones}, {Stefanon},
  {van der Werf}, \& {Schouws}}]{Witstok:2022}
{Witstok}, J., {Smit}, R., {Maiolino}, R., {et~al.} 2022, \mnras,
  \dodoi{10.1093/mnras/stac1905}

\bibitem[{{Witstok} {et~al.}(2025){Witstok}, {Smit}, {Baker}, {Rinaldi},
  {Hainline}, {Algera}, {Arribas}, {Bakx}, {Bunker}, {Carniani}, {Charlot},
  {Chevallard}, {Curti}, {Curtis-Lake}, {Eisenstein}, {Heintz}, {Helton},
  {Jones}, {Maiolino}, {Maseda}, {P{\'e}rez-Gonz{\'a}lez}, {Pollock},
  {Robertson}, {Saxena}, {Scholtz}, {Shivaei}, {Sun}, {Tacchella}, {{\"U}bler},
  {Watson}, {Willott}, \& {Wu}}]{Witstok:2025}
{Witstok}, J., {Smit}, R., {Baker}, W.~M., {et~al.} 2025, arXiv e-prints,
  arXiv:2507.22888, \dodoi{10.48550/arXiv.2507.22888}

\bibitem[{{Wolfire} {et~al.}(1995){Wolfire}, {Hollenbach}, {McKee}, {Tielens},
  \& {Bakes}}]{Wolfire:1995}
{Wolfire}, M.~G., {Hollenbach}, D., {McKee}, C.~F., {Tielens}, A.~G.~G.~M., \&
  {Bakes}, E.~L.~O. 1995, \apj, 443, 152, \dodoi{10.1086/175510}

\bibitem[{{Wong} {et~al.}(2022){Wong}, {Wang}, {Hashimoto}, {Takagi}, {Goto},
  {Kim}, {Wu}, {On}, {Santos}, {Lu}, {Kilerci-Eser}, {Ho}, \&
  {Hsiao}}]{Wong:2022}
{Wong}, Y. H.~V., {Wang}, P., {Hashimoto}, T., {et~al.} 2022, \apj, 929, 161,
  \dodoi{10.3847/1538-4357/ac5cc7}

\bibitem[{{Woosley} {et~al.}(2002){Woosley}, {Heger}, \&
  {Weaver}}]{Woosley:2002}
{Woosley}, S.~E., {Heger}, A., \& {Weaver}, T.~A. 2002, Reviews of Modern
  Physics, 74, 1015, \dodoi{10.1103/RevModPhys.74.1015}

\bibitem[{{Xiao} {et~al.}(2018){Xiao}, {Stanway}, \& {Eldridge}}]{Xiao:2018}
{Xiao}, L., {Stanway}, E.~R., \& {Eldridge}, J.~J. 2018, \mnras, 477, 904,
  \dodoi{10.1093/mnras/sty646}

\bibitem[{{Yanagisawa} {et~al.}(2024){Yanagisawa}, {Ouchi}, {Nakajima},
  {Harikane}, {Fujimoto}, {Ono}, {Umeda}, {Nakane}, {Yajima}, {Fukushima}, \&
  {Xu}}]{Yanagisawa:2024}
{Yanagisawa}, H., {Ouchi}, M., {Nakajima}, K., {et~al.} 2024, arXiv e-prints,
  arXiv:2411.19893, \dodoi{10.48550/arXiv.2411.19893}

\bibitem[{{Yoshida} {et~al.}(2003){Yoshida}, {Abel}, {Hernquist}, \&
  {Sugiyama}}]{Yoshida:2003}
{Yoshida}, N., {Abel}, T., {Hernquist}, L., \& {Sugiyama}, N. 2003, \apj, 592,
  645, \dodoi{10.1086/375810}

\bibitem[{{Zavala} {et~al.}(2024){Zavala}, {Bakx}, {Mitsuhashi}, {Castellano},
  {Calabro}, {Akins}, {Buat}, {Casey}, {Fernandez-Arenas}, {Franco}, {Fontana},
  {Hatsukade}, {Ho}, {Ikeda}, {Kartaltepe}, {Koekemoer}, {McKinney},
  {Napolitano}, {P{\'e}rez-Gonz{\'a}lez}, {Santini}, {Serjeant}, {Terlevich},
  {Terlevich}, \& {Yung}}]{Zavala:2024}
{Zavala}, J.~A., {Bakx}, T., {Mitsuhashi}, I., {et~al.} 2024, \apjl, 977, L9,
  \dodoi{10.3847/2041-8213/ad8f38}

\end{thebibliography}
\bibliographystyle{aasjournal}

\end{document}